\documentclass[12pt]{article}
\usepackage{epsf}
\def\bea{\begin{eqnarray}}
\def\eea{\end{eqnarray}}
\def\be{\begin{equation}}
\def\ee{\end{equation}}
\def\no{\nonumber}

\newcommand{\beq}{\begin{equation}}
\newcommand{\eeq}{\end{equation}}
\newcommand{\beqa}{\begin{eqnarray}}
\newcommand{\eeqa}{\end{eqnarray}}

\newcommand{\krig}[1]{\stackrel{\circ}{#1}}
\newcommand{\barr}[1]{\not\mathrel #1}

\newcommand{\dfrac}{\displaystyle \frac}
\newcommand{\vs}{\vspace{-0.20cm}}

\begin{document}
\newpage
\baselineskip 14pt plus 2pt minus 2pt

\thispagestyle{empty}

\par
\topmargin=-1cm      % distance from top of the page to first line
                     % of text plus one inch
%\vspace*{.5in}

{ \small

\hfill{FZJ-IKP(TH)-1999-15}

%\hfill{hep-ph/9703218}

}

\vspace{50.0pt}

\begin{centering}
{\Large\bf One--loop analysis of the reaction 
{\boldmath $\pi N \to \pi \pi N$}}\\
% above threshold in \\[0.3em]
%chiral perturbation theory}\\

\vspace{40.0pt}
{
{\bf Nadia~Fettes}$^a$\footnote{E-mail address: N.Fettes@fz-juelich.de},
{\bf V\'eronique~Bernard}$^b$\footnote{E-mail address: bernard@lpt1.u-strasbg.fr},
{\bf Ulf-G.~Mei{\ss}ner}$^a$}\footnote{E-mail address: Ulf-G.Meissner@fz-juelich.de}\\
\vspace{20.0pt}
{\it $^{a}$Institut f\"ur Kernphysik (Theorie), Forschungszentrum J\"ulich\\ 
D-52425 J\"ulich, Germany} \\
\vspace{15.0pt}
{\it $^{b}$Laboratoire de Physique Th\'eorique, Universit\'e Louis
  Pasteur\\ 3-5, rue de l'Universit\'e, F-67084 Strasbourg Cedex, France} \\
\end{centering}
\vspace{40.0pt}
\begin{center}
\begin{abstract}
\noindent Single pion production off nucleons is studied in
the framework of heavy  baryon chiral perturbation theory 
to third order in the chiral expansion. Using total and some older
differential cross section data to pin down the low--energy constants,
most of the recent differential cross sections and angular correlation
functions can be described as well as total cross sections at higher
energies. We show that the contributions from the one loop graphs are
essentially negligible and that the dominant terms at second and third
order are related to pion--nucleon scattering graphs with one pion added.
We also discuss the possibility of extracting the pion--pion S--wave
scattering lengths from the unpolarized data.
\end{abstract}

\vspace*{70pt}
%\noindent
\small{Keywords: {\it Inelastic pion production, chiral symmetry,
    effective field theory}}\\
\small{PACS nos.: 25.80.Hp , 12.39.Fe , 11.30.Rd}
\vfill
\end{center}
\newpage
%Body

%%%%%%%%%%%%%%%%%%%%%%%%%%%%%%%%%%%%%%%%%%%%%%%%%%%%%%%%%%%%%%%%%%%%%%%
\section{Introduction}
\def\theequation{\arabic{section}.\arabic{equation}}
\setcounter{equation}{0}
\subsection{General introductory remarks}

Single pion production off nucleons has been at the center of numerous
experimental and theoretical investigations for many years. One of
the original motivations of these works was the observation that
the elusive pion--pion threshold S--wave interaction could be deduced
from the pion--pole graph contribution of the pion production
reaction. A whole series of precision
experiments at PSI, Los Alamos, TRIUMF and CERN (and other laboratories)
has been performed over the last
decade and there is still on--going activity. On the theoretical side,
chiral perturbation theory has emerged as a precision tool in low
energy hadron physics. It not only allows to investigate low--energy
pion--pion scattering to high accuracy, see e.g.~refs.\cite{orsay,bern},
but also elastic pion--nucleon scattering and inelastic pion production can be
studied in the threshold region, for a review see~\cite{bkmrev}. Another 
important aspect of inelastic pion production is the excitation of resonances,
some of which couple much stronger to the $\pi\pi N$ final--state than to the
pion--nucleon continuum. In what follows, we will be concerned with the low
energy region which allows to address questions connected to the chiral structure of
QCD. Since there already exist a few studies based upon various approximations,
we first give an overview of the theoretical status and discuss in which
respects these existing calculations can be improved.

\subsection{Short review of existing calculations and objectives}
Beringer considered the reaction $\pi N \to \pi
\pi N$ to lowest order in relativistic baryon  chiral perturbation theory
\cite{bering}. Low--energy theorems for the threshold amplitudes
${\cal D}_1$
and ${\cal D}_2$\footnote{These are related to the more commonly used 
${\cal A}_{10}$ and ${\cal A}_{32}$ by ${\cal A}_{32} = \sqrt{10}{\cal
  D}_1$
and ${\cal A}_{10} = -2{\cal D}_1 - 3{\cal D}_2$, see also section~3.1.} were
derived in \cite{bkmplb}.
These are free of unknown parameters and not sensitive to the $\pi
\pi$--interaction beyond tree level. 
A direct comparison with the threshold data in
the  $\pi^+ p \to \pi^+ \pi^+ n$ channel, which is only sensitive to
${\cal D}_1$,
leads to a very satisfactory description, whereas in case of the process
$\pi^- p \to \pi^0 \pi^0 n$, which is only sensitive to ${\cal D}_2$, sizeable 
deviations were found from the total cross sections near threshold. These were 
originally attributed to the strong pionic final--state interactions in the
$I_{\pi\pi}=0$ channel. However, this conjecture turned out to be incorrect 
when a complete higher order calculation of the threshold amplitudes
${\cal D}_{1,2}$
was performed \cite{bkmppn}. In that paper,  the relation
between the threshold amplitudes ${\cal D}_1$ and ${\cal D}_2$ 
and the $\pi\pi$ S--wave scattering lengths $a_0^0$ and $a_0^2$
was investigated
in the framework of heavy baryon chiral perturbation theory to second
order in the pion mass (which is the only small
parameter at threshold). The pion loop and pionic counterterm 
corrections  only start  contributing to the $\pi\pi N$ threshold amplitudes at
third order in the chiral expansion. 
One of these counterterms, proportional to the low--energy
constant $\ell_3$, eventually allows to
measure the scalar quark condensate, i.e. the strength of the spontaneous 
chiral symmetry breaking in QCD. At that order, the largest 
contributions to ${\cal D}_{1,2}$ stem  from insertions of the dimension two
chiral pion--nucleon Lagrangian, which is characterized by a few low--energy
constants called $c_i$.  In particular this is the case for the amplitude
${\cal D}_2$. It appeared therefore natural to extend the same
calculation above threshold and to compare to the large body of data
for the various reaction channels. It was already shown by
Beringer \cite{bering} that taking simply the relativistic Born terms
does not suffice to describe the total cross section data for
incoming pion energies up to 400~MeV in most channels. In ref.\cite{bkmci}
the relativistic pion--nucleon Lagrangian including all dimension two
operators was used, their coefficients being fixed from pion--nucleon
scattering data and (sub)threshold parameters.
This parameter--free calculation lead to a satisfactory 
description of most existing total and
differential  cross section data as well as angular correlation functions. 
While being rather successful, that calculation could not
give a definite answer to the questions concerning the importance of loop
effects  (and thus  the sensitivity to the pion--pion interaction beyond 
tree level) and the convergence of the chiral expansion. Also, the 
amplitudes calculated from the  tree graphs in ref.\cite{bkmci} are, 
of course, purely real. How severe 
this approximation is can only be judged after a one loop calculation 
above threshold has been performed. A first attempt to include third
order contributions above threshold in the framework
of heavy baryon chiral perturbation theory was described in ref.\cite{bmz}.
There, the pertinent one loop graphs and insertions from the dimension three 
Lagrangian in the formulation of ref.\cite{em} were calculated. However,
only three of the five physical channels were considered and the finite
pieces of the unknown dimension three low--energy constants were not 
determined but rather varied within given generous bounds. It was claimed that the
contribution from the chiral dimension three amplitudes are large and 
play an essential role in the description of the data. Note that one
of the low--energy constants (LECs) varied in ref.\cite{bmz} was shown to have
a fixed coefficient in ref.\cite{fms}. Also, no explicit formulae for
the various third order contributions were given, so that it is not
possible to check the consistency of the amplitudes constructed there 
with the threshold
amplitudes obtained in~\cite{bkmppn}. For these reasons, we do not
consider the results  obtained in  ref.\cite{bmz} as conclusive.
In addition, there exist some model calculations which are only
partly constrained by chiral symmetry, see e.g. \cite{oset,jaeck}. 
The one closest in spirit to a chiral expansion is the one of 
ref.\cite{jemi}, in which Beringer's Born
terms where supplemented by explicit $\Delta$ and Roper (tree) contributions.
Clearly, the inclusion of the resonances as done in that paper is not based on 
a consistent power counting scheme but rather it is argued that
phenomenology demands the extension of the effective Lagrangian to include
these higher mass states. 

Our objective is to perform a complete third order calculation
using the {\it minimal} effective pion--nucleon Lagrangian. In addition,
we will use all available information  about the appearing
LECs, in particular from the comprehensive study of elastic pion--nucleon
scattering described in  ref.\cite{fms}. Throughout, we will work in
the so--called standard scenario of spontaneous chiral symmetry
breaking, i.e. assuming a large scalar quark condensate. While the
extension to the small condensate case is in principle
straightforward, we do not consider it in what follows. 
In particular, our study allows to address the following questions:
\begin{enumerate}
\item[i)]How sensitive are the data in the threshold region to the
$\pi\pi$ interaction beyond leading order? Previous investigations
seem to indicate that a better method to access the $\pi\pi$
interaction is the use of Chew--Low techniques (see e.g.~\cite{kermcl})
rather than a direct use of the chirally expanded $\pi\pi N$ threshold
amplitudes (see e.g.~refs.\cite{bkmppn,lange}).
\item[ii)]How quickly does the chiral series converge? The previous 
analyses have not yet given a unique answer to this, although
the study of ref.\cite{bmz} seems to indicate a slow convergence.
On the other hand, the results obtained in the relativistic tree
level calculation in ref.\cite{bkmci} could be taken as an indication
that loop effects are not very important.
\item[iii)]At third order, new LECs appear. These can be determined from
a fit to cross section data. 
Are their values of natural size? If that is the case, it would
also be interesting to develop an understanding of the numerical
values as it was done for the dimension two LECs in ref.\cite{bkmlec}.
Furthermore, in ref.\cite{bkmppn} resonance saturation was used to
pin down the order three LEC contribution to the threshold amplitude $D_2$ and
this procedure can be made more precise when all LECs are determined
from above threshold data. 
\item[iv)]Can the existing data be described consistently in chiral
perturbation theory? So far, mostly total cross section data and
some angular distributions have been studied (with the exception of
the detailed angular correlation functions in the $\pi^- p
\to \pi^- \pi^+ n$ channel). With the new TRIUMF data~\cite{lange} and  
the more recent ones from CHAOS~\cite{kermani}  on $\pi^\pm p \to 
\pi^\pm \pi^+ n$~\cite{lange,kermani}, this data base has
considerably increased and allows for detailed tests on 
chiral pion--nucleon dynamics. In particular, we now have 
differential cross sections with respect to the invariant dipion mass
squared $M_{\pi\pi}^2$,
the squared momentum transfer to the nucleon and the scattering 
angle between the two negative pions in the dipion restframe~\cite{kermani}. 
\end{enumerate}

\subsection{Organization of the paper}
This paper is organized as follows. In section~2 we briefly review the
effective Lagrangian underlying the calculation. Section~3 contains some formal
aspects including the definition of the T--matrix and of the
pertinent observables. The chiral expansion of the invariant
amplitudes is performed in section~4.
Section~5 contains the results and discussions thereof.  
In the appendix, we give the explicit expressions for the amplitudes.

%%%%%%%%%%%%%%%%%%%%%%%%%%%%%%%%%%%%%%%%%%%%%%%%%%%%%%%%%%%%%%%%%%%%
\section{Effective Lagrangian}
\label{sec:Leff}
\def\theequation{\arabic{section}.\arabic{equation}}
\setcounter{equation}{0}

At low energies, the relevant degrees of freedom are hadrons, in 
particular the Goldstone bosons linked to the spontaneous chiral symmetry
breaking (for a review, see e.g.~\cite{bkmrev}). We consider here 
the two flavor case and thus deal with the
iso--triplet of pions, collected in the unitary matrix $U(x) = u^2(x)$.
It is straightforward to build an effective Lagrangian
to describe their interactions, called ${\cal L}_{\pi\pi}$. This Lagrangian
admits a dual expansion in small (external) momenta and quark (meson) 
masses as detailed below. Matter fields such as nucleons can also be 
included in the effective field theory based on the
familiar notions of non--linearly realized chiral symmetry.
The pertinent effective Lagrangian  called ${\cal L}_{\pi N}$
consists of terms with  exactly one nucleon in the initial 
and the final state. The various terms contributing to a process
under consideration are organized according to their chiral dimension,
which counts the number of derivatives and/or meson mass insertions.
Here, we work to third order in the corresponding small parameter $q$
(which is a generic symbol for an external momentum or pion mass). 
Consequently, the effective Lagrangian consists of the following pieces:
\begin{equation}
{\cal L}_{\rm eff} = {\cal L}_{\pi\pi}^{(2)} + {\cal L}_{\pi\pi}^{(4)} + 
{\cal L}_{\pi N}^{(1)} + {\cal L}_{\pi N}^{(2)}+ {\cal L}_{\pi N}^{(3)} \,\, ,
\end{equation}
where the superscript $(i)$ gives the chiral dimension. The relevant terms
of the meson Lagrangian read
\beqa
 {\cal L}^{(2)}_{\pi\pi} &=&  {F^2 \over 4}\langle \nabla^\mu U
\nabla_\mu U^\dagger + \chi_+ \rangle  \, \, \, ,
\quad \chi_+ = u^\dagger \chi u^\dagger + u \chi^\dagger u \\ 
{\cal L}^{(4)}_{\pi \pi} &=& \dfrac{\ell_1}{4} 
\langle \nabla_\mu U \nabla^\mu U^\dagger \rangle^2 +
\dfrac{\ell_2}{4} \langle \nabla_\mu U \nabla_\nu U^\dagger \rangle 
\langle \nabla^\mu U \nabla^\nu U^\dagger \rangle
+ \dfrac{\ell_3}{16} \langle  \chi_+ \rangle^2 \nonumber \\
&+& \dfrac{\ell_4}{16} \biggl\lbrace 
2\langle \nabla_\mu U \nabla^\mu U^\dagger \rangle \langle \chi_+ \rangle + 
2\langle \chi^\dagger U \chi^\dagger U +
\chi U^\dagger \chi U^\dagger \rangle - 4 \langle\chi^\dagger \chi \rangle - 
\langle \chi_+ \rangle^2 \biggr\rbrace  
+ \ldots\label{Lpipi}
\eeqa
where the SU(2) matrix-valued field $U(x)$ collects the iso--triplet pions,
\beq 
U(x) = {1\over F} \biggl[ \sqrt{ F^2 - {\vec \pi} \, (x)^2 } + i \vec \tau
\cdot \vec \pi \, (x) \biggr] ~, 
%\quad u(x) =  \sqrt{U(x) }~,
\label{gauge}
\eeq
with $F$ the pion decay constant in the chiral limit and $\nabla_\mu = 
\partial_\mu + \ldots$ is the pion covariant derivative. Here, we only
need the partial derivatives since we are not concerned with left-- or
right--handed external vector currents. This representation
of the pion fields, the so-called $\sigma$-model gauge, is of particular 
convenience for calculations in the nucleon sector.
The quantity $\chi$ contains the light quark mass 
$m_u= m_d = \hat m$\footnote{Throughout we will work in the isospin 
limit apart from kinematical corrections to be discussed later.} and
external scalar and pseudoscalar sources. 
The ellipsis in eq.(\ref{Lpipi}) stands for other terms of order $q^4$ 
which do not  contribute in our case. 
The fourth order terms in the meson Lagrangian  serve to cancel some of
the divergences of the loop diagrams and contain the
mesonic low--energy constants $\ell_{1,2,3,4}$. The latter encode information
about the chiral corrections to the $\pi \pi$ scattering lengths.
The finite pieces $\ell^r_i$ of the low--energy constants
$\ell_i$ in eq.(\ref{Lpipi}) are renormalization
scale dependent and are related to the $\bar{\ell}_i$ of ref.\cite{gl84} via
(since both of these sets are used in the literature, it is pertinent to give
these relations)   
\beq
\bar{\ell}_i = 16\, \alpha_i\, \pi^2\, \ell^r_i (\lambda )
 - 2 \ln  \dfrac{M_\pi}{\lambda}  \, \, , \quad
\alpha_1 = 6~,\,\, \alpha_2 = 3~,\,\,  \alpha_3 = -4~,\,\,
 \alpha_4 = 1~,
\label{elli}
\eeq
and their actual values will be discussed later. Here, $\lambda$ is the
scale of dimensional regularization.
Let us now discuss in some more detail the terms appearing in the
various parts of the pion--nucleon Lagrangian, i.e. in 
$ {\cal L}_{\pi N}^{(1,2,3)}$. We make use of
 baryon chiral perturbation theory
in the heavy mass formulation~\cite{jm,bkkm} (HBCHPT). 
The nucleons are considered as
extremely heavy. This allows to decompose the nucleon Dirac spinor into ``large''
$(H)$ and ``small'' ($h)$ components 
\beq
\Psi(x) = e^{-i m v \cdot x } \{ H(x) + h(x)\}  \, \, ,
\label{heavy}
\eeq
with $v_\mu$ the nucleon four-velocity, $v^2 = 1$, and the velocity eigenfields
are defined via $ \barr v H = H$ 
and $\barr v h = - h$.\footnote{The role of $v_\mu$ is to
single out a particular reference frame, 
here the $\pi^a N$ center--of--mass frame.} 
Eliminating the ``small'' component field $h$ (which generates $1/m$
corrections), the leading order chiral $\pi N$ Lagrangian reads
\beq
{\cal L}_{\pi N}^{(1)} 
= {\bar H} ( i v\cdot D + {g}_A S \cdot u ) H \, \, \, .
\label{lpin1}
\eeq 
Here, $D_\mu = \partial_\mu +
\Gamma_\mu$ denotes the nucleon chiral covariant derivative, $S_\mu$ is a
covariant generalization of the Pauli spin vector, ${g}_A$ the
nucleon axial vector coupling constant  (in the chiral 
limit)\footnote{We do not display explicitly the ``$\krig{}$'' commonly
used to denote quantities in the chiral limit. The difference between
the physical and the chiral limit values has to be kept in mind since it
enters the renormalization discussed below.}
 and $u_\mu = i u^\dagger \nabla_\mu U u^\dagger $.
To leading one--loop accuracy, i.e. order ${\cal O}(q^3)$, 
one has to consider tree graphs from
\beq
{\cal L}_{\rm eff} = 
{\cal L}_{\pi N}^{(1)} + {\cal L}_{\pi N}^{(2)} + {\cal L}_{\pi N}^{(3)} 
\label{leff}
\eeq
and one--loop graphs with dimension one insertions only.
The pertinent terms of $ {\cal L}_{\pi N}^{(2)}$ read
\beqa
{\cal L}_{\pi N}^{(2)} &=& \bar{H} \biggl\{{(v\cdot D)^2\over 2m} - {D^2 \over
  2m} - {i g_A \over 2m} \{S\cdot D, v\cdot u\} + c_1 \langle \chi_+
  \rangle \nonumber \\
&& + \biggl( c_2 - \dfrac{g_A^2}{2m}\biggr) (v\cdot u)^2 + c_3 u\cdot
u +  \biggl( c_4 + \dfrac{1}{4m}\biggr) [S^\mu,S^\nu] u_\mu u_\nu +
\ldots \biggr\} H~. 
\eeqa
All terms in $ {\cal L}_{\pi N}^{(2)}$
are finite. The corresponding LECs have been determined in 
refs.\cite{bkmlec,mm,fms}. A physical interpretation of their numerical
values can be found in ref.\cite{bkmlec}, where it is shown that
resonance saturation (mostly due to the $\Delta$--resonance) can indeed
explain the size of these LECs.  
The first divergences appear at ${\cal O}(q^3)$ in HBCHPT,
the corresponding determinant has been worked out by
Ecker~\cite{ecker}. We use here the minimal form of the dimension
three $\pi N$ Lagrangian as constructed in ref.\cite{fms}
\beqa \label{lpin3f}
{\cal L}_{\pi N}^{(3)} &=& {\cal L}_{\pi N}^{(3),\, {\rm fixed}} +
\sum_{i=1}^{23} d_i \, \bar{H} O_i H 
+ \sum_{i=24}^{31} \tilde{d}_i \, \bar{H} \tilde{O}_i^{\rm div} H \,\,\, ,
\nonumber \\
& = & \bar{H} \left\{ {\cal O}^{(3)}_{\rm fixed} + {\cal O}^{(3)}_{\rm
    ct} + {\cal  O}^{(3)}_{\rm div}\right\} H \,\,\, ,
\eeqa
The first set contains terms which stem from the $1/m$ expansion of the various
dimension one and two operators and thus have fixed coefficients. Then
there are 23 terms, from which 14 are divergent. For our calculation, 
we need the following dimension three operators $O^{(3)}_{i}$,
\begin{eqnarray}
{\cal O}^{(3)}_{\rm ct} 
& = &
i\,d_1(\lambda)\,[u_\mu,[v\!\cdot\! D,u^\mu]]
+ i\left(d_2(\lambda)-\frac{1+8m\,c_4}{32m^2}\right)
[u_\mu,[D^\mu,v\!\cdot\! u]]
\nonumber\\%[0.5em]
& + &
i\left(d_3(\lambda)+\frac{g_A^2}{32m^2}\right)
[v\!\cdot\! u,[v\!\cdot\! D,v\!\cdot\! u]]  
+ 
i\left(d_4(\lambda)-\frac{g_A}{64m^2}\right)
\epsilon^{\mu\nu\alpha\beta}v_\alpha\langle u_\mu u_\nu
u_\beta\rangle
\nonumber\\%[0.5em]
&+& d_5(\lambda)\,[\chi_-,v\!\cdot\! u]
+ d_{10}(\lambda)\,S\!\cdot\! u\,\langle u\!\cdot\! u \rangle
+ d_{11}(\lambda)\,S^\mu u^\nu \langle u_\mu u_\nu\rangle
\nonumber\\%[0.5em]
& + &
\left(d_{12}(\lambda)-\frac{4g_A(1+4m\,c_4)+g_A^3}{32m^2}\right)
S\!\cdot\! u\,\langle (v\!\cdot\! u)^2 \rangle
\nonumber\\%[0.5em]
& + &
\left(d_{13}(\lambda)+\frac{2g_A(1+4m\,c_4)+g_A^3}{16m^2}\right)
S^\mu v\!\cdot\! u\,\langle u_\mu v\!\cdot\! u\rangle
\nonumber \\%[0.5em]
& + &
d_{14}(\lambda)\,\epsilon^{\mu\nu\alpha\beta}v_\alpha S_\beta \langle[v\!\cdot\!
D,u_\mu]\,u_\nu \rangle
+ \left(d_{15}+\frac{g_A^2}{16m^2}\right)
\epsilon^{\mu\nu\alpha\beta}v_\alpha S_\beta \langle u_\mu 
[D_\nu,v\!\cdot\! u]\rangle
\nonumber \\%[0.5em]
& + &
d_{16}(\lambda)\,S\!\cdot\! u\,\langle\chi_+\rangle
+ i\,d_{18}\,S^\mu[D_\mu,\chi_-]
\end{eqnarray}
For these divergent operators, we introduce scale--independent renormalized LECs,
\beq
\bar{d}_i = d_i - {\kappa_i \over (4\pi F)^2} \, \left( {\rm
  L}(\lambda ) + (4\pi)^2 \ln \frac{\lambda}{M} \right) \,\,  ,
\eeq
with
\beq
{\rm L} (\lambda ) = {\lambda^{d-4}\over 16\pi^2} 
\biggl\{ {1 \over d-4} - {1\over 2}
 \biggl[ \ln(4\pi ) + \Gamma '(1) +1 \biggr] \biggr\}  
\quad .
\eeq
Note also that the operators ${\cal O}_1$ and ${\cal O}_2$ (${\cal O}_{14}$  
and ${\cal O}_{15}) $ appear as a
sum (difference) for $\pi N \to \pi \pi N$. Four combinations of counterterms
($\bar{d}_1+\bar{d}_2, \bar{d}_3, \bar{d}_5, \bar{d}_{14}-\bar{d}_{15}$) have
already been fixed from pion--nucleon scattering data and $\bar{d}_{18}$ from
the so--called Goldberger--Treiman discrepancy.
In addition, there are eight terms which are only necessary for the
renormalization, and thus do not appear in matrix elements of
physical processes. To be precise, these terms stem solely from the
divergent part of the one--loop generating functional and have no
(finite) counterparts in the relativistic theory. They read
\begin{eqnarray}
{\cal O}^{(3)}_{\rm div} 
& = &
\tilde{d}_{24}(\lambda)\,i\left(v\!\cdot\!D\right)^3
+ \tilde{d}_{25}(\lambda)\,v\cdot\!\!\stackrel{\leftarrow}{D}
S\!\cdot\!u\,v\!\cdot\! D
+ \tilde{d}_{26}(\lambda)\left(i\,\langle u\!\cdot\!u\rangle\,v\!\cdot\!D +
  \mbox{h.c.}\right)
\nonumber \\[0.5em]
& + &
\tilde{d}_{27}(\lambda)\left(i\,\langle (v\!\cdot\!u)^2\rangle\,v\!\cdot\!D
 +
  \mbox{h.c.}\right)
+ \tilde{d}_{28}(\lambda)\left(i\,\langle\chi_+\rangle\,v\!\cdot\!D +
  \mbox{h.c.}\right)
\nonumber \\[0.5em]
& + &
\tilde{d}_{29}(\lambda)\left(S^\mu[v\!\cdot\!D,u_\mu]\,v\!\cdot\!D +
  \mbox{h.c.}\right) +
\tilde{d}_{30}(\lambda)\left(\epsilon^{\mu\nu\alpha\beta}v_\alpha S_\beta
[u_\mu,u_\nu] \,v\!\cdot\!D + \mbox{h.c.}\right)~.\nonumber \\ &&
\end{eqnarray}
All these terms are proportional to the nucleons equation of motion, 
$\sim v \cdot D \,H$, and thus can be
transformed away by appropriate field redefinitions. For more details,
see ref.\cite{fms}.

%%%%%%%%%%%%%%%%%%%%%%%%%%%%%%%%%%%%%%%%%%%%%%%%%%%%%%%%%%%%%%%%%%%%
\section{Single pion production: Formal aspects}
\def\theequation{\arabic{section}.\arabic{equation}}
\setcounter{equation}{0}

In this section, we outline the basic technical framework for the
reaction $\pi N \to \pi \pi N$ above threshold treating the nucleons
as very heavy fields. 

\subsection{Invariant amplitudes}
We seek the T--matrix for the process
\beq
\pi^a(q_1) + N(m v+p_1) \rightarrow \pi^b(q_2)+\pi^c(q_3)
+ N(m v+p_2)~,
\eeq
with $N$ denoting a nucleon (neutron or proton) and $\pi^a$ a pion of cartesian
isospin $a$. Note that this process is characterized by five independent
four--momenta (or five Mandelstam variables). It is most appropriate in
terms of the chiral expansion to choose as variables the energies of the
out--going pions and the three invariant momentum transfers $t_i$
$(i=1,2,3)$ defined below.  
%Throughout, we consider the momenta $q_1$ and
%$p_1$ as incoming and the other three as outgoing. 
Since we work
in the framework of HBCHPT, the nucleon four--momenta are written in terms of a 
conserved four--velocity $v$ and small residual momenta $p_{1,2}$,
where small means
$v\cdot p_{1,2} \ll m$ and $m$ is the nucleon mass. The $\pi N \to \pi\pi N$
transition matrix element can be expressed in terms of four invariant amplitudes,
denoted $A$, $B$, $C$ and $D$~\cite{bmz}, which depend on the five momenta 
$p_1,p_2,q_1,q_2$ and $q_3$\footnote{In what follows, we will suppress the 
dependence of the invariant amplitudes on the momenta.} as
\bea\label{T}
T^{abc}_{ss'} &=&
\langle N_{s'} (m v+p_2) \, \pi^b(q_2)\, \pi^c(q_3)\,|\, T\, | \, N_s (m
v+p_1)\, \pi^a(q_1) \rangle \nonumber\\
&=& {\cal N}_1 \, {\cal N}_2 \, \chi_{s'} \,
\biggl[ S \cdot q_1\, A+S \cdot q_2 \,B+S \cdot q_3 \,C+i\epsilon_{\mu\nu\alpha\beta}
q_1^\mu q_2^\nu q_3^\alpha v^\beta \, D\biggr]^{abc}\, \chi_s~,
\eea
with $\chi_s$ the conventional two--spinor and 
\beq
{\cal N}_i =  \sqrt{\frac{E_i+m}{2 m}}~, \quad (i = 1,2)~,
\eeq
the standard spinor normalization factors which are mandated by the matching
to the relativistic theory if one works in the so--called Pauli interpretation
of the heavy nucleon fields (a detailed discussion of this topic can be found
in ref.\cite{fmsz}). The invariant amplitudes are complex functions of the
momentum variables since in some of the one--loop graphs considered here, the intermediate
pion--nucleon state can go on--shell. The pertinent tree graphs lead, of course,
to real amplitudes. The isospin decomposition of the invariant amplitudes reads
\be
X^{abc} = \tau^a \delta^{bc} X_1+\tau^b \delta^{ac} X_2+\tau^c \delta^{ab}
X_3+i\epsilon^{abc} X_4~,
\hspace{0.7cm} \, X\in \lbrace A,B,C,D \rbrace.
\ee
The following calculations can be simplified considerably if one notices
that for most diagrams, the amplitudes have the following
symmetries under exchange of particles (note that
for these equations to hold, the momenta $p_1$ and $q_1$ have to be
chosen in--coming and $q_2$, $p_2$ and $p_3$ out--going):
\bea
A_1(q_1,q_2,q_3) & = & -B_2(-q_2,-q_1,q_3) \no\\
&=&-C_3(-q_3,q_2,-q_1)\\
A_2(q_1,q_2,q_3) & = & A_3(q_1,q_3,q_2)\no\\
& = & -B_1(-q_2,-q_1,q_3)\no\\
&= & -C_1(-q_3,-q_1,q_2)\\
B_3(q_1,q_2,q_3)& = & C_2(q_1,q_3,q_2)\\
A_4(q_1,q_2,q_3) & = & B_4(-q_2,-q_1,q_3)\no\\
& = & -C_4( -q_3,-q_1,q_2)\\
D_1(q_1,q_2,q_3) & = &-D_2(-q_2,-q_1,q_3)\no\\
&=&-D_3(-q_3,q_2,-q_1)~,
\eea
and in all these expressions the fixed values of $p_{1,2}$ are not made
explicit. Thus for these diagrams it is enough to give the formulae of six
amplitudes $A_1$, $A_2$, $A_4$, $B_3$, $D_1$ and $D_4$ in order to construct the
full T--matrix element. However, in the tree graphs with a
dimension two or three $\bar NN\pi\pi$ vertex, these symmetries are not present.
In these cases, there appear expressions proportional to 
$S\cdot p_1$ and $S\cdot p_2$.  In the center--of--mass (cm) frame these equal
$-S\cdot q_1$ and $-S\cdot(q_2+q_3)$ respectively, which obviously leads to
an inequivalent role played by the pion momenta $q_1$, $q_2$ and $q_3$.
For such diagrams the amplitudes $B_1$, $B_2$, $B_4$,$ D_2$ have to be given in
addition to the six discussed before in order to fully determine the T--matrix element.

The five experimentally accessible channels follow from the isospin amplitudes,
\beqa\label{iso1}
\pi^+ p \to \pi^+ \pi^+ n &:& X =\sqrt{2} \, (X_2+X_3)~,\\
\pi^+ p \to \pi^+ \pi^0 p &:& X =X_3+X_4~,\\
\pi^- p \to \pi^+ \pi^- n &:& X =\sqrt{2} \, (X_1+X_2)~,\\
\pi^- p \to \pi^0 \pi^0 n &:& X =\sqrt{2} X_1~,\\
\pi^- p \to \pi^0 \pi^- p &:& X =X_2+X_4~,
\label{iso5}
\eeqa
for $X\in \lbrace A,B,C,D \rbrace$. We will calculate the amplitudes $X_i$ in
the isospin limit with the charged pion mass ($M_\pi =139.57\,$MeV) and the
proton mass ($m=938.27\,$MeV). Isospin breaking is accounted
for in a minimal way by shifting the kinetic energy (in the laboratory
system) of the incoming pion,
called $T_\pi$, from the isospin symmetric threshold
\beq
T_\pi^{{\rm thr,iso}}= M_\pi \bigg(1+{3M_\pi \over 2m}\bigg) =170.71\,\, {\rm
  MeV}
\eeq 
to the correct threshold of the corresponding reaction. 
For the five channels given above, the corresponding
shift is $\delta T_\pi = +1.68$, $-5.95$, $+1.68$, $-10.21$ and $-5.95\,$MeV in
the same order as in eqs.(\ref{iso1}-\ref{iso5}). Finally, we need the expressions
of the two threshold quantities ${\cal D}_{1,2}$  in terms of
the invariant amplitudes defined above. At threshold, $\vec{q}_2 = \vec{q}_3 =0$
and there are only two amplitudes, 
\beqa\label{D12}
{\cal D}_1 &=& -\frac{i}{2}\, {\cal N}_1^{\rm thr}\, A_2^{\rm thr} =
 -\frac{i}{2}\, {\cal N}_1^{\rm thr}\, A_3^{\rm thr}~,\nonumber\\
{\cal D}_2 &=& -\frac{i}{2}\, {\cal N}_1^{\rm thr}\, A_1^{\rm thr}~,
\eeqa
with ${\cal N}_1^{\rm thr} = 1.006$. The chiral expansion of ${\cal D}_{1}$ and
${\cal D}_{2}$ has already been considered to third order, i.e. including
all terms of order $M_\pi^2$ since the pion mass is the only dimensionful quantity
at threshold and the lowest order terms are non--vanishing in the chiral
limit. To end this paragraph, we note that from now on we use the energies
of the outgoing pions and the three invariant momentum transfers squared as
kinematical variables, because $\omega_2 = v\cdot q_2,\omega_3 = v\cdot q_3
= {\cal O}(q)$ and $t_1, t_2,t_3 = {\cal O}(q^2)$ have the expected chiral dimensions and
consequently the residual nucleon energies $v\cdot p_i$ are of second
order.

\subsection{Observables}
To calculate the total (unpolarized) cross section, we
need the invariant matrix--element squared multiplied by the appropriate
weight functions. It reads
\bea
\vert {\cal M} \vert^2 &=&
\frac{1}{2} \sum_{s,s'}^{} T^\dagger_{ss'} T_{ss'} \no\\&=&
\frac{1}{4}\Bigg[
A^*A \,(\omega_1^2-q_1^2)
+B^*B \,(\omega_2^2-q_2^2)
+C^*C \,(\omega_3^2-q_3^2)
+(A^*B+A B^*) \,(\omega_1\omega_2-q_1\cdot q_2)\no\\&&
+(A^*C+A C^*) \,(\omega_1\omega_3-q_1\cdot q_3)
+(B^*C+B C^*) \, (\omega_2\omega_3-q_2\cdot q_3)\no\\&&
+4 D^*D \,(w_1^2-q_1^2)(w_2^2-q_2^2)(w_3^2-q_3^2)(1-x_1^2)(1-x_2^2)
\left(1-\frac{(z-x_1 x_2)^2}{(1-x_1^2)(1-x_2^2)}\right) \Bigg]~,\no\\ &&
\eea
where the angular variables $x_{1,2}$ and $z$ are defined below.
The formulae for total and the double and triple differential cross sections 
then become:
\bea\label{XStot}
\sigma_{\rm tot}(T_\pi) &=& \frac{4 m {\cal S}}{(4\pi)^4 \sqrt{T_\pi(T_\pi+2
M_\pi)}}
\int \int_{z^2\le 1}^{} d\omega_2 d\omega_3 \int_{-1}^{1} d x_1 \int_0^\pi d
\phi
\, \vert {\cal M} \vert^2~, \\
\frac{d^2 \sigma}{d \omega_2 d \Omega_2} & = & \frac{8 m^2 {\cal S}}{(4\pi)^5
\sqrt{s}
\vert \vec{q}_1 \vert} \int_{\omega_3^-}^{\omega_3^+} d\omega_3
\int_0^\pi d\phi\,
\vert {\cal M} \vert^2~, \\
\frac{d^3 \sigma}{d \omega_2 d\Omega_2 d\Omega_3} & = &
\frac{4 m^2 \vert \vec{q}_2 \vert \vert \vec{q}_3 \vert {\cal S}}
{(4\pi)^5 \sqrt{s} \vert \vec{q}_1 \vert \tilde{E}_3 }\, \vert {\cal M} \vert^2~,
\eea
with $s=(m+M_\pi)^2 + 2mT_\pi$ the total center--of--mass energy
squared and $\omega_i = \sqrt{\vec{q}_i^{\,2} + M_\pi^2}$.
${\cal S}$ is a Bose symmetry factor, ${\cal S}=1/2$ for identical pions
and ${\cal S} =1$ otherwise. 
Here, $\phi$ is the (auxiliary) angle between the planes spanned by $\vec q_2$
and $\vec q_1$ as well as $\vec q_2$ and $\vec q_3$. In accordance with the
experimentalists convention, we have chosen the coordinate frame such that the
incoming pion momentum $\vec q_1$ defines the z--direction, whereas $\vec q_2$ 
lies in the  xz--plane. The polar angles $\theta_1$
and $\theta_2$ of the outgoing pions (with $x_i = \cos\theta_i$) are in general
non--vanishing and so is the azimuthal angle $\varphi_2$ of $\pi^c$. By
construction, the azimuthal angle of $\pi^b$ is zero.
Furthermore, 
\begin{equation} 
\tilde E_3 = E_3\biggl(1+{\partial E_3\over \partial
\omega_3}\biggr) = {\omega_3({1\over2}(s-m^2)-\sqrt{s}\,\omega_2)+M_\pi^2(
\omega_2 +\omega_3 -\sqrt{s})\over \omega_3^2-M_\pi^2} \,\, ,
\end{equation}
The cosine of the angle between the two three--momenta of the outgoing pions,
$\vec q_2$ and $\vec q_3$, respectively, is given by 
\begin{equation} 
z = \cos \theta_1\cos \theta_2+ \sin \theta_1\sin \theta_2
\cos \varphi_2   \,\, ,
\end{equation}
and it can be used  to express the energy $\omega_3$ as
\begin{eqnarray} 
\omega_3 &=& {1\over 2[(\sqrt{s}-\omega_1)^2-z^2|\vec q_2|^2]}
\biggl\{ (\sqrt{s}-\omega_2)(s-2\sqrt{s}\,\omega_1-m^2+2M_\pi^2) \nonumber \\ &
& - z |\vec q_2| \sqrt{(s-2\sqrt{s}\,\omega_1-m^2)^2-4
  M_\pi^2(m^2+(1-z^2)|\vec q_2|^2)}\biggr\} \,\, .
\end{eqnarray}
In the formula for the total cross section, eq.(\ref{XStot}), the
restriction $z^2\le 1$ is equivalent to $\omega_2 \in [I_1^-,I_1^+]$,
$\omega_3 \in [\omega_3^-,\omega_3^+]$, with
\bea
I_1^- &=& M_\pi~, \quad I_1^+ =\frac{(\sqrt{s}-M_\pi)^2-m^2+M_\pi^2}{2 (\sqrt{s}-M_\pi)}~
, \\
\omega_3^\pm &=& {1 \over 2(s-2\sqrt{s} \omega_2 +M_\pi^2)}
\Big[ (\sqrt{s}-\omega_2)(s-2\sqrt{s} \omega_2 -m^2+2M_\pi^2) \nonumber \\ & &
\pm |\vec q_2\,| \sqrt{(s-2\sqrt{s} \omega_2 -m^2)^2 -4m^2 M_\pi^2} \Big] \, .
\eea
To compare our calculation to the most recent  TRIUMF data, other combinations of 
differential cross sections are needed. Consider first the
differential cross section  with respect
to the invariant mass of the final dipion system,
\beq\label{dsdM}
\frac{d \sigma}{d M_{\pi\pi}^2} =
\frac{m {\cal S}}{(4\pi)^4 \sqrt{s}  |\vec q_1\,| |\vec q_2 +\vec q_3\,|
\sqrt{T_\pi(T_\pi+2 M_\pi)}}
\int d t \int d\omega_2 
\int_{0}^{\pi} d \phi'
%\int_{-1}^{1} \frac{d \cos\phi'}{\sin\phi'}
\, \vert {\cal M} \vert^2~.
\eeq
%
%\frac{d \sigma}{d M_{\pi\pi}^2} =
%\frac{m {\cal S}}{(4\pi)^4 \sqrt{s}  |\vec q_1\,| \sqrt{T_\pi(T_\pi+2 M_\pi)}}
%\int d t \int d\omega_2 \int_{-1}^{1} d \cos\phi'
%\, \vert {\cal M} \vert^2~.
%\eeq
The expression for $d^2 \sigma/(d M_{\pi\pi}^2 d t)$ can be easily 
deduced from eq.(\ref{dsdM}). 
Here, the integration boundaries of $t$ are given by
\beq
t^{\pm} = M_\pi^2 + M_{\pi\pi}^2 - 2 \omega_1 \omega_{23} \pm 2 |\vec q_1| |\vec
q_2+\vec q_3|~,
\eeq
and $\omega_{23}$ is fully determined by $M_{\pi\pi}^2$ via
\beq
\omega_{23} = \omega_2+\omega_3 = \frac{M_{\pi\pi}^2 + s -m^2}{2 \sqrt{s}}~.
\eeq
Consequently, $\omega_2$  is restricted to the overlap of the intervals
$[I_i^-,I_i^+]$~$(i=1,2)$ with
\be
I_2^\pm = \frac{\omega_{23}}{2} \pm \sqrt 
{\frac{(\omega_{23}^2-M_{\pi\pi}^2)(M_{\pi\pi}^2 - 4M_\pi^2)}{4M_{\pi\pi}^2}}~.
\ee
Denoting by  $\alpha$ the angle between $\vec q_1$ and $\vec q_2 + \vec q_3$ and
by $\beta$ the one between $\vec q_2$ and $\vec q_2 + \vec q_3$ ,
the azimuthal angle $\phi'$ is defined via
\be
x_1 = \cos\alpha \cos\beta + \sqrt{(1-\cos^2\alpha)(1-\cos^2\beta)} \cos\phi'~.
\ee
The differential cross section with respect to $t=(q_1-q_2-q_3)^2$ reads
\beq
\frac{d \sigma}{d t} =
\frac{m {\cal S}}{(4\pi)^4 \sqrt{s}  |\vec q_1\,| \sqrt{T_\pi(T_\pi+2 M_\pi)}}
\int \frac{d M_{\pi\pi}^2}{|\vec q_2+\vec q_3\,|} \int d\omega_2 \int_{0}^{\pi} d \phi'
%\frac{d \cos\phi'}{\sin\phi'}
\, \vert {\cal M} \vert^2~. %% \no\\
\eeq
%\frac{d \sigma}{d t} & = &
%\frac{m {\cal S}}{(4\pi)^4 \sqrt{s}  |\vec q_1\,| \sqrt{T_\pi(T_\pi+2 M_\pi)}}
%\int d M_{\pi\pi}^2 \int d\omega_2 \int_{-1}^{1} d \cos\phi'
%\, \vert {\cal M} \vert^2 \no\\
%\eea
The kinematically allowed region for $M_{\pi\pi}^2$ is
\beq
M_{\pi\pi}^2  =  \frac{1}{m^2} \Biggl\{
t/2\,(s+m^2-M_\pi^2) +m^2 M_\pi^2  \pm |\vec q_1| \sqrt{-s\,t\,(4m^2-t)}
\Biggr\}~,
\eeq
and also $M_{\pi\pi}^2 \in ( 4M_\pi^2 , (\sqrt{s}-m)^2)$.
Finally, the last measured quantity is the single differential cross section
with respect to the scattering angle of the two outgoing pions, 
in the center-of-mass frame of the outgoing two-pion system:
\bea
\frac{d \sigma}{d \cos\theta} & = & \frac{2mS}{(4\pi)^4 \sqrt{T_\pi (T_\pi+2M_\pi)}}
\, \frac{1}{\sqrt{s} |\vec{q}_1  |}\,
\int d M_{\pi\pi}^2 \int d\cos\alpha \int d\omega_2\,
\, \frac{\vert {\cal M} \vert^2}{\sin \phi '}\no \\ &&
\qquad\qquad\quad \times
\frac{| \vec{q_1'}|| \vec{q_2'}|}{|\vec{q}_2  | \, \sqrt{
(1-\cos^2\alpha)(1-\cos^2\beta )}}~,
\eea
%with $\vec{q_1'}  = (\vec{q}_1)_{\rm CM} = (M_{\pi\pi}^4 -
%2M_{\pi\pi}^2(t+M_\pi^2)+ (t-M_\pi^2)^2)/4M_{\pi\pi}^2$ and 
%$\vec{q_2'}  = (\vec{q}_2)_{\rm  CM} = (M_{\pi\pi}^2-4M_\pi^2)/4$ 
%the pion momenta in the dipion cms.
with $\vec{q_1'}  = (\vec{q}_1)_{\rm CM}$ and $\vec{q_2'}  = (\vec{q}_2)_{\rm  CM} $
the pion momenta in the dipion cms, with magnitude
$|\vec{q_1'}|^2 = (M_{\pi\pi}^4 -2M_{\pi\pi}^2(t+M_\pi^2)+ (t-M_\pi^2)^2)/4M_{\pi\pi}^2$
and $|\vec{q_2'}|^2 = (M_{\pi\pi}^2-4M_\pi^2)/4$ respectively.
The necessary kinematic constraints are given by the two cosine functions.
This completes the necessary formalism for our calculation.

%%%%%%%%%%%%%%%%%%%%%%%%%%%%%%%%%%%%%%%%%%%%%%%%%%%%%%%%%%%%%%%%%%%%
\section{Chiral expansion}
\def\theequation{\arabic{section}.\arabic{equation}}
\setcounter{equation}{0}

The effective Lagrangian described in section~\ref{sec:Leff} can now be used
to work out the chiral expansion of the invariant amplitudes defined in
eq.(\ref{T}). Before discussing the explicit contributions, we first display
the general structure of the chiral expansion. All the invariant amplitudes
take the form
\beq
X = X^{\rm tree} + X^{\rm loop}~,
\eeq
where the tree terms are of dimension one, two and three whereas the loop
graphs are of third order only. Corrections not calculated here start at order
$q^4$. The corresponding tree graphs are shown in figs.~1,2. Here, diagram~3p 
denotes a genuine third order counterterm specific to pion production and
the graph~3s embodies the next--to--leading order pion--pion interaction. 
We note
that many of the diagrams are simply elastic pion--nucleon scattering graphs
with one additional pion attached in all possible topologies. The one--loop
graphs are split into tadpoles (t1--t14 in fig.3) and so--called self--energy
diagrams (s1--s35 in figs.4,5). A typical one--loop $\pi\pi$ interaction is
e.g.~t13. We note again that many diagrams can be split into a $\pi N$
graph with an additional pion. This leads one to expect that a precise
description of elastic pion--nucleon scattering is an important
ingredient to the calculation performed here.  As an important check of our calculation
we recover the explicit expressions for the two threshold amplitudes worked
out to this order in ref.\cite{bkmppn}. The contribution
from the tree graphs proportional to the dimension three LECs was estimated
via resonance saturation in  ref.\cite{bkmppn}. We are in the position to
test this assumption (see below). Of course, many of the loop graphs
shown in figs.4,5 simply amount to mass and coupling constant renormalization.
This is by now a standard procedure and we refrain from discussing it
here in detail (see the appendix). Also, the appearing 
divergences $\sim 1/(d-4)$, with $d$ the
number of space--time dimensions are of course cancelled by the appropriate
infinite contributions from the dimension four mesonic LECs and the dimension three 
pion--nucleon LECs, first given in an
overcomplete basis by Ecker~\cite{ecker}, see section~2.
This procedure closely parallels the calculations described in detail 
in ref.\cite{bkmppn}. It serves of course as an important check of the 
calculation.

%%%%%%%%%%%%%%%%%%%%%%%%%%%%%%%%%%%%%%%%%%%%%%%%%%%%%%%%%%%%%%%%%%%%%%%%%%%
\section{Results and discussion}
\label{sec:res}

We are now in the position to analyze the existing data within the
framework laid out in the previous section. We first discuss the fits
which determine the LECs and then show predictions and analyze in detail
the chiral expansion, in terms of sensitivity to the $\pi\pi$ and $\pi
N$ interactions, its convergence and related issues.

\subsection{The fitting procedure}
\label{sub:sec:fit} 

As stated in section~\ref{sec:Leff}, we have to deal in total with 19 independent
combinations of LECs. We take the values of the four mesonic LECs 
from ref.\cite{LECval},
\beq
{\ell}_1^r (1\,{\rm GeV}) = -5.95~, \,\,\,
{\ell}_2^r (1\,{\rm GeV}) =  4.35~, \,\,\,
{\ell}_3^r (1\,{\rm GeV}) =  1.64~, \,\,\,
{\ell}_4^r (1\,{\rm GeV}) =  2.29~,
\eeq
in units of $10^{-3}$. Although we did not perform a systematic
calculation within the framework of generalized CHPT (i.e. assuming a
small value of the quark condensate), we can get an idea about that
approach by setting $\bar{\ell}_3 \simeq -70$~\cite{gl84}, i.e. 
$\ell_3^r (1~{\rm GeV}) = 117\cdot10^{-3}$. Clearly,
this is by no means a substitute for  a complete calculation but
should allow us to discuss some trends. A full analysis based on
GCHPT is, however, not the topic of this paper.
Turning now to the pion--nucleon sector, we note that
 the four dimension two LECs as well as five (combinations
of) dimension three meson--baryon LECs can e.g. be taken from Fit~2 of
ref.\cite{fms},
\beqa\label{values}
c_1 &=& -1.42~, \,\, c_2 = 3.13~, \,\,  c_3 = -5.85~, \,\,
c_4 = 3.50~, \nonumber \\
\bar{d}_1+\bar{d}_2 &=& 3.31~, \,\,
\bar{d}_3= -2.75~, \,\, \bar{d}_5 = -0.48~, \,\,
\bar{d}_{14}-\bar{d}_{15} = -5.69~, \,\,
\bar{d}_{18} = -0.78~.
\eeqa
The values for the $c_i$ and $\bar{d}_i$ are in GeV$^{-1}$ and
GeV$^{-2}$, respectively.
Therefore, we are left with six genuine LECs which have to be 
fitted. These are $\bar{d}_4$, $\bar{d}_{10}$, $\bar{d}_{11}$,
$\bar{d}_{12}$, $\bar{d}_{13}$ and $\bar{d}_{16}$.
For the fitting procedure, we only use data with $T_\pi <
250\,$MeV. This rather large value is mainly motivated by the fact that in the
very near threshold region, say in the first 30 MeV above the
respective thresholds, there are essentially no data in the 
two channels $\pi^- p \to \pi^0 \pi^- p$ and $\pi^+ p \to \pi^0 \pi^+
p$. The total cross section data are taken from
refs.\cite{se91}-\cite{kr75}. Moreover, there exist double
differential cross section data from Los Alamos~\cite{manl,man2} and
the Erlangen group~\cite{mue,bohn,malz}.\footnote{As described in
  \cite{bkmci}, neither chiral perturbation theory nor any of the 
  resonance models can describe the double differential cross section
  data at $\sqrt{s} = 1.301\,$GeV from ref.\cite{mue}. We therefore
  decided not to use these in the fit. Also, we do not use the
  unpublished data of refs.\cite{bohn,malz}.} In ref.\cite{mue},
a large amount of angular correlation functions are also given. 
All these data refer
to the $\pi^- p \to \pi^+ \pi^- n$ channel. In addition, there are
the recent TRIUMF data listed in the introduction.  We have performed
a series of fits based on different input material. We refrain from
describing all these in detail. What we will present here as 
results is based on fits to the total (including the recent TRIUMF data)
as well as the double
differential cross sections of refs.\cite{manl,man2}. In that way,
we can predict the angular correlation functions of ref.\cite{mue},
the new TRIUMF data~\cite{kermcl} and the total cross sections for
$T_\pi > 250\,$MeV. In these fits, we have found almost perfect
anticorrelations between the values of the LECs $\bar{d}_{10}$ and
$\bar{d}_{12}$ as well as $\bar{d}_{9}$ and $\bar{d}_{11}$. Therefore,
we can only determine four independent LECs. In table~\ref{tab:di},
we give the results for various fits with all six LECs free and with
one or two of them being fixed.
\begin{table}[h]
\begin{center}
\begin{tabular}{|c|c|c|c|}
\hline
$\bar{d}_i$ & None fixed  & $\bar{d}_{11} = -5$ & $\bar{d}_{11} 
= \bar{d}_{10} = -5$ \\ \hline
4 & $0.26 \pm 2.29$ & $1.60 \pm 2.00$ & $0.97 \pm 2.00$  \\
10 &   $-3.60 \pm 4.52$ & $-12.81 \pm 3.11$ & fixed  \\
11 &   $-20.58 \pm 5.90$ & fixed & fixed  \\
12 & $4.11 \pm 4.32$ & $12.25 \pm 3.27$ & $4.28 \pm 0.45$  \\
13 & $22.00 \pm 5.78$ & $6.60 \pm 0.36$ & $6.08 \pm 0.32$  \\
16 & $-4.70 \pm 0.84$ & $-3.67 \pm 0.74$ & $-3.84 \pm 0.70$  \\
\hline 
$\chi^2/$dof & 2.20 & 2.25  & 2.29 \\
\hline
\end{tabular}
\caption{The dimension three LECs $\bar{d}_{i}$ in GeV$^{-2}$ 
from the various fits as described in the text.\label{tab:di}}
\end{center}
\end{table}
We observe that due to the large anticorrelations, the fit with all
six LECs left free leads to rather large values for some of the LECs.
If one fixes two to  any value of natural size, which in our
normalization is of order one, the remaining LECs come out to be of
natural size too, compare the last column in table~\ref{tab:di}. The
uncertainties given in that table are the parabolic errors of the
MINUIT package and have to be taken with some caution. In what
follows, we will always use the values for the $\bar{d}_i$ as given in
the first column of table~\ref{tab:di}.
In fig.\ref{fig6} we show the fit to the total cross sections in all
five physical channels and in fig.\ref{fig7} the corresponding double
differential cross sections. We note that there are some inconsistencies
between the older and more recent data, most pronounced in the two
channels $\pi^+ p \to \pi^+ \pi^+ n$ and $\pi^- p \to \pi^+ \pi^- n$.
Note also that the data of the OMICRON collaboration~\cite{ke89,ke90} have
been criticized concerning the normalization. If we perform fits without these
data, the $\chi^2$/dof only improves by one permille. The LECs $\bar d_i$ change
mildly, the only appreciable difference is found for $\bar d_4$. This LEC
only appears in the amplitude $D_4$, which contributes to the channels
with one neutral pion in the final state. The data from the OMICRON
collaboration feature prominently in the process $\pi^- p \to \pi^0 \pi^- p$
which explains the sensitivity of the LEC $\bar d_4$. In what follows, we
keep the data of ref.\cite{ke89,ke90}  in our data base.
We have also performed fits using the LECs $c_i$ and $\bar d_i$
from fits~1 and 3 of ref.\cite{fms}.
The resulting dimension three LECs only change moderately and the $\chi^2$/dof
is 2.21, 2.20 and 2.20 for fits~1,2 and 3, in order. None of
the conclusions drawn in the following paragraphs depends on this choice, we
always use the values from fit~2 given in eq.(\ref{values}).

\subsection{Predictions, further results and discussion}
\label{sub:sec:res} 

We now turn to the predictions. In fig.\ref{fig8} we show the total
cross sections for incoming pion momenta up to 400~MeV. Despite the
large momentum transfers involved, the chiral description works 
fairly well, in particular in the $\pi^- p \to \pi^+ \pi^- n$ channel.
On the other hand, for the two reactions $\pi^+ p \to \pi^+ \pi^+ n$
and $\pi^- p \to \pi^0 \pi^- p$ the chiral prediction overshoots the
data for energies larger than 300~MeV. The angular correlation functions
as defined and given in ref.\cite{mue} are mostly well described, 
see figs.\ref{fig9},\ref{fig10},
with the exception of the smaller $\theta_2$ values, a trend already observed
in ref.\cite{bkmci}. If one insists that these data are also fitted, the
$\chi^2$/dof worsens considerably. The comparison
to the recent TRIUMF data is shown in fig.\ref{fig11} for 
$d\sigma/dM_{\pi\pi}^2$ and $d\sigma/dt$.  Whereas the data in the 
$\pi^- p$ channel are mostly well reproduced, there are some deviations 
for the $\pi^+ p$ process, most notably in  $d\sigma/dM_{\pi\pi}^2$. 
The description of $d^2\sigma/dt dM_{\pi\pi}^2$ and
$d\sigma/d\cos\theta$ is less good, we refrain from showing these. 
Again, if one includes these
data in the fit, the $\chi^2$/dof becomes intolerably large. This means that 
the presently existing data base shows some inconsistencies, a fact which also
limits the precision of our description based on  the chiral symmetry of
QCD.

\medskip

\noindent We now consider the chiral expansion in more detail.
We remark that in three channels the series seems to converge 
well ($\pi^+ p \to \pi^+ \pi^+ n$, $\pi^+ p \to \pi^+ \pi^0 p$ and
$\pi^- p \to \pi^0 \pi^- p$), in the two others ($\pi^- p \to \pi^+ \pi^- n$
and $\pi^- p \to \pi^0 \pi^0 n$) the third order corrections become large even 
close to threshold. As two representatives of these classes, we
show in the upper part of fig.\ref{fig13} the close to threshold
region ($T_\pi \le 210\,$MeV) for $\pi^+ p \to \pi^+ \pi^+ n$ (good
convergence) and $\pi^- p \to \pi^+ \pi^- n$ (poor convergence).
However, it is of importance to further analyze these third order 
contributions. As shown in the lower part of fig.\ref{fig13},
in all cases the contributions from the loops and (in most cases) 
from the terms
proportional to the dimension three LECs are small (this holds also
for the channels not shown in the figure). If the third order
correction is large, it comes entirely from the $1/m$ corrections
to the tree graphs proportional to the dimension two LECs $c_i$ and the ones
with fixed coefficients.
The smallness of the unitarity corrections is a very important
result. At first, it appears to be surprising since in  elastic pion--nucleon
scattering, the loop contributions are sizeable (in some of the invariant
amplitudes) already in the threshold region. We note, however, that such
terms appear with an additional pion line attached before or after the
$\pi N \to \pi N$ subprocess, which leads to sizeable cancelations in
the imaginary parts for the reactions considered here. This finding 
explains why the relativistic calculation
based on dimension one and two tree graphs of ref.\cite{bkmci} worked so
well, since in that calculation all imaginary parts were neglected. Similarly,
the commonly used resonance models like e.g. the one of the Valencia
group~\cite{oset} only acquire small imaginary parts from the finite
width of the resonances, which is built in by modifying the corresponding
propagators. Our analysis for the first time gives a reason why
such a seemingly ad hoc procedure can work. Another consequence concerns
polarization observables, which usually stem from the interference
of real and imaginary parts of certain invariant amplitudes. 
In order to differentiate between
different approaches to a given reaction, one often has to analyze polarization
data, one prominent example being the multipole separation in pion
photoproduction off nucleons. Due to the intrinsic smallness of the
imaginary parts for inelastic pion production, one should find
out in which polarization observables the small imaginary part is most
efficiently amplified by the real part of a large amplitude. In principle,
the framework outlined here can be used to do that, but due to the absence
of polarization data, we refrain from discussing such an analysis here.

\medskip

\noindent We now consider briefly the threshold amplitudes ${\cal D}_{1,2}$,
as defined in eq.(\ref{D12}). In ref.\cite{bkmppn} it was found that no tree
graphs with dimension three LECs generated from resonance excitation
(via the dominant Roper resonance) contribute to  ${\cal D}_{1}$, whereas
two distinct terms were considered for  ${\cal D}_{2}$. The first is
linked to the deviation from the Goldberger--Treiman relation and corresponds
to our term $\sim \bar{d}_{18}$, the second one was the contribution due
to the Roper decay $N^\star (1440) \to N (\pi\pi)_S$, where the two pions
are in a relative S--wave. The uncertainty
related to this decay in fact was the largest source of theoretical uncertainty
for the chiral description of ${\cal D}_{2}$. In our more general approach,
we also have a contribution from the dimension three LECs to  ${\cal D}_{1}$.
If we convert our central
values and errors of the corresponding  $\bar{d}_{i}$ ($i\neq 18)$
into the contribution to  ${\cal D}_{1,2}$, we find
\beq\label{D12val}
{\cal D}_{1}^{\bar{d}_{i}} = (0.27 \pm 0.07)~{\rm fm}^3~, \quad
{\cal D}_{2}^{\bar{d}_{i}} = (-1.60 \pm 0.15)~{\rm fm}^3~,
\eeq
which is within 1.5 $\sigma$ of  the result of ref.\cite{bkmppn} based
on Roper excitation, ${\cal D}_{2}^{N^\star} = (-0.40\pm 0.90)~{\rm
  fm}^3$. In addition, we have a smaller contribution to ${\cal D}_1$,
which is however of the same size as the other contributions form the
loops, $c_i$ and the $\bar{\ell}_i$ terms, cf. table~2 of  ref.\cite{bkmppn}.
The theoretical uncertainties in eq.(\ref{D12val}) have been obtained from
the MINUIT uncertainties in the $\bar{d}_{i}$ taking into account the correlation
matrix.

\medskip
\noindent Finally, we consider the sensitivity to the LEC $\bar{\ell}_3$.
In fig.\ref{fig14} we show total cross sections for the two channels 
$\pi^+ p\to \pi^+ \pi^+ n$ and  $\pi^- p\to \pi^0 \pi^0 n$ at $T_\pi \le 220\,$MeV
(which have already been investigated in the study of the low--energy 
theorems~\cite{bkmplb}).
In both cases the error bars of the recent data are smaller than the difference
in the curves with  $\bar{\ell}_3 = 2.9$ (standard case) and $-70$
(generalized case), in order. A similar trend
is observed for the differential cross sections $d\sigma/dt$
(at $T_\pi = 220, 240\,$MeV) , which have been used
in ref.\cite{kermcl} to extract $a_0^0$ by Chew--Low techniques. Our analysis
tends to support the finding of ref.\cite{bkmppn}, namely that one can extract
$a_0^0$ from the threshold data, but with an uncertainty which includes both the standard
as well as the generalized scenario. Presumably, a study of polarization observables
would give a better handle on this question. In  ref.\cite{bkmppn}, a
smaller theoretical uncertainty on $a_0^2$ was given, based on the
observation that there is no $N^* (1440)$ contribution to the
threshold amplitude ${\cal D}_{1}$. Taking into account the additional contribution
given in eq.(\ref{D12val}), we conclude that the theoretical
uncertainty for $a_0^2$ was underestimated in ref.\cite{bkmppn}.

\section{Summary}

We have performed a complete third order calculation of the reaction $\pi N\to
\pi\pi N$ in heavy baryon chiral perturbation theory based on the minimal Lagrangian
developed in ref.\cite{fms}. We had to consider 26 different tree graphs
and 49 one loop topologies as displayed in figs.~1-5. Note that the tree
contributions with fixed coefficients are obtained from the $1/m$
expansion of the relativistic amplitudes. 
The pertinent results of this investigation can be summarized as follows:
\begin{enumerate}
\item[i)] We have used the total cross sections in all five physical
channels (for $T_\pi \le 250\,$MeV, with $T_\pi$ the kinetic energy
of the incoming pion in the laboratory frame) and the older double differential cross 
sections $d^2\sigma / d\Omega dT$ for the process 
$\pi^- p \to \pi^+\pi^-n$ to fit the six new dimension three LECs. The
other four  dimension two and five (combinations of) dimension
 three LECs were taken from 
the study of elastic $\pi N$ scattering in ref.\cite{fms}. We observe that
the values of two pairs of these LECs are almost perfectly anticorrelated,
so that only four LECs can be determined. They come out of natural size.
\smallskip
\item[ii)] Using this input, we predict the total cross sections for
energies up to $T_\pi = 400\,$MeV. We find an excellent description
of the $\pi^-p \to\pi^+\pi^-n$ channel whereas the largest deviations
are seen for the
process $\pi^+p \to\pi^+\pi^+n$, see fig.\ref{fig8}.  
The angular correlation functions for
$\pi^-p \to\pi^+\pi^-n$ can be satisfactorily reproduced, with the exception of the
small $\theta_2$ angles, cf. figs.\ref{fig9},\ref{fig10}. 
In addition, most of the recent TRIUMF data
on $d\sigma / dM_{\pi\pi}^2$, $d\sigma / dt$,  $d\sigma / d\cos \theta$ and
 $d^2\sigma / dtdM_{\pi\pi}^2$ can also be reproduced. 
\smallskip
\item[iii)] At third order, the contribution from the loop graphs is essentially
negligible. Unitarity corrections therefore play no role. This allows one
to understand why resonance models like in refs.\cite{oset,jaeck} work fairly well
even in the threshold region (although these are not as precise as the calculation
presented here). The effect of the terms proportional to the dimension three
LECs is somewhat more pronounced. By far the largest contribution at this
order comes from the $1/m$ corrections to the dimension two LECs
$c_i$ and the $1/m^2$ corrections with fixed coefficients. This together
with the smallness of the loop contributions explains why the tree calculation 
in ref.\cite{bkmci} works so well. It also means that by far the most important
terms are the pion--nucleon subgraphs with an additional pion added in all
possible topologies.
\smallskip
\item[iv)] Our study is based on the standard scenario of chiral symmetry
breaking where a quark mass insertion is counted as second order in the
chiral expansion. This reflects itself in the natural size of the
mesonic LECs $\bar{\ell}_{1,2,3,4}$ used here. The generalized scenario with $m_q 
\sim {\cal O}(q)$ can be modeled by setting $\bar{\ell}_3 \simeq 
-70$~\cite{gl84}.\footnote{Clearly, this can not substitute for a detailed study within
generalized CHPT. Such a procedure should, however, be sufficient for
simply getting an idea about the sensitivity to the
$\pi\pi$ interaction.} 
Keeping all other mesonic and baryonic LECs fixed, we have studied the
sensitivity of the total and differential cross sections as well as the
angular correlation functions to the value of $\bar{\ell}_3$. 
We conclude that one is not able to pin down the LEC $\bar{\ell}_3$ with sufficient
accuracy to discriminate between the two scenarios of chiral symmetry breaking
by just comparing the observables directly with chiral analysis
(which is of course different from applying Chew--Low techniques).
We have also evaluated the dimension three counterterm contributions
to the threshold amplitudes and discussed the resonance saturation
approximation used in ref.\cite{bkmppn}.
\medskip

\end{enumerate}

There are two directions in which the study presented here should be extended. First,
it would be interesting to study the sensitivity to polarization. In particular,
it is conceivable that some polarization observables will be more sensitive
to the pion--pion interaction than the unpolarized observables studied here.
Second, a complete one--loop calculation requires to include the fourth order
terms. For doing that, one first has to work out elastic pion--nucleon scattering
to that order. Such investigations are underway and we hope to report on their
results soon.

\vfill
%\pagebreak

%%%%%%%%%%%%%%%%%%%%%%%%%%%%%%%%%%%%%%%%%%%%%%%%%%%%%%%
\appendix
\def\theequation{\Alph{section}.\arabic{equation}}
\setcounter{equation}{0}
\section{Contributions to the invariant amplitudes}

In this appendix, we explicitly give the contributions from the
tree and loop graphs shown in figs~1--5 to the invariant amplitudes
$A,B,C$ and $D$. Only the minimal number of non--vanishing amplitudes
as explained in section~3 is given. 

\medskip
\noindent
The calculation of the Born amplitude with fixed coefficients has been
done by 1/m expansion of the relativistic amplitude. In this expansion,
$\omega_2$ and $\omega_3$ have been treated as quantities of order $q$,
$t_1=(q_1-q_2)^2,t_2=(q_1-q_3)^2$ and $t_3=(q_2+q_3)^2= M_{\pi\pi}^2$ 
count as ${\cal O}(q^2)$. As a consequence, 
\bea
v\cdot p_1 &=& \frac{(\omega_2+\omega_3)^2-M_\pi^2}{2 m} + \ldots 
= {\cal O}(q^2)~,\\
v\cdot p_2 &=& \frac{(\omega_2+\omega_3)^2-t_3}{2 m} + \ldots 
= {\cal O}(q^2)~,\\
\omega_1 &=& (\omega_2+\omega_3)+{\cal O}(q^2)~,
\eea
where the ellipsis denotes higher order terms.
In diagrams 3a, 3b, 3n and 3o, there appear expressions in
$S\cdot p_1=-S\cdot q_1$ and $S\cdot p_2=-S\cdot(q_2+q_3)$. 
As stated before,
for these diagrams the amplitudes $B_1$,$B_2$,$B_4$,$D_2$ have to be given in addition
to the six usual ones, in order to fully determine the T-matrix element.
%Through an interchange of $q_2$ and $q_3$, the amplitude $A_3$ can be
%trivially recovered from $B_2$, $C_1$ from $B_1$, $C_2$ from $B_3$;
%$B_2$ directly gives $C_3$, $B_4$ gives $C_4$ and $D_2$ determines $D_3$.
\begin{enumerate}
\item{\underline{Counterterm amplitudes}}\\[0.3em]
Diagrams 2a+2b:\\
\bea
A_1 & = & -i \frac{g_A}{F^3} v\cdot(p_1 + p_2) \left[
-\frac{1}{\omega_1^2} (-4 c_1 M^2 - 2 c_2 \omega_2 \omega_3 -2 c_3 q_2\cdot q_3) \right.
\no\\&&\left.
+ c_4 (\omega_2 \omega_3 - q_2\cdot q_3) 
\left( \frac{1}{\omega_2^2}+\frac{1}{\omega_3^2} \right) 
\right]\no\\
A_2 & = & i \frac{g_A}{F^3} c_4 (\omega_2 \omega_3 - q_2\cdot q_3) 
\frac{v\cdot(p_1 + p_2)}{\omega_3^2}  \no\\
A_4 & = & i \frac{g_A}{F^3} c_4 (\omega_2 \omega_3 - q_2\cdot q_3) \left[
\frac{2}{\omega_2}-\frac{2}{\omega_3} + 
v\cdot(p_1 - p_2) \left( \frac{1}{\omega_2^2} -\frac{1}{\omega_3^2} \right) \right]\no\\
B_3 & = & i \frac{g_A}{F^3} c_4 (\omega_1 \omega_3 - q_1\cdot q_3) 
\frac{v\cdot(p_1 + p_2)}{\omega_1^2}  \no\\
D_1 & = & i \frac{g_A}{2 F^3} c_4 \left[
\frac{2}{\omega_2}-\frac{2}{\omega_3} + 
v\cdot(p_1 - p_2) \left( \frac{1}{\omega_2^2} -\frac{1}{\omega_3^2} \right) \right]\no\\
D_4 & = & -i \frac{g_A}{2 F^3} c_4 v\cdot(p_1 + p_2)
\left( \frac{1}{\omega_1^2} + \frac{1}{\omega_2^2}+\frac{1}{\omega_3^2} \right) 
\eea
Diagrams 3a+3b:\\
\bea
A_1 & = & -i \frac{g_A}{m F^3} c_4 [\omega_1 (\omega_2+\omega_3) -
\omega_2^2-\omega_3^2 -q_1\cdot (q_2+q_3)+2 M^2 ]\no\\
A_2 & = & -i \frac{g_A}{m F^3} [ -4 c_1 M^2 + 2 c_2 \omega_1 \omega_3 +
2 c_3 q_1\cdot q_3- c_4 (\omega_1 \omega_2-\omega_2^2 - q_1\cdot q_2+M^2)]\no\\
A_4 & = & i \frac{g_A}{m F^3} c_4 [\omega_1 (\omega_3-\omega_2)+\omega_3^2
-\omega_2^2-q_1\cdot( q_3-q_2) ]\no\\
B_1 & = & i \frac{g_A}{m F^3} [ -4 c_1 M^2 - 2 c_2 \omega_2 \omega_3 -
2 c_3 q_2\cdot q_3- c_4 (\omega_1 \omega_2-\omega_1^2 - q_1\cdot q_2+M^2)]\no\\
B_2 & = & -i \frac{g_A}{m F^3} c_4 [\omega_2 (\omega_3-\omega_1) +
\omega_3^2+\omega_1^2 -q_2\cdot (q_3-q_1)-2 M^2 ]\no\\
B_3 & = & i \frac{g_A}{m F^3} [ -4 c_1 M^2 + 2 c_2 \omega_1 \omega_2 +
2 c_3 q_1\cdot q_2+ c_4 (\omega_2 \omega_3+\omega_3^2 - q_2\cdot q_3-M^2)]\no\\
B_4 & = & i \frac{g_A}{m F^3} c_4 [(\omega_1+\omega_3) (\omega_1+\omega_2+\omega_3) 
-(q_1+q_3)\cdot (q_1+q_2+q_3)]\no\\
D_2 & = & i \frac{g_A}{m F^3} c_4\no\\
D_1 & = & D_4 = 0\no \\ 
\eea
Diagrams 3c+3d:\\
\bea
A_1 & = & -i \frac{g_A}{2 m F^3} \left[
\frac{(\omega_1+\omega_2+\omega_3)^2-(q_1+q_2+q_3)^2}{\omega_1^2} 
(-4 c_1 M^2 - 2 c_2 \omega_2 \omega_3 -2 c_3 q_2\cdot q_3) 
\right.\no\\&&\left.
-c_4 (\omega_2 \omega_3 - q_2\cdot q_3) 
\left( \frac{(\omega_1+\omega_2)^2+\omega_3^2-(q_1+q_2)^2-M^2}{\omega_2^2} 
\right.\right.\no\\&&\left.\left.+
\frac{(\omega_1+\omega_3)^2+\omega_2^2-(q_1+q_3)^2-M^2}{\omega_3^2} \right) 
\right]\no\\
A_2 & = & -i \frac{g_A}{2 m F^3} c_4 (\omega_2 \omega_3 - q_2\cdot q_3) 
\frac{(\omega_1+\omega_3)^2+\omega_2^2-(q_1+q_3)^2-M^2}{\omega_3^2} \no\\
A_4 & = & -i \frac{g_A}{2 m F^3} c_4 (\omega_2 \omega_3 - q_2\cdot q_3) 
\left( \frac{(\omega_1+\omega_2)^2-\omega_3^2-(q_1+q_2)^2+M^2}{\omega_2^2} 
\right.\no\\&&\left.-
\frac{(\omega_1+\omega_3)^2-\omega_2^2-(q_1+q_3)^2+M^2}{\omega_3^2} \right)\no\\
B_3 & = &  -i \frac{g_A}{2 m F^3} c_4 (\omega_1 \omega_3 - q_1\cdot q_3) 
\frac{(\omega_1+\omega_2+\omega_3)^2-(q_1+q_2+q_3)^2}{\omega_1^2}\no\\
D_1 & = & -i \frac{g_A}{4 m F^3} c_4 \no\\ 
&&\left( \frac{(\omega_1+\omega_2)^2-\omega_3^2-(q_1+q_2)^2+M^2}{\omega_2^2} -
\frac{(\omega_1+\omega_3)^2-\omega_2^2-(q_1+q_3)^2+M^2}{\omega_3^2} \right)\no\\
D_4 & = & i \frac{g_A}{4 m F^3} c_4 \left( 
\frac{(\omega_1+\omega_2+\omega_3)^2-(q_1+q_2+q_3)^2}{\omega_1^2} 
\right.\no\\&&\left.+
\frac{(\omega_1+\omega_2)^2+\omega_3^2-(q_1+q_2)^2-M^2}{\omega_2^2} +
\frac{(\omega_1+\omega_3)^2+\omega_2^2-(q_1+q_3)^2-M^2}{\omega_3^2} \right)\no\\&& 
\eea
Diagrams 3e+3f+3g:\\
\bea
A_1 & = & -i \frac{g_A^2}{2 F^3} (\omega_2 \omega_3 - q_2\cdot q_3)\no
\\&& \Bigg[\frac{1}{\omega_1 \omega_3} (12 M^2 d_{16}(\lambda)-6 M^2 d_{18} 
     + \tilde{d}_{25}(\lambda) \omega_1 \omega_3 
     +\tilde{d}_{29} (\lambda) (\omega_1^2+\omega_2^2+\omega_3^2) )\no
\\&&-\frac{1}{\omega_2 \omega_3} (12 M^2 d_{16}(\lambda)-6 M^2 d_{18} 
     - \tilde{d}_{25}(\lambda) \omega_2 \omega_3 
     +\tilde{d}_{29} (\lambda) (\omega_1^2+\omega_2^2+\omega_3^2) )\no
\\&&+\frac{1}{\omega_1 \omega_2} (12 M^2 d_{16}(\lambda)-6 M^2 d_{18} 
     + \tilde{d}_{25}(\lambda) \omega_1 \omega_2 
     +\tilde{d}_{29} (\lambda) (\omega_1^2+\omega_2^2+\omega_3^2) )
\Bigg]\no\\
A_2 & = & -i \frac{g_A^2}{2 F^3} (\omega_2 \omega_3 - q_2\cdot q_3)\no
\\&& \Bigg[-\frac{1}{\omega_1 \omega_3} (12 M^2 d_{16}(\lambda)-6 M^2 d_{18} 
     + \tilde{d}_{25}(\lambda) \omega_1 \omega_3 
     +\tilde{d}_{29} (\lambda) (\omega_1^2+\omega_2^2+\omega_3^2) )\no
\\&&+\frac{1}{\omega_2 \omega_3} (12 M^2 d_{16}(\lambda)-6 M^2 d_{18} 
     - \tilde{d}_{25}(\lambda) \omega_2 \omega_3 
     +\tilde{d}_{29} (\lambda) (\omega_1^2+\omega_2^2+\omega_3^2) )\no
\\&&+\frac{1}{\omega_1 \omega_2} (12 M^2 d_{16}(\lambda)-6 M^2 d_{18} 
     + \tilde{d}_{25}(\lambda) \omega_1 \omega_2 
     +\tilde{d}_{29} (\lambda) (\omega_1^2+\omega_2^2+\omega_3^2) )
\Bigg]\no\\
B_3 & = & i \frac{g_A^2}{2 F^3} (\omega_1 \omega_3 - q_1\cdot q_3)\no
\\&& \Bigg[\frac{1}{\omega_1 \omega_3} (12 M^2 d_{16}(\lambda)-6 M^2 d_{18} 
     + \tilde{d}_{25}(\lambda) \omega_1 \omega_3 
     +\tilde{d}_{29} (\lambda) (\omega_1^2+\omega_2^2+\omega_3^2) )\no
\\&&+\frac{1}{\omega_2 \omega_3} (12 M^2 d_{16}(\lambda)-6 M^2 d_{18} 
     - \tilde{d}_{25}(\lambda) \omega_2 \omega_3 
     +\tilde{d}_{29} (\lambda) (\omega_1^2+\omega_2^2+\omega_3^2) )\no
\\&&+\frac{1}{\omega_1 \omega_2} (12 M^2 d_{16}(\lambda)-6 M^2 d_{18} 
     + \tilde{d}_{25}(\lambda) \omega_1 \omega_2 
     +\tilde{d}_{29} (\lambda) (\omega_1^2+\omega_2^2+\omega_3^2) )
\Bigg]\no\\
D_4 & = & -i \frac{g_A^2}{4 F^3}\no
\\&& \Bigg[\frac{1}{\omega_1 \omega_3} (12 M^2 d_{16}(\lambda)-6 M^2 d_{18} 
     + \tilde{d}_{25}(\lambda) \omega_1 \omega_3 
     +\tilde{d}_{29} (\lambda) (\omega_1^2+\omega_2^2+\omega_3^2) )\no
\\&&-\frac{1}{\omega_2 \omega_3} (12 M^2 d_{16}(\lambda)-6 M^2 d_{18} 
     - \tilde{d}_{25}(\lambda) \omega_2 \omega_3 
     +\tilde{d}_{29} (\lambda) (\omega_1^2+\omega_2^2+\omega_3^2) )\no
\\&&+\frac{1}{\omega_1 \omega_2} (12 M^2 d_{16}(\lambda)-6 M^2 d_{18} 
     + \tilde{d}_{25}(\lambda) \omega_1 \omega_2 
     +\tilde{d}_{29} (\lambda) (\omega_1^2+\omega_2^2+\omega_3^2) )
\Bigg]\no\\
A_4 & = & D_1 = 0
\eea
Diagrams 3h+3i:\\
\bea
A_1 & = & -i \frac{g_A^3}{2 F^3} (\omega_2\omega_3-q_2\cdot q_3) \Bigg[
\frac{1}{\omega_1\omega_3}(\tilde{d}_{24}(\lambda) (\omega_1^2+\omega_3^2)
                          +16 M^2 \tilde{d}_{28}(\lambda))\no\\&&
+\frac{1}{\omega_1\omega_2}(\tilde{d}_{24}(\lambda) (\omega_1^2+\omega_2^2)
                          +16 M^2 \tilde{d}_{28}(\lambda))
-\frac{1}{\omega_2\omega_3}(\tilde{d}_{24}(\lambda) (\omega_2^2+\omega_3^2)
                          +16 M^2 \tilde{d}_{28}(\lambda)) \Bigg]\no\\
A_2 & = & -i \frac{g_A^3}{2 F^3} (\omega_2\omega_3-q_2\cdot q_3) \Bigg[
-\frac{1}{\omega_1\omega_3}(\tilde{d}_{24}(\lambda) (\omega_1^2+\omega_3^2)
                          +16 M^2 \tilde{d}_{28}(\lambda))\no\\&&
+\frac{1}{\omega_1\omega_2}(\tilde{d}_{24}(\lambda) (\omega_1^2+\omega_2^2)
                          +16 M^2 \tilde{d}_{28}(\lambda))
+\frac{1}{\omega_2\omega_3}(\tilde{d}_{24}(\lambda) (\omega_2^2+\omega_3^2)
                          +16 M^2 \tilde{d}_{28}(\lambda)) \Bigg]\no\\
A_4 & = & 0\no\\
B_3 & = & i \frac{g_A^3}{2 F^3} (\omega_1\omega_3-q_1\cdot q_3) \Bigg[
\frac{1}{\omega_1\omega_3}(\tilde{d}_{24}(\lambda) (\omega_1^2+\omega_3^2)
                          +16 M^2 \tilde{d}_{28}(\lambda))\no\\&&
+\frac{1}{\omega_1\omega_2}(\tilde{d}_{24}(\lambda) (\omega_1^2+\omega_2^2)
                          +16 M^2 \tilde{d}_{28}(\lambda))
+\frac{1}{\omega_2\omega_3}(\tilde{d}_{24}(\lambda) (\omega_2^2+\omega_3^2)
                          +16 M^2 \tilde{d}_{28}(\lambda)) \Bigg]\no\\
D_4 & = & -i \frac{g_A^3}{4 F^3} \Bigg[
\frac{1}{\omega_1\omega_3}(\tilde{d}_{24}(\lambda) (\omega_1^2+\omega_3^2)
                          +16 M^2 \tilde{d}_{28}(\lambda))\no\\&&
+\frac{1}{\omega_1\omega_2}(\tilde{d}_{24}(\lambda) (\omega_1^2+\omega_2^2)
                          +16 M^2 \tilde{d}_{28}(\lambda))
-\frac{1}{\omega_2\omega_3}(\tilde{d}_{24}(\lambda) (\omega_2^2+\omega_3^2)
                          +16 M^2 \tilde{d}_{28}(\lambda)) \Bigg]\no\\
D_1 & = & 0
\eea
Diagrams 3j+3k:\\
\bea
A_2 & = & -i \frac{1}{2 F^3} \frac{\omega_2-\omega_3}{\omega_1}
(4 M^2 d_{16}(\lambda) - 2 M^2 d_{18})\no\\
B_3 & = & -i \frac{1}{2 F^3} \frac{\omega_1+\omega_3}{\omega_2}
(4 M^2 d_{16}(\lambda) - 2 M^2 d_{18})\no\\
A_1 & = & A_4 = D_1 = D_4 = 0
\eea
Diagrams 3l+3m:\\
\bea
A_2 & = & -i \frac{g_A}{2 F^3} \frac{\omega_2-\omega_3}{\omega_1} 
(\tilde{d}_{24}(\lambda) \omega_1^2 - 8 M^2 \tilde{d}_{28}(\lambda) )\no\\
B_3 & = &  -i \frac{g_A}{2 F^3} \frac{\omega_1+\omega_3}{\omega_2} 
(\tilde{d}_{24}(\lambda) \omega_2^2 - 8 M^2 \tilde{d}_{28}(\lambda) )\no\\
A_1 & = & A_4 = D_1 = D_4 =  0
\eea
Diagrams 3n+3o:\\ 
\bea
A_1 & = & i \frac{g_A}{F^3} \Bigg[ 
2 \frac{c_2}{m} \frac{\omega_2 q_1\cdot q_3 + \omega_3 q_1\cdot q_2}{\omega_1}
-\frac{c_4}{m} \left( \frac{\omega_3 (\omega_2^2-M^2)}{\omega_2} + 
\frac{\omega_2 (\omega_3^2-M^2)}{\omega_3} \right) \no\\&&
+8 \tilde{d}_{26}(\lambda)q_2\cdot q_3 + 
8 \tilde{d}_{27}(\lambda)\omega_2 \omega_3 
+8 M^2 \tilde{d}_{28}(\lambda)
-8 \tilde{d}_{30}(\lambda) (\omega_2\omega_3 - q_2\cdot q_3)
\Bigg]\no\\
A_2 & = & i \frac{g_A}{F^3} \Bigg[ 
\frac{c_4}{m} \left( 
-\frac{(\omega_2-\omega_3)(q_2\cdot q_3+M^2)+\omega_3 q_1\cdot q_2 
   - \omega_2 q_1\cdot q_3}{\omega_1}+\frac{\omega_2(\omega_3^2-M^2)}{\omega_3}\right)
\no\\&&
+4 (d_1(\lambda)+d_2) \frac{(\omega_2-\omega_3) q_2\cdot q_3}{\omega_1}
+4 d_3(\lambda) \frac{(\omega_2-\omega_3)\omega_2\omega_3}{\omega_1}
-8 M^2 d_5(\lambda) \frac{(\omega_2-\omega_3)}{\omega_1}\no\\&&
+2(d_{14}(\lambda)-d_{15})\frac{(\omega_1+\omega_3)(\omega_2\omega_3-q_2\cdot q_3)}
{\omega_2} 
+\frac{1}{2} \tilde{d}_{24}(\lambda) \omega_1 (\omega_2-\omega_3)\no\\&&
-4 M^2\tilde{d}_{28}(\lambda) \frac{\omega_2-\omega_3}{\omega_1}
+4 \tilde{d}_{30}(\lambda) (\omega_2\omega_3-q_2\cdot q_3)
\Bigg]\no\\
A_4 & = & i \frac{g_A}{m F^3} c_4  \Bigg[
\frac{\omega_3 q_1\cdot q_2-\omega_2 q_1\cdot q_3}{\omega_1}
+\frac{(\omega_1+\omega_3)(\omega_2 \omega_3-q_2\cdot q_3) 
+ \omega_3 (\omega_2^2-M^2)}{\omega_2} \no\\&& 
-\frac{(\omega_1+\omega_2)(\omega_2 \omega_3-q_2\cdot q_3) 
+ \omega_2 (\omega_3^2-M^2)}{\omega_3} \Bigg]\no\\
B_1 & = & i \frac{g_A}{F^3} \Bigg[ 
\frac{c_4}{m} \left( 
\frac{-(\omega_1+\omega_3)(q_1\cdot q_3-M^2)-\omega_3 q_1\cdot q_2 
   + \omega_1 q_2\cdot q_3}{\omega_2}+\frac{\omega_1(\omega_3^2-M^2)}{\omega_3}\right)
\no\\&&
+4 (d_1(\lambda)+d_2) \frac{(\omega_1+\omega_3) q_1\cdot q_3}{\omega_2}
+4 d_3(\lambda) \frac{(\omega_1+\omega_3)\omega_1\omega_3}{\omega_2}
+8 M^2 d_5(\lambda) \frac{(\omega_1+\omega_3)}{\omega_2}\no\\&&
+2(d_{14}(\lambda)-d_{15})\frac{(\omega_2-\omega_3)(\omega_1\omega_3-q_1\cdot q_3)}
{\omega_1} 
-\frac{1}{2} \tilde{d}_{24}(\lambda) \omega_2 (\omega_1+\omega_3)\no\\&&
+4 M^2\tilde{d}_{28}(\lambda) \frac{\omega_1+\omega_3}{\omega_2}
+4 \tilde{d}_{30}(\lambda) (\omega_1\omega_3-q_1\cdot q_3)
\Bigg]\no\\
B_2 & = & i \frac{g_A}{F^3} \Bigg[ 
2 \frac{c_2}{m} \frac{\omega_1 q_2\cdot q_3 + \omega_3 q_1\cdot q_2}{\omega_2}
-\frac{c_4}{m} \left( \frac{\omega_3 (\omega_1^2-M^2)}{\omega_1} + 
\frac{\omega_1 (\omega_3^2-M^2)}{\omega_3} \right) \no\\&&
+8 \tilde{d}_{26}(\lambda)q_1\cdot q_3 + 
8 \tilde{d}_{27}(\lambda)\omega_1 \omega_3 
-8 M^2 \tilde{d}_{28}(\lambda)
-8 \tilde{d}_{30}(\lambda) (\omega_1\omega_3 - q_1\cdot q_3)
\Bigg]\no\\
B_3 & = & i \frac{g_A}{F^3} \Bigg[ 
\frac{c_4}{m} \left( 
\frac{\omega_3(\omega_1^2-M^2)}{\omega_1}+
\frac{(\omega_1+\omega_3)(q_1\cdot q_3-M^2)+\omega_3 q_1\cdot q_2 
   - \omega_1 q_2\cdot q_3}{\omega_2}\right)\no
\\&&
-4 (d_1(\lambda)+d_2) \frac{(\omega_1+\omega_3) q_1\cdot q_3}{\omega_2}
-4 d_3(\lambda) \frac{(\omega_1+\omega_3)\omega_1\omega_3}{\omega_2}
-8 M^2 d_5(\lambda) \frac{(\omega_1+\omega_3)}{\omega_2}\no\\&&
-2(d_{14}(\lambda)-d_{15})\frac{(\omega_1+\omega_2)(\omega_1\omega_3-q_1\cdot q_3)}
{\omega_3} 
+\frac{1}{2} \tilde{d}_{24}(\lambda) \omega_2 (\omega_1+\omega_3)\no\\&&
-4 M^2\tilde{d}_{28}(\lambda) \frac{\omega_1+\omega_3}{\omega_2}
+4 \tilde{d}_{30}(\lambda) (\omega_1\omega_3-q_1\cdot q_3)
\Bigg]\no\\
B_4 & = & i \frac{g_A}{m F^3} c_4  \Bigg[
\frac{(\omega_2+\omega_3)(\omega_1 \omega_3-q_1\cdot q_3) 
+ \omega_3 (\omega_1^2-M^2)}{\omega_1}  
+\frac{\omega_3 q_1\cdot q_2-\omega_1 q_2\cdot q_3}{\omega_2}\no\\&&
+\frac{(\omega_1+\omega_2)(\omega_1 \omega_3-q_1\cdot q_3) 
+ \omega_1 (\omega_3^2-M^2)}{\omega_3} \Bigg]\no\\
D_1 & = & i \frac{g_A}{2 m F^3} c_4  \Bigg[
\frac{\omega_1+\omega_3}{\omega_2} - \frac{\omega_1+\omega_2}{\omega_3}\Bigg]\no\\
D_2 & = & i \frac{g_A}{2 m F^3} c_4  \Bigg[
\frac{\omega_2+\omega_3}{\omega_1} + \frac{\omega_1+\omega_2}{\omega_3}\Bigg]\no\\
D_4 & = & -6 i \frac{g_A}{F^3} \tilde{d}_{30}(\lambda)
\eea
Diagram 3p:\\
\bea
A_1 & = & \frac{i}{F^3} \Bigg[
2\frac{c_4}{m}g_A \omega_2 \omega_3 -4 d_{10}(\lambda) q_2\cdot q_3 
-4 d_{12}(\lambda) \omega_2\omega_3-4 M^2 d_{16}(\lambda)+2 M^2 d_{18} \Bigg]\no\\
A_2 & = & \frac{i}{F^3} \Bigg[
-\frac{c_4}{m}g_A \omega_2 \omega_3 -2 d_{11}(\lambda) q_2\cdot q_3
-2 d_{13}(\lambda) \omega_2\omega_3+2 M^2 d_{16}(\lambda)- M^2 d_{18} \no\\&&
+\frac{1}{2} \tilde{d}_{29}(\lambda) (\omega_2^2-\omega_3^2) \Bigg]\no\\
B_3 & = & \frac{i}{F^3} \Bigg[
-\frac{c_4}{m}g_A \omega_1 \omega_3 -2 d_{11}(\lambda) q_1\cdot q_3
-2 d_{13}(\lambda) \omega_1\omega_3-2 M^2 d_{16}(\lambda)+ M^2 d_{18} \no\\&&
+\frac{1}{2} \tilde{d}_{29}(\lambda) (\omega_1^2-\omega_3^2) \Bigg]\no\\
D_4 & = & \frac{i}{F^3} 12 d_4(\lambda)\no\\
A_4 & = & D_1 = 0
\eea
Diagram 3q:\\
\bea
A_1 & = & - \frac{i}{F^3} \frac{1}{(q_1-q_2-q_3)^2-M^2}
[4 M^2 d_{16}(\lambda) - 2 M^2 d_{18}] (t_3 - M^2)\no\\
A_2 & = & - \frac{i}{F^3} \frac{1}{(q_1-q_2-q_3)^2-M^2}
[4 M^2 d_{16}(\lambda) - 2 M^2 d_{18}] (t_2 - M^2)\no\\
B_3 & = & \frac{i}{F^3} \frac{1}{(q_1-q_2-q_3)^2-M^2}
[4 M^2 d_{16}(\lambda) - 2 M^2 d_{18}] (t_1 - M^2)\no\\
A_4 & = & D_1 = D_4 = 0
\eea
Diagram 3r: Combine with diagrams R1e and s35.\\
Diagram 3s:\\
\bea
A_1 & = & -i \frac{g_A}{F^5} \frac{1}{(q_1-q_2-q_3)^2-M^2}
\Bigg[ 8 {\ell}_1 q_1\cdot (q_2+q_3-q_1) q_2\cdot q_3 \no\\&&
+4 {\ell}_2 [q_3\cdot (q_2+q_3-q_1) q_1\cdot q_2 + q_2\cdot(q_2+q_3-q_1) q_1\cdot q_3]
\no\\&&
+2 M^2 {\ell}_4 [q_1\cdot(q_2+q_3-q_1)+ q_2\cdot q_3 +(q_2+q_3)^2] \Bigg]\no\\
A_2 & = & -i \frac{g_A}{F^5} \frac{1}{(q_1-q_2-q_3)^2-M^2}
\Bigg[ 8 {\ell}_1 q_2\cdot (q_2+q_3-q_1) q_1\cdot q_3 \no\\&&
+4 {\ell}_2 [q_3\cdot (q_2+q_3-q_1) q_1\cdot q_2 + q_1\cdot(q_2+q_3-q_1) q_2\cdot q_3]
\no\\&&
-2 M^2 {\ell}_4 [q_2\cdot(q_2+q_3-q_1) + q_1\cdot q_3 -(q_1-q_3)^2] \Bigg]\no\\
B_3 & = & i \frac{g_A}{F^5} \frac{1}{(q_1-q_2-q_3)^2-M^2}
\Bigg[ 8 {\ell}_1 q_3\cdot (q_2+q_3-q_1) q_1\cdot q_2 \no\\&&
+4 {\ell}_2 [q_2\cdot (q_2+q_3-q_1) q_1\cdot q_3 + q_1\cdot(q_2+q_3-q_1) q_2\cdot q_3]
\no\\&&
-2 M^2 {\ell}_4 [q_3\cdot(q_2+q_3-q_1)+ q_1\cdot q_2 -(q_1-q_2)^2] \Bigg]\no\\
A_4 & = & D_1 = D_4  = 0
\eea
%\vfill
\item{\underline{Tadpoles}}\\[0.3em]
We use the loop functions as defined in ref.\cite{mm}.\\
Diagrams t1+t2+t3:\\
\bea
A_1 & = & -\frac{3 i }{4}\frac{g_A^3}{F^5} (\omega_2\omega_3-q_2 \cdot q_3) \Delta_\pi
\left(\frac{1}{\omega_1\omega_2}+\frac{1}{\omega_1\omega_3}-\frac{1}{\omega_2\omega_3}
\right)\no\\
A_2 & = & -\frac{3 i }{4}\frac{g_A^3}{F^5} (\omega_2\omega_3-q_2\cdot q_3) \Delta_\pi
\left(\frac{1}{\omega_1\omega_2}-\frac{1}{\omega_1\omega_3}+\frac{1}{\omega_2\omega_3}
\right)\no\\
B_3 & = & \frac{3 i }{4}\frac{g_A^3}{F^5} (\omega_1\omega_3-q_1\cdot q_3) \Delta_\pi
\left(\frac{1}{\omega_1\omega_2}+\frac{1}{\omega_1\omega_3}+\frac{1}{\omega_2\omega_3}
\right)\no\\
D_4 & = & -\frac{3 i }{8}\frac{g_A^3}{F^5} \Delta_\pi
\left(\frac{1}{\omega_1\omega_2}+\frac{1}{\omega_1\omega_3}-\frac{1}{\omega_2\omega_3}
\right)\no\\
A_4 & = & D_1  =  0
\eea
Diagrams t4+t5:\\
\bea
A_2 & = & -\frac{i}{4}\frac{g_A}{F^5} \Delta_\pi \frac{\omega_2-\omega_3}{\omega_1}\no\\
B_3 & = & -\frac{i}{4}\frac{g_A}{F^5} \Delta_\pi \frac{\omega_1+\omega_3}{\omega_2}\no\\
A_1 & = & A_4 = D_1 = D_4 = 0
\eea
Diagrams t6+t7:\\
\bea
A_2 & = & -\frac{5 i}{8}\frac{g_A}{F^5} \Delta_\pi \frac{\omega_2-\omega_3}{\omega_1}
\no\\
B_3 & = & -\frac{5 i}{8}\frac{g_A}{F^5} \Delta_\pi \frac{\omega_1+\omega_3}{\omega_2}
\no\\
A_1 & = & A_4 = D_1 = D_4 = 0
\eea
Diagrams t8+t9:\\
\bea
A_2 & = & i \frac{g_A}{F^5} \frac{\omega_2-\omega_3}{\omega_1} I_2(t_3)\no\\
B_3 & = & i \frac{g_A}{F^5} \frac{\omega_1+\omega_3}{\omega_2} I_2(t_2)\no\\
A_1 & = & A_4 = D_1 = D_4 = 0
\eea
Diagram t10:\\
\bea
A_1 & = & \frac{i}{4} \frac{g_A}{F^5} \Bigg\{ 
2 (t_3-M^2) I_0(t_3)  \no\\&&
+\frac{1}{3}\left[ \left(3 q_1\cdot q_3-\frac{t_2}{2} + 2 M^2 \right) I_0(t_2)
-4 \Delta_\pi +\frac{1}{16 \pi^2} \left(2 M^2-\frac{t_2}{3} \right) \right]  \no\\&&
+\frac{1}{3}\left[ \left(3 q_1 \cdot q_2-\frac{t_1}{2} + 2 M^2 \right) I_0(t_1)
-4 \Delta_\pi +\frac{1}{16 \pi^2} \left(2 M^2-\frac{t_1}{3} \right) \right]
\Bigg\}\no\\
A_2 & = & \frac{i}{4} \frac{g_A}{F^5} \Bigg\{ 
(2 M^2-t_3) I_0(t_3)-2 \Delta_\pi
+(3 t_2-2 M^2) I_0(t_2)-2 \Delta_\pi \no\\&&
-\frac{1}{3}\left[ \left(-3 q_1 \cdot q_2-\frac{t_1}{2} + 2 M^2 \right) I_0(t_1)
+2 \Delta_\pi +\frac{1}{16 \pi^2} \left(2 M^2-\frac{t_1}{3} \right) \right]
\Bigg\}\no\\
B_3 & = &  \frac{i}{4} \frac{g_A}{F^5} \Bigg\{ 
\frac{1}{3}\left[ \left(3 q_2 \cdot q_3-\frac{t_3}{2} + 2 M^2 \right) I_0(t_3)
+2 \Delta_\pi +\frac{1}{16 \pi^2} \left(2 M^2-\frac{t_3}{3} \right) \right] \no\\&&
-(2 M^2-\frac{t_2}{2}) I_0(t_2)+2 \Delta_\pi
-(3 t_1-2 M^2) I_0(t_1)+2 \Delta_\pi
\Bigg\}\no\\
A_4 & = & D_1 = D_4 = 0
\eea
Diagram t11:\\
\bea
A_1 & = & \frac{3 i }{4}\frac{g_A}{F^5} \Delta_\pi\no\\
A_2 & = & \frac{21 i }{8}\frac{g_A}{F^5} \Delta_\pi\no\\
B_3 & = & -\frac{21 i }{8}\frac{g_A}{F^5} \Delta_\pi\no\\
A_4 & = & D_1 = D_4 = 0
\eea
Diagram t12:\\
\bea
A_1 & = & -i\frac{g_A}{2 F^5} \frac{1}{(q_1-q_2-q_3)^2-M^2} \Delta_\pi (t_3-M^2)\no\\
A_2 & = & -i\frac{g_A}{2 F^5} \frac{1}{(q_1-q_2-q_3)^2-M^2} \Delta_\pi (t_2-M^2)\no\\
B_3 & = & i\frac{g_A}{2 F^5} \frac{1}{(q_1-q_2-q_3)^2-M^2} \Delta_\pi (t_1-M^2)\no\\
A_4 & = & D_1 = D_4 = 0
\eea
Diagram t13:\\
\bea
A_1 & = & -i\frac{g_A}{2 F^5}\frac{1}{(q_1-q_2-q_3)^2-M^2} \no\\&&
\Bigg\{
(t_3-M^2) [I_0(t_3)(2 t_3+M^2-2 q_1\cdot(q_2+q_3)) -4 \Delta_\pi]\no\\&&
+I_0(t_2)[M^2(2 M^2-t_2)+\frac{q_1\cdot q_2}{3}(4 t_2 -10 M^2) 
-\frac{q_2\cdot q_3}{3}(2 t_2 -2 M^2)]\no\\&&
+\Delta_\pi(-4 M^2 +\frac{8}{3} q_1\cdot q_2
-\frac{4}{3} q_2\cdot q_3)-\frac{1}{72 \pi^2}(6 M^2 -  t_2) q_2\cdot(q_1+q_3)\no\\&&
+I_0(t_1)[M^2(2 M^2-t_1)+\frac{q_1\cdot q_3}{3}(4 t_1 -10 M^2) 
-\frac{q_2\cdot q_3}{3}(2 t_1 -2 M^2)]\no\\&&
+\Delta_\pi(-4 M^2 +\frac{8}{3} q_1\cdot q_3
-\frac{4}{3} q_2\cdot q_3)-\frac{1}{72 \pi^2}(6 M^2 -  t_1) q_3\cdot(q_1+q_2)
\Bigg\}\no\\
A_2 & = & -i\frac{g_A}{2 F^5}\frac{1}{(q_1-q_2-q_3)^2-M^2} \no\\&&\Bigg\{
I_0(t_3)[M^2(2 M^2-t_3)+\frac{q_1\cdot q_2}{3}(4 t_3 -10 M^2) 
+\frac{q_1\cdot q_3}{3}(2 t_3 -2 M^2)]\no\\&&
+\Delta_\pi(-4 M^2 +\frac{8}{3} q_1\cdot q_2
+\frac{4}{3} q_1\cdot q_3)-\frac{1}{72 \pi^2}(6 M^2 -  t_3) q_1\cdot(q_2-q_3)\no\\&&
+(t_2-M^2) [I_0(t_2)(2 t_2+M^2-2 q_2\cdot(q_1-q_3)) -4 \Delta_\pi]\no\\&&
+I_0(t_1)[M^2(2 M^2-t_1)-\frac{q_2\cdot q_3}{3}(4 t_1 -10 M^2) 
+\frac{q_1\cdot q_3}{3}(2 t_1 -2 M^2)]\no\\&&
+\Delta_\pi(-4 M^2 -\frac{8}{3} q_2\cdot q_3
+\frac{4}{3} q_1\cdot q_3)+\frac{1}{72 \pi^2}(6 M^2 -  t_1) q_3\cdot(q_1+q_2)
\Bigg\}\no\\
B_3 & = & i\frac{g_A}{2 F^5}\frac{1}{(q_1-q_2-q_3)^2-M^2} \no\\&&\Bigg\{
I_0(t_3)[M^2(2 M^2-t_3)+\frac{q_1\cdot q_3}{3}(4 t_3 -10 M^2) 
+\frac{q_1\cdot q_2}{3}(2 t_3 -2 M^2)]\no\\&&
+\Delta_\pi(-4 M^2 +\frac{8}{3} q_1\cdot q_3
+\frac{4}{3} q_1\cdot q_2)-\frac{1}{72 \pi^2}(6 M^2 -  t_3) q_1\cdot(q_3-q_2)\no\\&&
+I_0(t_2)[M^2(2 M^2-t_2)-\frac{q_2\cdot q_3}{3}(4 t_2 -10 M^2) 
+\frac{q_1\cdot q_2}{3}(2 t_2 -2 M^2)]\no\\&&
+\Delta_\pi(-4 M^2 -\frac{8}{3} q_2\cdot q_3
+\frac{4}{3} q_1\cdot q_2)+\frac{1}{72 \pi^2}(6 M^2 -  t_2) q_2\cdot(q_1+q_3)\no\\&&
+(t_1-M^2) [I_0(t_1)(2 t_1+M^2-2 q_3\cdot(q_1-q_2)) -4 \Delta_\pi]
\Bigg\}\no\\
A_4 & = & D_1 = D_4 = 0
\eea
Diagram t14:\\
\bea
A_1 & = & i\frac{g_A}{F^5} \frac{1}{(q_1-q_2-q_3)^2-M^2} \Delta_\pi 
\left[-5 t_3 + \frac{7}{2} M^2 -(q_2+q_3-q_1)^2\right]\no\\
A_2 & = & i\frac{g_A}{F^5} \frac{1}{(q_1-q_2-q_3)^2-M^2} \Delta_\pi 
\left[-5 t_2 + \frac{7}{2} M^2 -(q_2+q_3-q_1)^2\right]\no\\
B_3 & = & -i\frac{g_A}{F^5} \frac{1}{(q_1-q_2-q_3)^2-M^2} \Delta_\pi 
\left[-5 t_1 + \frac{7}{2} M^2 -(q_2+q_3-q_1)^2\right]\no\\
A_4 & = & D_1 = D_4 = 0
\eea
\item{\underline{Self--energies}}\\[0.3em]
We use the loop functions as defined in ref.\cite{mm}.\\
Diagrams s1+s3:\\
\bea
A_1 & = & -\frac{i}{4} \left( \frac{g_A}{F}\right)^5 (\omega_2 \omega_3-q_2 \cdot q_3)
\left(S^2+\frac{1}{2}\right) \left[ 
\frac{J_2(\omega_1)-J_2(-\omega_1)}{\omega_1^2} 
\left( \frac{1}{\omega_2}+\frac{1}{\omega_3}\right)\right.\no\\
&&\left.+\frac{J_2(\omega_2)-J_2(-\omega_2)}{\omega_2^2} 
\left( \frac{1}{\omega_1}-\frac{1}{\omega_3}\right)
+ \frac{J_2(\omega_3)-J_2(-\omega_3)}{\omega_3^2} 
\left( \frac{1}{\omega_1}-\frac{1}{\omega_2}\right) \right]\no\\
A_2 & = & -\frac{i}{4} \left( \frac{g_A}{F}\right)^5 (\omega_2 \omega_3-q_2 \cdot q_3)
\left(S^2+\frac{1}{2}\right) \left[ 
\frac{J_2(\omega_1)-J_2(-\omega_1)}{\omega_1^2} 
\left( \frac{1}{\omega_2}-\frac{1}{\omega_3}\right)\right.\no\\&&\left.
+\frac{J_2(\omega_2)-J_2(-\omega_2)}{\omega_2^2} 
\left( \frac{1}{\omega_1}+\frac{1}{\omega_3}\right) 
+ \frac{J_2(\omega_3)-J_2(-\omega_3)}{\omega_3^2} 
\left( -\frac{1}{\omega_1}+\frac{1}{\omega_2}\right) \right]\no\\
A_4 & = & -\frac{i}{4} \left( \frac{g_A}{F}\right)^5 (\omega_2 \omega_3-q_2 \cdot q_3)
\left(S^2+\frac{1}{2}\right) \left[ 
\frac{J_2(\omega_1)+J_2(-\omega_1)-2 J_2(0)}{\omega_1^2} 
\left( \frac{1}{\omega_2}-\frac{1}{\omega_3}\right)\right.\no\\
&&\left.+\frac{J_2(\omega_2)+J_2(-\omega_2)-2 J_2(0)}{\omega_2^2} 
\left( \frac{1}{\omega_1}-\frac{1}{\omega_3}\right)\right.\no\\&&\left.
+ \frac{J_2(\omega_3)+J_2(-\omega_3)-2 J_2(0)}{\omega_3^2} 
\left( -\frac{1}{\omega_1}+\frac{1}{\omega_2}\right) \right]\no\\
B_3 & = & \frac{i}{4} \left( \frac{g_A}{F}\right)^5 (\omega_1 \omega_3-q_1 \cdot q_3)
\left(S^2+\frac{1}{2}\right) \left[ 
\frac{J_2(\omega_1)-J_2(-\omega_1)}{\omega_1^2} 
\left( \frac{1}{\omega_2}+\frac{1}{\omega_3}\right)\right.\no\\
&&\left.+\frac{J_2(\omega_2)-J_2(-\omega_2)}{\omega_2^2} 
\left( \frac{1}{\omega_1}+\frac{1}{\omega_3}\right)
+ \frac{J_2(\omega_3)-J_2(-\omega_3)}{\omega_3^2} 
\left( \frac{1}{\omega_1}+\frac{1}{\omega_2}\right) \right]\no\\
D_1 & = & -\frac{i}{8} \left( \frac{g_A}{F}\right)^5 
\left(S^2+\frac{1}{2}\right) \left[ 
\frac{J_2(\omega_1)+J_2(-\omega_1)-2 J_2(0)}{\omega_1^2} 
\left( \frac{1}{\omega_2}-\frac{1}{\omega_3}\right)\right.\no\\
&&\left.+\frac{J_2(\omega_2)+J_2(-\omega_2)-2 J_2(0)}{\omega_2^2} 
\left( \frac{1}{\omega_1}-\frac{1}{\omega_3}\right)\right.\no\\
&&\left. + \frac{J_2(\omega_3)+J_2(-\omega_3)-2 J_2(0)}{\omega_3^2} 
\left( -\frac{1}{\omega_1}+\frac{1}{\omega_2}\right) \right]\no\\
D_4 & = & -\frac{i}{8} \left( \frac{g_A}{F}\right)^5 
\left(S^2+\frac{1}{2}\right) \left[ 
\frac{J_2(\omega_1)-J_2(-\omega_1)}{\omega_1^2} 
\left( \frac{1}{\omega_2}+\frac{1}{\omega_3}\right)\right.\no\\
&&\left.+\frac{J_2(\omega_2)-J_2(-\omega_2)}{\omega_2^2} 
\left( \frac{1}{\omega_1}-\frac{1}{\omega_3}\right)
+ \frac{J_2(\omega_3)-J_2(-\omega_3)}{\omega_3^2} 
\left( \frac{1}{\omega_1}-\frac{1}{\omega_2}\right) \right]
\eea
Diagram s2:\\
\bea
A_1 & = & -\frac{i}{2 \omega_1 \omega_2 \omega_3} \left( \frac{g_A}{F}\right)^5 
(\omega_2 \omega_3-q_2 \cdot q_3) \left(S^2+\frac{1}{2}\right) \no\\&&
\Big[ 
(J_2(\omega_1)-J_2(-\omega_1))
-(J_2(\omega_2)-J_2(-\omega_2))
- (J_2(\omega_3)-J_2(-\omega_3)) \Big]\no\\
A_2 & = & -\frac{i}{2 \omega_1 \omega_2 \omega_3} \left( \frac{g_A}{F}\right)^5 
(\omega_2 \omega_3-q_2 \cdot q_3) \left(S^2+\frac{1}{2}\right)  
(J_2(\omega_3)-J_2(-\omega_3))\no\\
A_4 & = & -\frac{i}{2 \omega_1 \omega_2 \omega_3} \left( \frac{g_A}{F}\right)^5 
(\omega_2 \omega_3-q_2 \cdot q_3) \left(S^2+\frac{1}{2}\right) \no\\
&&\Big[ 
-(J_2(\omega_2)+J_2(-\omega_2))
+ (J_2(\omega_3)+J_2(-\omega_3)) \Big]\no\\
B_3 & = & \frac{i}{2 \omega_1 \omega_2 \omega_3} \left( \frac{g_A}{F}\right)^5 
(\omega_1 \omega_3-q_1 \cdot q_3) \left(S^2+\frac{1}{2}\right)  
(J_2(\omega_1)-J_2(-\omega_1))\no\\
D_1 & = & -\frac{i}{4 \omega_1 \omega_2 \omega_3} \left( \frac{g_A}{F}\right)^5 
\left(S^2+\frac{1}{2}\right) \no\\
&&\Big[ 
-(J_2(\omega_2)+J_2(-\omega_2))
+ (J_2(\omega_3)+J_2(-\omega_3)) \Big]\no\\
D_4 & = & -\frac{i}{4 \omega_1 \omega_2 \omega_3} \left( \frac{g_A}{F}\right)^5 
\left(S^2+\frac{1}{2}\right) \no\\&&
\Big[ 
(J_2(\omega_1)-J_2(-\omega_1))-(J_2(\omega_2)-J_2(-\omega_2))
- (J_2(\omega_3)-J_2(-\omega_3)) \Big]
\eea
Diagrams s4+s5:\\
\bea
A_1 & = & \frac{3 i }{4} \left( \frac{g_A}{F}\right)^5 S^2 
(\omega_2\omega_3 - q_2 \cdot q_3)
\left[ 
\frac{J_2(\omega_1)-J_2(-\omega_1)}{\omega_1^2}
\left(\frac{1}{\omega_2}+\frac{1}{\omega_3}\right)\right.\no\\
&&\left.
+\frac{J_2(\omega_2)-J_2(-\omega_2)}{\omega_2^2}
\left(\frac{1}{\omega_1}-\frac{1}{\omega_3}\right)
+\frac{J_2(\omega_3)-J_2(-\omega_3)}{\omega_3^2}
\left(\frac{1}{\omega_1}-\frac{1}{\omega_2}\right)\right]\no\\
A_2 & = & \frac{3 i }{4} \left( \frac{g_A}{F}\right)^5 S^2 
(\omega_2\omega_3 - q_2 \cdot q_3)
\left[ 
\frac{J_2(\omega_1)-J_2(-\omega_1)}{\omega_1^2}
\left(\frac{1}{\omega_2}-\frac{1}{\omega_3}\right)\right.\no\\
&&\left.
+\frac{J_2(\omega_2)-J_2(-\omega_2)}{\omega_2^2}
\left(\frac{1}{\omega_1}+\frac{1}{\omega_3}\right)
+\frac{J_2(\omega_3)-J_2(-\omega_3)}{\omega_3^2}
\left(-\frac{1}{\omega_1}+\frac{1}{\omega_2}\right)\right]\no\\
A_4 & = & \frac{3 i }{4} \left( \frac{g_A}{F}\right)^5 S^2 
(\omega_2\omega_3 - q_2 \cdot q_3)
\left[ 
\frac{J_2(\omega_1)+J_2(-\omega_1)}{\omega_1^2}
\left(\frac{1}{\omega_2}-\frac{1}{\omega_3}\right)\right.\no\\
&&\left.
+\frac{J_2(\omega_2)+J_2(-\omega_2)}{\omega_2^2}
\left(\frac{1}{\omega_1}-\frac{1}{\omega_3}\right)
+\frac{J_2(\omega_3)+J_2(-\omega_3)}{\omega_3^2}
\left(-\frac{1}{\omega_1}+\frac{1}{\omega_2}\right)\right]\no\\
B_3 & = & -\frac{3 i }{4} \left( \frac{g_A}{F}\right)^5 S^2 
(\omega_1\omega_3 - q_1 \cdot q_3)
\left[ 
\frac{J_2(\omega_1)-J_2(-\omega_1)}{\omega_1^2}
\left(\frac{1}{\omega_2}+\frac{1}{\omega_3}\right)\right.\no\\
&&\left.
+\frac{J_2(\omega_2)-J_2(-\omega_2)}{\omega_2^2}
\left(\frac{1}{\omega_1}+\frac{1}{\omega_3}\right)
+\frac{J_2(\omega_3)-J_2(-\omega_3)}{\omega_3^2}
\left(\frac{1}{\omega_1}+\frac{1}{\omega_2}\right)\right]\no\\
D_1 & = & \frac{3 i }{8} \left( \frac{g_A}{F}\right)^5 S^2 
\left[ 
\frac{J_2(\omega_1)+J_2(-\omega_1)}{\omega_1^2}
\left(\frac{1}{\omega_2}-\frac{1}{\omega_3}\right)\right.\no\\
&&\left.
+\frac{J_2(\omega_2)+J_2(-\omega_2)}{\omega_2^2}
\left(\frac{1}{\omega_1}-\frac{1}{\omega_3}\right)
+\frac{J_2(\omega_3)+J_2(-\omega_3)}{\omega_3^2}
\left(-\frac{1}{\omega_1}+\frac{1}{\omega_2}\right)\right]\no\\
D_4 & = & \frac{3 i }{8} \left( \frac{g_A}{F}\right)^5 S^2 
\left[ 
\frac{J_2(\omega_1)-J_2(-\omega_1)}{\omega_1^2}
\left(\frac{1}{\omega_2}+\frac{1}{\omega_3}\right)\right.\no\\
&&\left.
+\frac{J_2(\omega_2)-J_2(-\omega_2)}{\omega_2^2}
\left(\frac{1}{\omega_1}-\frac{1}{\omega_3}\right)
+\frac{J_2(\omega_3)-J_2(-\omega_3)}{\omega_3^2}
\left(\frac{1}{\omega_1}-\frac{1}{\omega_2}\right)\right]
\eea
Diagrams s6+s7:\\
\bea
A_1 & = & \frac{i}{4 \omega_1}\frac{g_A}{F^5}
\Big[ (\omega_2-\omega_3)\left(\omega_3(J_0(\omega_3)-J_0(-\omega_3)) 
-\omega_2(J_0(\omega_2)-J_0(-\omega_2))\right)\no\\
&& -(\omega_2+\omega_3) \Delta_\pi \Big]
\no\\
A_2 & = & -\frac{i}{8 \omega_1}\frac{g_A}{F^5} (\omega_2-\omega_3)
\Big[ \omega_3(J_0(\omega_3)-J_0(-\omega_3))
+\omega_2(J_0(\omega_2)-J_0(-\omega_2))+3 \Delta_\pi \Big]\no\\
A_4 & = & -\frac{i}{8 \omega_1}\frac{g_A}{F^5} (\omega_2-\omega_3)
\Big[ \omega_3(J_0(\omega_3)+J_0(-\omega_3))
+\omega_2(J_0(\omega_2)+J_0(-\omega_2)) \Big]\no\\
B_3 & = & -\frac{i}{8 \omega_2}\frac{g_A}{F^5}
(\omega_1+\omega_3)\left(\omega_3(J_0(\omega_3)-J_0(-\omega_3))
  +\omega_1(J_0(\omega_1)-J_0(-\omega_1))+3 \Delta_\pi  \right)
\no\\
D_1 & = & D_4 = 0
\eea
Diagram s8+s9:\\
\bea
A_1 & = & \frac{i}{4} \left( \frac{g_A}{F}\right)^5 
(\omega_2 \omega_3-q_2 \cdot q_3)\no\\
&&\left[ 
\frac{J_2(\omega_1)-J_2(-\omega_1)}{\omega_1}
\left(\frac{3 S^2}{\omega_1\omega_2} +\frac{3 S^2}{\omega_1\omega_3}
-\frac{2(S^2+1)}{\omega_2\omega_3}\right)
\right.\no\\&&\left.
-\frac{J_2(\omega_2)-J_2(-\omega_2)}{\omega_2}
\left(\frac{S^2+1}{\omega_1\omega_2} -\frac{S^2+1}{\omega_2\omega_3}
+\frac{2 S^2-1}{\omega_1\omega_3}\right)
\right.\no\\&&\left.
-\frac{J_2(\omega_3)-J_2(-\omega_3)}{\omega_3}
\left(\frac{S^2+1}{\omega_1\omega_3} -\frac{S^2+1}{\omega_2\omega_3}
+\frac{2 S^2-1}{\omega_1\omega_2}\right) \right]\no\\
A_2 & = & \frac{i}{4} \left( \frac{g_A}{F}\right)^5 
(\omega_2 \omega_3-q_2 \cdot q_3)\no\\
&&\left[ 
\frac{J_2(\omega_1)-J_2(-\omega_1)}{\omega_1}
\left(\frac{S^2}{\omega_1\omega_2} -\frac{S^2}{\omega_1\omega_3}
+\frac{4(S^2+1)}{\omega_2\omega_3}\right)
\right.\no\\&&\left.
-\frac{J_2(\omega_2)-J_2(-\omega_2)}{\omega_2}
\left(\frac{3(S^2+1)}{\omega_1\omega_2} +\frac{3(S^2+1)}{\omega_2\omega_3}
+\frac{1}{\omega_1\omega_3}\right)
\right.\no\\&&\left.
+\frac{J_2(\omega_3)-J_2(-\omega_3)}{\omega_3}
\left(\frac{S^2+1}{\omega_1\omega_3} -\frac{S^2+1}{\omega_2\omega_3}
+\frac{2 S^2+3}{\omega_1\omega_2}\right) \right]\no\\
A_4 & = & \frac{i}{4} \left( \frac{g_A}{F}\right)^5 
(\omega_2 \omega_3-q_2 \cdot q_3)\no\\
&&\left[ 
\frac{J_2(\omega_1)+J_2(-\omega_1)-2 J_2(0)}{\omega_1}
\left(\frac{S^2}{\omega_1\omega_2} -\frac{S^2}{\omega_1\omega_3}\right)
\right.\no\\&&\left.
-\frac{J_2(\omega_2)+J_2(-\omega_2)-2 J_2(0)}{\omega_2}
\left(\frac{S^2+1}{\omega_1\omega_2} -\frac{S^2+1}{\omega_2\omega_3}
-\frac{2 S^2+1}{\omega_1\omega_3}\right)
\right.\no\\&&\left.
+\frac{J_2(\omega_3)+J_2(-\omega_3)-2 J_2(0)}{\omega_3}
\left(\frac{S^2+1}{\omega_1\omega_3} -\frac{S^2+1}{\omega_2\omega_3}
-\frac{2 S^2+1}{\omega_1\omega_2}\right) \right]\no\\
B_3 & = & \frac{i}{4} \left( \frac{g_A}{F}\right)^5 
(\omega_1 \omega_3-q_1 \cdot q_3)\no\\
&&\left[ 
\frac{J_2(\omega_1)-J_2(-\omega_1)}{\omega_1}
\left(\frac{S^2+1}{\omega_1\omega_2} +\frac{S^2+1}{\omega_1\omega_3}
-\frac{2 S^2+3}{\omega_2\omega_3}\right)
\right.\no\\&&\left.
-\frac{J_2(\omega_2)-J_2(-\omega_2)}{\omega_2}
\left(\frac{S^2}{\omega_1\omega_2} +\frac{S^2}{\omega_2\omega_3}
+\frac{4( S^2+1)}{\omega_1\omega_3}\right)
\right.\no\\&&\left.
+\frac{J_2(\omega_3)-J_2(-\omega_3)}{\omega_3}
\left(\frac{3(S^2+1)}{\omega_1\omega_3} +\frac{3(S^2+1)}{\omega_2\omega_3}
-\frac{1}{\omega_1\omega_2}\right) \right]\no\\
D_1 & = & \frac{i}{8} \left( \frac{g_A}{F}\right)^5 (S^2+1)\no\\
&&\left[ 
-\frac{J_2(\omega_1)+J_2(-\omega_1)-2 J_2(0)}{\omega_1}
\left(\frac{3}{\omega_1\omega_2} -\frac{3}{\omega_1\omega_3}\right)
\right.\no\\&&\left.
-\frac{J_2(\omega_2)+J_2(-\omega_2)-2 J_2(0)}{\omega_2}
\left(\frac{1}{\omega_1\omega_2} -\frac{1}{\omega_2\omega_3}
+\frac{2 }{\omega_1\omega_3}\right)
\right.\no\\&&\left.
+\frac{J_2(\omega_3)+J_2(-\omega_3)-2 J_2(0)}{\omega_3}
\left(\frac{1}{\omega_1\omega_3} -\frac{1}{\omega_2\omega_3}
+\frac{2}{\omega_1\omega_2}\right) \right]\no\\
D_4 & = & \frac{i}{8} \left( \frac{g_A}{F}\right)^5 (S^2+1)\no\\
&&\left[ 
-\frac{J_2(\omega_1)-J_2(-\omega_1)}{\omega_1}
\left(\frac{1}{\omega_1\omega_2} +\frac{1}{\omega_1\omega_3}
+\frac{2}{\omega_2\omega_3}\right)
\right.\no\\&&\left.
-\frac{J_2(\omega_2)-J_2(-\omega_2)}{\omega_2}
\left(\frac{1}{\omega_1\omega_2} -\frac{1}{\omega_2\omega_3}
-\frac{2}{\omega_1\omega_3}\right)
\right.\no\\&&\left.
-\frac{J_2(\omega_3)-J_2(-\omega_3)}{\omega_3}
\left(\frac{1}{\omega_1\omega_3} -\frac{1}{\omega_2\omega_3}
-\frac{2}{\omega_1\omega_2}\right) \right]
\eea
Diagram s10:\\
\bea
A_2 & = & \frac{i}{8 \omega_1}\frac{g_A}{F^5}
\Big[ (\omega_2-\omega_3)\left(\omega_3(J_0(\omega_3)-J_0(-\omega_3))
-\omega_2(J_0(\omega_2)-J_0(-\omega_2))\right) \no\\
&& -(\omega_2+\omega_3) \Delta_\pi \Big]
\no\\
A_4 & = & -\frac{i}{8 \omega_1}\frac{g_A}{F^5} (\omega_2-\omega_3)
\Big[\omega_3(J_0(\omega_3)+J_0(-\omega_3))
+\omega_2(J_0(\omega_2)+J_0(-\omega_2))\Big]\no\\
B_3 & = & \frac{i}{8 \omega_2}\frac{g_A}{F^5}
\Big[ (\omega_1+\omega_3)\left(-\omega_3(J_0(\omega_3)-J_0(-\omega_3))
+\omega_1(J_0(\omega_1)-J_0(-\omega_1))\right) \no\\&&
 +(\omega_1-\omega_3) \Delta_\pi \Big]
\no\\
A_1 & = & D_1 = D_4 = 0
\eea
Diagram s11:\\
\bea
A_1 & = & -\frac{i}{4} \left( \frac{g_A}{F}\right)^5 (\omega_2 \omega_3-q_2 \cdot q_3)
\left(S^2+\frac{1}{2}\right)\no\\
&&\left[ 
\frac{J_2(\omega_1)-J_2(0)}{\omega_1^2} 
\left( \frac{1}{\omega_2}+\frac{1}{\omega_3}\right)
- \frac{J_2(-\omega_1)-J_2(-\omega_2)}{\omega_1} 
\left( \frac{2}{\omega_2\omega_3}-\frac{1}{\omega_1\omega_2}-\frac{1}{\omega_1\omega_3}
\right)\right.\no\\
&&\left.
- \frac{J_2(\omega_2)-J_2(0)}{ \omega_2}
\left( \frac{2}{\omega_1\omega_3}+\frac{1}{\omega_1\omega_2}-\frac{1}{\omega_2\omega_3}
\right)
- \frac{J_2(-\omega_2)-J_2(0)}{\omega_2^2} 
\left( \frac{1}{\omega_1}-\frac{1}{\omega_3}\right)\right.\no\\
&&\left.
-\frac{J_2(\omega_3)-J_2(0)}{\omega_3}
\left( \frac{2}{\omega_1\omega_2}+\frac{1}{\omega_1\omega_3}-\frac{1}{\omega_2\omega_3}
\right)
- \frac{J_2(-\omega_3)-J_2(0)}{\omega_3^2} 
\left( \frac{1}{\omega_1}-\frac{1}{\omega_2}\right)\right.\no\\
&&\left.
+2 J^\prime_2(0) \left( \frac{1}{\omega_1\omega_2}+\frac{1}{\omega_1\omega_3}-
\frac{1}{\omega_2\omega_3}\right)\right]\no\\
A_2 & = & -\frac{i}{4} \left( \frac{g_A}{F}\right)^5 (\omega_2 \omega_3-q_2 \cdot q_3)
\left(S^2+\frac{1}{2}\right)\no\\
&&\left[ 
-\frac{J_2(\omega_1)-J_2(-w_1)}{\omega_1^2} 
\left( \frac{1}{\omega_2}-\frac{1}{\omega_3}\right)
- \frac{J_2(\omega_2)-J_2(0)}{ \omega_2^2}
\left( \frac{1}{\omega_1}+\frac{1}{\omega_3}
\right)\right.\no\\
&&\left.
+\frac{J_2(\omega_3)-J_2(0)}{\omega_3}
\left( \frac{2}{\omega_1\omega_2}+\frac{1}{\omega_1\omega_3}-\frac{1}{\omega_2\omega_3}
\right)
+ \frac{J_2(-\omega_3)-J_2(0)}{\omega_3^2} 
\left( \frac{1}{\omega_1}-\frac{1}{\omega_2}\right)\right.\no\\
&&\left.
+2 J^\prime_2(0) \left( \frac{1}{\omega_1\omega_2}-\frac{1}{\omega_1\omega_3}+
\frac{1}{\omega_2\omega_3}\right)\right]\no\\
A_4 & = & -\frac{3 i}{4} \left( \frac{g_A}{F}\right)^5 (\omega_2 \omega_3-q_2 \cdot q_3)
\left(S^2+\frac{1}{2}\right)\no\\
&&\left[ 
\frac{J_2(\omega_1)+J_2(-\omega_1)-2 J_2(0)}{\omega_1^2} 
\left( \frac{1}{\omega_2}-\frac{1}{\omega_3}\right)
\right.\no\\
&&\left.
+ \frac{J_2(\omega_2)-J_2(0)}{ \omega_2}
\left( \frac{2}{\omega_1\omega_3}+\frac{1}{\omega_1\omega_2}-\frac{1}{\omega_2\omega_3}
\right)
- \frac{J_2(-\omega_2)-J_2(0)}{\omega_2^2} 
\left( \frac{1}{\omega_1}-\frac{1}{\omega_3}\right)\right.\no\\
&&\left.
-\frac{J_2(\omega_3)-J_2(0)}{\omega_3}
\left( \frac{2}{\omega_1\omega_2}+\frac{1}{\omega_1\omega_3}-\frac{1}{\omega_2\omega_3}
\right)
+ \frac{J_2(-\omega_3)-J_2(0)}{\omega_3^2} 
\left( \frac{1}{\omega_1}-\frac{1}{\omega_2}\right)\right]\no\\
B_3 & = & -\frac{i}{4} \left( \frac{g_A}{F}\right)^5 (\omega_2 \omega_3-q_2 \cdot q_3)
\left(S^2+\frac{1}{2}\right)\no\\
&&\left[ 
-\frac{J_2(\omega_1)-J_2(0)}{\omega_1^2} 
\left( \frac{1}{\omega_2}+\frac{1}{\omega_3}\right)
+ \frac{J_2(-\omega_1)-J_2(0)}{\omega_1} 
\left( \frac{2}{\omega_2\omega_3}-\frac{1}{\omega_1\omega_2}-\frac{1}{\omega_1\omega_3}
\right)\right.\no\\
&&\left.
+ \frac{J_2(\omega_2)-J_2(-\omega_2)}{ \omega_2^2}
\left( \frac{1}{\omega_1}+\frac{1}{\omega_3}
\right)
+\frac{J_2(\omega_3)-J_2(-\omega_3)}{\omega_3^2}
\left(\frac{1}{\omega_1}+\frac{1}{\omega_2}
\right)
\right.\no\\
&&\left.
-2 J^\prime_2(0) \left( \frac{1}{\omega_1\omega_2}+\frac{1}{\omega_1\omega_3}+
\frac{1}{\omega_2\omega_3}\right)\right]\no\\
D_1 & = & -\frac{i}{4} \left( \frac{g_A}{F}\right)^5 \left(S^2+\frac{3}{2}\right)\no\\
&&\left[ 
-\frac{J_2(\omega_1)+J_2(-\omega_1)-2 J_2(0)}{\omega_1^2} 
\left( \frac{1}{\omega_2}-\frac{1}{\omega_3}\right)
\right.\no\\
&&\left.
- \frac{J_2(\omega_2)-J_2(0)}{ \omega_2}
\left( \frac{2}{\omega_1\omega_3}+\frac{1}{\omega_1\omega_2}-\frac{1}{\omega_2\omega_3}
\right)
+ \frac{J_2(-\omega_2)-J_2(0)}{\omega_2^2} 
\left( \frac{1}{\omega_1}-\frac{1}{\omega_3}\right)\right.\no\\
&&\left.
+\frac{J_2(\omega_3)-J_2(0)}{\omega_3}
\left( \frac{2}{\omega_1\omega_2}+\frac{1}{\omega_1\omega_3}-\frac{1}{\omega_2\omega_3}
\right)
- \frac{J_2(-\omega_3)-J_2(0)}{\omega_3^2} 
\left( \frac{1}{\omega_1}-\frac{1}{\omega_2}\right)\right]\no\\
D_4 & = & -\frac{3 i}{4} \left( \frac{g_A}{F}\right)^5 \left(S^2+\frac{3}{2}\right)\no\\
&&\left[ 
-\frac{J_2(\omega_1)-J_2(0)}{\omega_1^2} 
\left( \frac{1}{\omega_2}+\frac{1}{\omega_3}\right)
+ \frac{J_2(-\omega_1)-J_2(0)}{\omega_1} 
\left( \frac{2}{\omega_2\omega_3}-\frac{1}{\omega_1\omega_2}-\frac{1}{\omega_1\omega_3}
\right)\right.\no\\
&&\left.
+ \frac{J_2(\omega_2)-J_2(0)}{ \omega_2}
\left( \frac{2}{\omega_1\omega_3}+\frac{1}{\omega_1\omega_2}-\frac{1}{\omega_2\omega_3}
\right)
+ \frac{J_2(-\omega_2)-J_2(0)}{\omega_2^2} 
\left( \frac{1}{\omega_1}-\frac{1}{\omega_3}\right)\right.\no\\
&&\left.
+\frac{J_2(\omega_3)-J_2(0)}{\omega_3}
\left( \frac{2}{\omega_1\omega_2}+\frac{1}{\omega_1\omega_3}-\frac{1}{\omega_2\omega_3}
\right)
+ \frac{J_2(-\omega_3)-J_2(0)}{\omega_3^2} 
\left( \frac{1}{\omega_1}-\frac{1}{\omega_2}\right)\right.\no\\
&&\left.
-2 J^\prime_2(0) \left( \frac{1}{\omega_1\omega_2}+\frac{1}{\omega_1\omega_3}-
\frac{1}{\omega_2\omega_3}\right)\right]
\eea
Diagrams s12+s13: add up to 0.\\
Diagrams s14+s15:\\
\bea
A_1 & = & -i\frac{g_A^3}{F^5}S^2\frac{J_2(\omega_1)-J_2(-\omega_1)}{\omega_1}\no\\
A_2 & = & A_4 = B_3 = D_1 = D_4 = 0
\eea
Diagrams s16+s17:\\
\bea
A_2 & = & \frac{i}{4}\frac{g_A^3}{F^5} S^2 \frac{\omega_2-\omega_3}{\omega_1^2}
(J_2(\omega_1)-J_2(-\omega_1))\no\\
A_4 & = & \frac{i}{4}\frac{g_A^3}{F^5} S^2 \frac{\omega_2-\omega_3}{\omega_1^2}
(J_2(\omega_1)+J_2(-\omega_1)-2 J_2(0))\no\\
B_3 & = & \frac{i}{4}\frac{g_A^3}{F^5} S^2 \frac{\omega_1+\omega_3}{\omega_2^2}
(J_2(\omega_2)-J_2(-\omega_2))\no\\
A_1 & = & D_1 = D_4 = 0
\eea
Diagrams s18+s19:\\
\bea
A_2 & = & -\frac{i}{4}\frac{g_A^3}{F^5} \left(S^2+\frac{1}{2}\right) 
\frac{\omega_2-\omega_3}{\omega_1^2}
(J_2(\omega_1)-J_2(-\omega_1))\no\\
A_4 & = & -\frac{i}{4}\frac{g_A^3}{F^5} \left(S^2+\frac{1}{2}\right) 
\frac{\omega_2-\omega_3}{\omega_1^2}
(J_2(\omega_1)+J_2(-\omega_1)-2 J_2(0))\no\\
B_3 & = & -\frac{i}{4}\frac{g_A^3}{F^5} \left(S^2+\frac{1}{2}\right) 
\frac{\omega_1+\omega_3}{\omega_2^2}
(J_2(\omega_2)-J_2(-\omega_2))\no\\
A_1 & = & D_1 = D_4 = 0
\eea
Diagrams s20+s21:\\
\bea
A_2 & = & \frac{3i}{4}\frac{g_A^3}{F^5}S^2\frac{\omega_2-\omega_3}{\omega_1^2}
(J_2(\omega_1)-J_2(-\omega_1))\no\\
A_4 & = & \frac{3i}{4}\frac{g_A^3}{F^5}S^2\frac{\omega_2-\omega_3}{\omega_1^2}
(J_2(\omega_1)+J_2(-\omega_1))\no\\
B_3 & = & \frac{3i}{4}\frac{g_A^3}{F^5}S^2\frac{\omega_1+\omega_3}{\omega_2^2}
(J_2(\omega_2)-J_2(-\omega_2))\no\\
A_1 & = & D_1 = D_4 = 0
\eea
Diagrams s22+s23:\\
\bea
A_1 & = & -\frac{i}{4}\frac{g_A}{F^5}
\Big[\omega_2(J_0(\omega_2)-J_0(-\omega_2))
+\omega_3(J_0(\omega_3)-J_0(-\omega_3))+2\Delta_\pi\Big]\no\\
A_2 & = & \frac{3 i}{4} \frac{g_A}{F^5}
\Big[\omega_2(J_0(\omega_2)-J_0(-\omega_2))+\Delta_\pi\Big]\no\\
A_4 & = & -\frac{i}{4}\frac{g_A}{F^5}
\Big[\omega_2(J_0(\omega_2)+J_0(-\omega_2))
-\omega_3(J_0(\omega_3)+J_0(-\omega_3))\Big]\no\\
B_3 & = & -\frac{3 i}{4} \frac{g_A}{F^5}
\Big[\omega_3(J_0(\omega_3)-J_0(-\omega_3))+\Delta_\pi\Big]\no\\
D_1 & = & D_4 = 0
\eea
Diagrams s24+s25:\\
\bea
A_1 & = & i \frac{g_A^3}{F^5} \left( S^2+\frac{1}{2}\right) 
\frac{J_2(\omega_1)-J_2(-\omega_1)}{\omega_1}\no\\
A_2 & = & -i \frac{g_A^3}{F^5} \left( S^2+\frac{1}{2}\right) 
\frac{J_2(\omega_1)-J_2(-\omega_1)}{\omega_1}\no\\
B_3 & = &i \frac{g_A^3}{F^5} \left( S^2+\frac{1}{2}\right) 
\frac{J_2(\omega_2)-J_2(-\omega_2)}{\omega_2}\no\\
A_4 & = & D_1 = D_4 = 0
\eea
Diagram s26+s27:\\
\bea
A_2 & = & \frac{i}{4} \frac{g_A^3}{F^5} \left( S^2+\frac{1}{2}\right)
\frac{\omega_2-\omega_3}{\omega_1} \left[\frac{J_2(\omega_1)-J_2(-\omega_1)}{\omega_1}
-2 J_2^\prime(0)\right]\no\\
A_4 & = & -\frac{3 i}{4} \frac{g_A^3}{F^5} \left( S^2+\frac{1}{2}\right)
\frac{\omega_2-\omega_3}{\omega_1^2} \Big[J_2(\omega_1)+J_2(-\omega_1)-2 J_2(0)\Big]
\no\\
B_3 & = & \frac{i}{4} \frac{g_A^3}{F^5} \left( S^2+\frac{1}{2}\right)
\frac{\omega_1+\omega_3}{\omega_2} \left[\frac{J_2(\omega_2)-J_2(-\omega_2)}{\omega_2}
-2 J_2^\prime(0)\right]\no\\
A_1 & = & D_1 = D_4 = 0 
\eea
Diagrams s28+s29:\\
\bea
A_1 &= & i \frac{g_A^3}{F^5} \Bigg\{
\frac{K_0(t_3,\omega_1)-K_0(t_3,-\omega_1)}{8 \omega_1}(2 t_3^2-5 M^2 t_3 +2 M^4)\no\\&&
+\frac{J_0(\omega_1)-J_0(-\omega_1)}{8 \omega_1}(-2 t_3+2 \omega_1^2-M^2)
+I_0(t_3) \frac{-2 t_3 + M^2}{4}
+\frac{\Delta_\pi}{2}\no\\&&
+\frac{\omega_2 \omega_3 - q_2\cdot q_3}{t_2-\omega_2^2} \left[
-\frac{K_0(t_2,\omega_2)-K_0(t_2,-\omega_2)}{8 \omega_2}
(t_2^2-4 M^2 t_2 + 4 M^2 \omega_2^2)\right.\no\\&&\left.
+\frac{J_0(\omega_2)-J_0(-\omega_2)}{8 \omega_2}(t_2-2 \omega_2^2)
+I_0(t_2) \frac{t_2}{2}+\frac{t_2-\omega_2^2}{16 \pi^2} \right]\no\\&&
+\frac{\omega_2 \omega_3 - q_2\cdot q_2}{t_1-\omega_3^2} \left[
-\frac{K_0(t_1,\omega_3)-K_0(t_1,-\omega_3)}{8 \omega_3}
(t_1^2-4 M^2 t_1 + 4 M^2 \omega_3^2)\right.\no\\&&\left.
+\frac{J_0(\omega_3)-J_0(-\omega_3)}{8 \omega_3}(t_1-2 \omega_3^2)
+I_0(t_1) \frac{t_1}{2}+\frac{t_1-\omega_3^2}{16 \pi^2} \right]
\Bigg\}\no\\
A_2 & = & i \frac{g_A^3}{F^5} \Bigg\{
\frac{\omega_2 - \omega_3}{8(t_3-\omega_1^2)} \Bigg[ 
(K_0(t_3,\omega_1)-K_0(t_3,-\omega_1))t_3  (t_3-2 M^2).\no\\&&
-(J_0(\omega_1)-J_0(-\omega_1))(t_3+2 M^2-2 \omega_1^2)
-\frac{20}{3 \omega_1} \Delta_\pi \no\\&&
-I_0(t_3) \frac{2}{3 \omega_1} (5 t_3^2-8 M^2 t_3+\omega_1^2 t_3 - 4 M^2 \omega_1^2) \no\\&&
-\frac{1}{36 \pi^2 \omega_1} (6 M^2-t_3)(t_3-\omega_1^2) \Bigg]\no\\&&
-\frac{\omega_2 \omega_3 - q_2\cdot q_3}{t_1-\omega_3^2} \left[
-\frac{K_0(t_1,\omega_3)-K_0(t_1,-\omega_3)}{8 \omega_3}
(t_1^2-4 M^2 t_1 + 4 M^2 \omega_3^2)\right.\no\\&&\left.
+\frac{J_0(\omega_3)-J_0(-\omega_3)}{8 \omega_3}(t_1-2 \omega_3^2)
+I_0(t_1) \frac{t_1}{2}+\frac{t_1-\omega_3^2}{16 \pi^2} \right]
\Bigg\}\no\\
A_4 & = & i \frac{g_A^3}{F^5} \Bigg\{
\frac{(\omega_2-\omega_3)}{8 (t_3-\omega_1^2)} \left[
(K_0(t_3,\omega_1)+K_0(t_3,-\omega_1)) t_3(t_3-2 M^2)\right.\no\\&&\left.
-(J_0(\omega_1)+J_0(-\omega_1))(t_3+2 M^2-2 \omega_1^2) 
-J_0(0) 2 (t_3-2 M^2) \right]\no\\&&
+\frac{(\omega_2 \omega_3 -q_2\cdot q_3)}{8 \omega_2 (t_2-\omega_2^2)}
\left[-(K_0(t_2,\omega_2)+K_0(t_2,-\omega_2))(t_2^2-4 M^2 t_2+4 M^2 \omega_2^2)
\right.\no\\&&\left.
+(J_0(\omega_2)+J_0(-\omega_2))(t_2-2\omega_2^2)+J_0(0) 2 t_2 \right]\no\\&&
-\frac{(\omega_2 \omega_3 -q_2\cdot q_3)}{8 \omega_3 (t_1-\omega_3^2)}
\left[-(K_0(t_1,\omega_3)+K_0(t_1,-\omega_3))(t_1^2-4 M^2 t_1+4 M^2 \omega_3^2)
\right.\no\\&&\left.
+(J_0(\omega_3)+J_0(-\omega_3))(t_1-2\omega_3^2)+J_0(0) 2 t_1 \right]
\Bigg\}\no\\
B_3 & = & i \frac{g_A^3}{F^5} \Bigg\{
-\frac{\omega_1 \omega_3 - q_1\cdot q_3}{t_3-\omega_1^2} \left[
-\frac{K_0(t_3,\omega_1)-K_0(t_3,-\omega_1)}{8 \omega_1}
(t_3^2-4 M^2 t_2 + 4 M^2 \omega_1^2)\right.\no\\&&\left.
+\frac{J_0(\omega_1)-J_0(-\omega_1)}{8 \omega_1}(t_3-2 \omega_1^2)
+I_0(t_3) \frac{t_3}{2}+\frac{t_3-\omega_1^2}{16 \pi^2} \right]\no\\&&
+\frac{\omega_1 + \omega_3}{8(t_2-\omega_2^2)} \Bigg[ 
(K_0(t_2,\omega_2)-K_0(t_2,-\omega_2))t_2  (t_2-2 M^2)\no\\&&
-(J_0(\omega_2)-J_0(-\omega_2))(t_2+2 M^2-2 \omega_2^2)
-\frac{20}{3 \omega_2} \Delta_\pi \no\\&&
-I_0(t_2) \frac{2}{3 \omega_2} (5 t_2^2-8 M^2 t_2+\omega_2^2 t_2 - 4 M^2 \omega_2^2) \no\\&&
-\frac{1}{36 \pi^2 \omega_2} (6 M^2-t_2)(t_2-\omega_2^2) \Bigg]\Bigg\}\no\\
D_1 & = & i \frac{g_A^3}{F^5} \Bigg\{
-\frac{1}{16 \omega_2 (t_2-\omega_2^2)} \left[
(K_0(t_2,\omega_2)+K_0(t_2,-\omega_2))(t_2^2-4 M^2 t_2 + 4 M^2 \omega_2^2)
\right.\no\\&&\left.
-(J_0(\omega_2)+J_0(-\omega_2))(t_2-2 \omega_2^2)-2 J_0(0) t_2 \right]\no\\&&
+\frac{1}{16 \omega_3 (t_1-\omega_3^2)} \left[
(K_0(t_1,\omega_3)+K_0(t_1,-\omega_3))(t_1^2-4 M^2 t_1 + 4 M^2 \omega_3^2)
\right.\no\\&&\left.
-(J_0(\omega_3)+J_0(-\omega_3))(t_1-2 \omega_3^2)-2 J_0(0) t_1 \right]\Bigg\}\no\\
D_4 & = & i \frac{g_A^3}{16 F^5} \Bigg\{
\frac{1}{t_3-\omega_1^2} \left[
-\frac{K_0(t_3,\omega_1)-K_0(t_2,-\omega_2)}{\omega_1}
(t_3^2-4 M^2 t_3 + 4 M^2 \omega_1^2)\right.\no\\&&\left.
+\frac{J_0(\omega_1)-J_0(-\omega_1)}{\omega_1}(t_3-2 \omega_1^2)
+I_0(t_3) 4 t_3+\frac{t_3-\omega_1^2}{2 \pi^2} \right]\no\\&&
+\frac{1}{t_2-\omega_2^2} \left[
-\frac{K_0(t_2,\omega_2)-K_0(t_2,-\omega_2)}{\omega_2}
(t_2^2-4 M^2 t_2 + 4 M^2 \omega_2^2)\right.\no\\&&\left.
+\frac{J_0(\omega_2)-J_0(-\omega_2)}{\omega_2}(t_2-2 \omega_2^2)
+I_0(t_2) 4 t_2+\frac{t_2-\omega_2^2}{2 \pi^2} \right]\no\\&&
+\frac{1}{t_1-\omega_3^2} \left[
-\frac{K_0(t_1,\omega_3)-K_0(t_1,-\omega_3)}{\omega_3}
(t_1^2-4 M^2 t_1 + 4 M^2 \omega_3^2)\right.\no\\&&\left.
+\frac{J_0(\omega_3)-J_0(-\omega_3)}{\omega_3}(t_1-2 \omega_3^2)
+I_0(t_1) 4 t_1+\frac{t_1-\omega_3^2}{2 \pi^2} \right]
\Bigg\}
\eea
Diagrams s30+s31:\\
\bea
A_1 & = & i \frac{g_A}{4 F^5} \Bigg\{
\frac{1}{(t_2-\omega_2^2)^2} \left(
\frac{\omega_2 (K_0(t_2,\omega_2)-K_0(t_2,-\omega_2))}{2} \right.\no\\&&\left.
\left[(q_1\cdot (q_1-q_3)-\omega_1\omega_2)(-4 t_2 M^2 + 4 M^2 \omega_2^2 + 3 t_2^2)
\right.\right.\no\\&&\left.\left.
-(t_2-\omega_2^2) (3 t_2^2-6 t_2 M^2 + 4 M^2 \omega_2^2)\right]\right.\no\\&&\left.
+\frac{\omega_2 (J_0(\omega_2)-J_0(-\omega_2))}{2} 
\left[(q_1\cdot (q_1-q_3)-\omega_1\omega_2)(5 t_2 - 2 \omega_2^2)
\right.\right.\no\\&&\left.\left.
-(t_2-\omega_2^2) (t_2 - 2 M^2+2\omega_2^2)\right]\right.\no\\&&\left.
+ I_0(t_2)\left[2 \omega_1\omega_2 t_2 (2 t_2 + \omega_2^2)
+\frac{t_2-\omega_2^2}{3}(4 t_2^2-7 t_2 M^2 + 14 t_2 \omega_2^2-5 M^2 \omega_2^2)
\right.\right.\no\\&&\left.\left.
-\frac{2}{3 t_2}q_1\cdot (q_1-q_3) (11 t_2^2 \omega_2^2 - 4 t_2 \omega_2^4 
- 2 M^2 t_2^2 + 4 M^2 t_2 \omega_2^2 -2 M^2 \omega_2^4 + 2 t_2^3)\right]
\right.\no\\&&\left.
+\Delta_\pi (t_2 -\omega_2)^2
\left[\frac{4}{3 t_2}q_1\cdot (q_1-q_3) +\frac{2}{3}\right]\right.\no\\&&\left.
+\frac{t_2-\omega_2^2}{16 \pi^2} 
\left[-\frac{2}{9 t_2} q_1\cdot (q_1-q_3) 
(t_2^2-6 t_2 M^2 + 17 t_2 \omega_2^2 + 6 M^2 \omega_2^2)
+4 \omega_1 \omega_2^3 \right.\right.\no\\&&\left.\left.
+\frac{2(t_2-\omega_2^2)}{9} (t_2-6 M^2+18 \omega_2^2)\right]
\right)\no\\&&
+\frac{1}{(t_1-\omega_3^2)^2} \left(
\frac{\omega_3 (K_0(t_1,\omega_3)-K_0(t_1,-\omega_3))}{2} \right.\no\\&&\left.
\left[(q_1\cdot (q_1-q_2)-\omega_1\omega_3)(-4 t_1 M^2 + 4 M^2 \omega_3^2 + 3 t_1^2)
\right.\right.\no\\&&\left.\left.
-(t_1-\omega_3^2) (3 t_1^2-6 t_1 M^2 + 4 M^2 \omega_3^2)\right]\right.\no\\&&\left.
+\frac{\omega_3 (J_0(\omega_3)-J_0(-\omega_3))}{2} 
\left[(q_1\cdot (q_1-q_2)-\omega_1\omega_3)(5 t_1 - 2 \omega_3^2)
\right.\right.\no\\&&\left.\left.
-(t_1-\omega_3^2) (t_1 - 2 M^2+2\omega_3^2)\right]\right.\no\\&&\left.
+ I_0(t_1)
\left[2 \omega_1\omega_3 t_1 (2 t_1 + \omega_3^2)
+\frac{t_1-\omega_3^2}{3}(4 t_1^2-7 t_1 M^2 + 14 t_1 \omega_3^2-5 M^2 \omega_3^2)
\right.\right.\no\\&&\left.\left.
+-\frac{2}{3 t_1}q_1\cdot (q_1-q_2) (11 t_1^2 \omega_3^2 - 4 t_1 \omega_3^4 
- 2 M^2 t_1^2 + 4 M^2 t_1 \omega_3^2 -2 M^2 \omega_3^4 + 2 t_1^3)\right]
\right.\no\\&&\left.
+\Delta_\pi (t_1 -\omega_3)^2
\left[\frac{4}{3 t_1}q_1\cdot (q_1-q_2) +\frac{2}{3}\right]\right.\no\\&&\left.
+\frac{t_1-\omega_3^2}{16 \pi^2} 
\left[-\frac{2}{9 t_1} q_1\cdot (q_1-q_2) 
(t_1^2-6 t_1 M^2 + 17 t_1 \omega_3^2 + 6 M^2 \omega_3^2)
+4 \omega_1 \omega_3^3 \right.\right.\no\\&&\left.\left.
+\frac{2(t_1-\omega_3^2)}{9} (t_1-6 M^2+18 \omega_3^2)\right]
\right)
\Bigg\}\no\\
A_2 &= & i \frac{g_A}{4 F^5} \Bigg\{
\frac{1}{t_2-\omega_2^2} \left[
-(K_0(t_2,\omega_2)-K_0(t_2,-\omega_2))\omega_2 t_2^2\right.\no\\
&&\left. -(J_0(\omega_2)-J_0(-\omega_2)) 
\omega_2 (3 t_2-2 \omega_2^2)\right.\no\\&&\left.
+I_0(t_2) t_2(t_2+3 \omega_2^2)
-\Delta_\pi 2(t_2-\omega_2^2) \right]\no\\&&
+\frac{1}{(t_1-\omega_3^2)^2} \left(
\frac{\omega_3 (K_0(t_1,\omega_3)-K_0(t_1,-\omega_3))}{2} \right.\no\\&&\left.
\left[(-q_2\cdot (q_1-q_2)+\omega_2\omega_3)(-4 t_1 M^2 + 4 M^2 \omega_3^2 + 3 t_1^2)
\right.\right.\no\\&&\left.\left.
-2(t_1-\omega_3^2)t_1 (t_1- M^2) \right]\right.\no\\&&\left.
+\frac{\omega_3 (J_0(\omega_3)-J_0(-\omega_3))}{2} 
\left[(-q_2\cdot (q_1-q_2)+\omega_2\omega_3)(5 t_1 - 2 \omega_3^2)
\right.\right.\no\\&&\left.\left.
-2(t_1-\omega_3^2) (t_1 - M^2)\right]\right.\no\\&&\left.
+ I_0(t_1)
\left[\frac{2}{3 t_1}q_2\cdot (q_1-q_2) (11 t_1^2 \omega_3^2 - 4 t_1 \omega_3^4 
- 2 M^2 t_1^2 + 4 M^2 t_1 \omega_3^2 -2 M^2 \omega_3^4 + 2 t_1^3)
\right.\right.\no\\&&\left.\left.
-2 \omega_2\omega_3 t_1 (2 t_1 + \omega_3^2)
+2(t_1-\omega_3^2)(t_1-M^2)(t_1+3 \omega_3^2)\right]
\right.\no\\&&\left.
-\Delta_\pi (t_1 -\omega_3)^2
\frac{4}{3 t_1}q_2\cdot (q_1-q_2)\right.\no\\&&\left.
+\frac{t_1-\omega_3^2}{16 \pi^2} 
\left[\frac{2}{9 t_1} q_2\cdot (q_1-q_2) 
(t_1^2-6 t_1 M^2 + 17 t_1 \omega_3^2 + 6 M^2 \omega_3^2)
-4 \omega_2 \omega_3^3 \right]
\right)
\Bigg\}\no\\
A_4 & = & i \frac{g_A}{4 F^5} \Bigg\{
\frac{\omega_2}{2(t_2-\omega_2^2)^2} \Bigg(
(K_0(t_2,\omega_2)+K_0(t_2,-\omega_2))\no\\&&
\left[-(\omega_1+\omega_3)\omega_2 (-4 t_2 M^2+4 M^2 \omega_2^2+3 t_2^2)
-(t_2-\omega_2^2) (t_2^2-4 t_2 M^2+4 M^2 \omega_2^2)\right]\no\\&&
+(J_0(\omega_2)+J_0(-\omega_2))
\left[-(\omega_1+\omega_3)\omega_2(5 t_2-2\omega_2^2)
+(t_2-\omega_2^2)(t_2-2 \omega_2^2)\right]\no\\&&
+J_0(0)t_2\left[6 (\omega_1+\omega_3)\omega_2 +2  (t_2-\omega_2^2)\right] \Bigg)\no\\&&
+\frac{\omega_3}{2(t_1-\omega_3^2)^2} \Bigg(
(K_0(t_1,\omega_3)+K_0(t_1,-\omega_3))\no\\&&
\left[(\omega_1+\omega_2)\omega_3 (-4 t_1 M^2+4 M^2 \omega_3^2+3 t_1^2)
+(t_1-\omega_3^2) (t_1^2-4 t_1 M^2+4 M^2 \omega_3^2)\right]\no\\&&
+(J_0(\omega_3)+J_0(-\omega_3))
\left[(\omega_1+\omega_2)\omega_3(5 t_1-2\omega_3^2)
-(t_1-\omega_3^2)(t_1-2 \omega_3^2)\right]\no\\&&
-J_0(0)t_1 \left[6 (\omega_1+\omega_2)\omega_3+2 (t_1-\omega_3^2)\right] \Bigg)
\Bigg\}\no\\
B_3 & = & i \frac{g_A}{4 F^5} \Bigg\{
\frac{1}{(t_3-\omega_1^2)^2} \left(
\frac{\omega_1 (K_0(t_3,\omega_1)-K_0(t_3,-\omega_1))}{2} \right.\no\\&&\left.
\left[(-q_3\cdot (q_2+q_3)+\omega_1\omega_3) (-4 t_3 M^2 + 4 M^2 \omega_1^2 + 3 t_3^2)
+2(t_3-\omega_1^2)t_3 (t_3- M^2) \right]\right.\no\\&&\left.
+\frac{\omega_1 (J_0(\omega_1)-J_0(-\omega_1))}{2} 
\left[(-q_3\cdot (q_2+q_3)+\omega_1\omega_3 ) (5 t_3 - 2
  \omega_1^2)\right.\right. \no\\
&& \left.\left.
+2(t_3-\omega_1^2) (t_3 - M^2)\right]\right.\no\\&&\left.
+ I_0(t_3)
\left[\frac{2}{3 t_3}q_3\cdot (q_2+q_3) (11 t_3^2 \omega_1^2 - 4 t_3 \omega_1^4 
- 2 M^2 t_3^2 + 4 M^2 t_3 \omega_1^2 -2 M^2 \omega_1^4 + 2 t_3^3)
\right.\right.\no\\&&\left.\left.
-2 \omega_1\omega_3 t_3 (2 t_3 + \omega_1^2)
-2(t_3-\omega_1^2)(t_3-M^2)(t_3+3 \omega_1^2)\right]
\right.\no\\&&\left.
-\Delta_\pi (t_3 -\omega_1)^2
\frac{4}{3 t_3}q_2\cdot (q_2+q_3)\right.\no\\&&\left.
+\frac{t_3-\omega_1^2}{16 \pi^2} 
\left[\frac{2}{9 t_3} q_3\cdot (q_2+q_3) 
(t_3^2-6 t_3 M^2 + 17 t_3 \omega_1^2 + 6 M^2 \omega_1^2)
-4 \omega_1^3 \omega_3 \right]
\right)\no\\&&
+\frac{1}{t_1-\omega_3^2} \left[
(K_0(\omega_3)-K_0(-\omega_3))\omega_3 t_1^2
+(J_0(\omega_3)-J_0(-\omega_3)) \omega_3 (3 t_1-2 \omega_3^2)\right.\no\\&&\left.
-I_0(t_1) t_1(t_1+3 \omega_3^2)
+\Delta_\pi 2(t_1-\omega_3^2) \right]
\Bigg\}\no\\
D_1 & = & D_4 = 0
\eea
Diagram s32:\\
\bea
A_2 & = & i\frac{g_A^3}{2 F^5} \left( S^2+\frac{1}{2}\right) J_2^\prime(0)\no\\
B_3 & = & -i\frac{g_A^3}{2 F^5} \left( S^2+\frac{1}{2}\right) J_2^\prime(0)\no\\
A_1 & = & A_4 = D_1 = D_4 = 0
\eea
Diagram s33:\\
\bea
A_1 & = & -i \frac{g_A^3}{F^5} \Bigg\{
\frac{1}{\omega_1 (t_3-\omega_1^2)} \left[
(K_0(t_3,\omega_1)-K_0(t_3,-\omega_1)) \frac{M^2 t_3 (t_3-2 \omega_1^2)}{16}
\right.\no\\&&\left.
+(J_0(\omega_1)-J_0(-\omega_1))\frac{4 \omega_1^2 t_3 -7 M^2 t_3 -4 \omega_1^4 + 
4 M^2 \omega_1^2}{48}
+I_0(t_3) \frac{M^2 \omega_1^3}{4}\right.\no\\&&\left.
+\Delta_\pi \frac{\omega_1 (t_3-\omega_1^2)}{6}
-\frac{\omega_1}{16 \pi^2} \frac{-21 M^2+8 \omega_1^2}{18}\right]\no\\&&
-\frac{\omega_2^2-q_2\cdot (q_1-q_3)}{16 \omega_2 (t_2-\omega_2^2)^2} \no\\&&\Bigg[ 
(K_0(t_2,\omega_2)-K_0(t_2,-\omega_2))(t_2^3+2 M^2 t_2^2 -4 t_2^2 \omega_2^2 -
8 t_2 M^4 + 4 M^2 \omega_2^2 t_2 + 8 M^4 \omega_2^2)\no\\&&
-(J_0(\omega_2)-J_0(-\omega_2))(t_2^2-2 M^2 t_2 + 2 \omega_2^2 t_2 -4 M^2 \omega_2^2)
\no\\&&
+I_0(t_2) 4 \omega_2 (t_2^2-2 M^2 t_2 + 2 \omega_2^2 t_2 -4 M^2 \omega_2^2)
+\frac{\omega_2 (t_2-2 M^2)(t_2-\omega_2^2)}{2\pi^2} \Bigg]\no\\&&
-\frac{\omega_3^2-q_3\cdot (q_1-q_2)}{16 \omega_3 (t_1-\omega_3^2)^2} \no\\&& \Bigg[ 
(K_0(t_1,\omega_3)-K_0(t_1,-\omega_3))(t_1^3+2 M^2 t_1^2 -4 t_1^2 \omega_3^2 -
8 t_1 M^4 + 4 M^2 \omega_3^2 t_1 + 8 M^4 \omega_3^2)\no\\&&
-(J_0(\omega_3)-J_0(-\omega_3))(t_1^2-2 M^2 t_1 + 2 \omega_3^2 t_1 -4 M^2 \omega_3^2)
\no\\&&
+I_0(t_1) 4 \omega_3 (t_1^2-2 M^2 t_1 + 2 \omega_3^2 t_1 -4 M^2 \omega_3^2)
+\frac{\omega_3 (t_1-2 M^2)(t_1-\omega_3^2)}{2\pi^2} \Bigg]
\Bigg\}\no\\
A_2 & = & -i \frac{g_A^3}{F^5} \Bigg\{
\frac{1}{\omega_1 (t_3-\omega_1^2)} \left[
(K_0(t_3,\omega_1)-K_0(t_3,-\omega_1))
\frac{t_3(t_3^2-2 \omega_1^2 t_3-2 M^2 t_3 + 4 M^2 \omega_1^2)}{16}\right.\no\\&&\left.
-(J_0(\omega_1)-J_0(-\omega_1))\frac{3 t_3^2-10 M^2 t_3 + 4 \omega_1^2 t_3 -4 \omega_1^4
+4 M^2 \omega_1^2}{48}\right.\no\\&&\left.
+I_0(t_3) \frac{\omega_1^3 (t_3-2 M^2)}{4}
-\Delta_\pi \frac{\omega_1 (t_3-\omega_1^2)}{6}
+\frac{\omega_1}{16 \pi^2}\frac{9 t_3 -30 M^2 + 8 \omega_1^2}{18} \right]\no\\&&
-\frac{M^2(\omega_2^2-q_2\cdot (q_1-q_3))}{16 \omega_2 (t_2-\omega_2^2)^2} \Bigg[
-(J_0(\omega_2)-J_0(-\omega_2)) (t_2+2\omega_2^2)
+I_0(t_2) 4 \omega_2 (t_2+2\omega_2^2)\no\\&&
+(K_0(t_2,\omega_2)-K_0(t_2,-\omega_2))  
(-4\omega_2^2 t_2 + t_2^2 -4 M^2 \omega_2^2+4 M^2 t_2)
+\frac{\omega_2(t_2-\omega_2^2)}{2\pi^2} \Bigg] \no\\&&
-\frac{\omega_3^2-q_3\cdot (q_1-q_2)}{16 \omega_3 (t_1-\omega_3^2)^2} \no\\&& \Bigg[ 
(K_0(t_1,\omega_3)-K_0(t_1,-\omega_3))(t_1^3+2 M^2 t_1^2 -4 t_1^2 \omega_3^2 -
8 t_1 M^4 + 4 M^2 \omega_3^2 t_1 + 8 M^4 \omega_3^2)\no\\&&
-(J_0(\omega_3)-J_0(-\omega_3))(t_1^2-2 M^2 t_1 + 2 \omega_3^2 t_1 -4 M^2 \omega_3^2)
\no\\&&
+I_0(t_1) 4 \omega_3 (t_1^2-2 M^2 t_1 + 2 \omega_3^2 t_1 -4 M^2 \omega_3^2)
+\frac{\omega_3 (t_1-2 M^2)(t_1-\omega_3^2)}{2\pi^2} \Bigg]
\Bigg\}\no\\
A_4 & = & -i \frac{g_A^3}{F^5} \Bigg\{
\frac{\omega_2-\omega_3}{16 (t_3-\omega_1^2)^2} \Bigg[
(K_0(t_3,\omega_1)+K_0(t_3,-\omega_1))t_3^2 (t_3-2\omega_1^2)\no\\&&
-(J_0(\omega_1)+J_0(-\omega_1))\frac{1}{3} 
(-4\omega_1^2 t_3 + 3 t_3^2 -4 M^2\omega_1^2 + 4 M^2 t_3 + 4 \omega_1^4)
\no\\&&
-J_0(0) \frac{2}{3} (-6 \omega_2^2 t_3+3 t_3^2 +4 M^2 \omega_1^2-4 M^2 t_3) \Bigg]
\no\\&&
+\frac{(\omega_2^2-q_2\cdot(q_1-q_3))(\omega_1+\omega_3)}{16 (t_2-\omega_2^2)^3} 
\no\\&&\Bigg[
-(K_0(t_2,\omega_2)+K_0(t_2,-\omega_2))
(4 M^2\omega_2^2 t_2 + 4 M^2 t_2^2 -8 M^2 \omega_2^4 -6 t_2^2\omega_2^2+t_2^3)
\no\\&&
+(J_0(\omega_2)+J_0(-\omega_2))\frac{1}{3} (16 M^2 \omega_2^2-16 M^2 t_2 -4 \omega_2^4
+16 \omega_2^2 t_2 + 3 t_2^2)\no\\&&
+J_0(0)\frac{2}{3} (-16 M^2 \omega_2^2 + 16 M^2 t_2 -18 \omega_2^2 t_2 + 3 t_2^2)
\Bigg]\no\\&&
+\frac{\omega_1\omega_2 -q_1 \cdot q_2}{24 \omega_2 (t_2-\omega_2^2)^2} \left[
(K_0(t_2,\omega_2)+K_0(t_2,-\omega_2))3 \omega_2^2 (4 M^2 \omega_2^2+t_2^2-4 M^2 t_2)
\right.\no\\&&\left.
+(J_0(\omega_2)+J_0(-\omega_2))(8 M^2 \omega_2^2 -8 M^2 t_2+5\omega_2^2 t_2-2\omega_2^4)
\right.\no\\&&\left.
-J_0(0) 2 (-8 M^2 t_2+8 M^2 \omega_2^2+3 \omega_2^2 t_2) \right]\no\\&&
-\frac{(\omega_3^2-q_3\cdot(q_1-q_2))(\omega_1+\omega_2)}{16 (t_1-\omega_3^2)^3} 
\no\\&&\Bigg[
-(K_0(t_1,\omega_3)+K_0(t_1,-\omega_3))
(4 M^2\omega_3^2 t_1 + 4 M^2 t_1^2 -8 M^2 \omega_3^4 -6 t_1^2\omega_3^2+t_1^3)
\no\\&&
+(J_0(\omega_3)+J_0(-\omega_3))\frac{1}{3} (16 M^2 \omega_3^2-16 M^2 t_1 -4 \omega_3^4
+16 \omega_3^2 t_1 + 3 t_1^2)\no\\&&
+J_0(0)\frac{2}{3} (-16 M^2 \omega_3^2 + 16 M^2 t_1 -18 \omega_3^2 t_1 + 3 t_1^2)
\Bigg]\no\\&&
-\frac{\omega_1\omega_3 -q_1 \cdot q_3}{24 \omega_3 (t_1-\omega_3^2)^2} \left[
(K_0(t_1,\omega_3)+K_0(t_1,-\omega_3))3 \omega_3^2 (4 M^2 \omega_3^2+t_1^2-4 M^2 t_1)
\right.\no\\&&\left.
+(J_0(\omega_3)+J_0(-\omega_3))(8 M^2 \omega_3^2 -8 M^2 t_1+5\omega_3^2 t_1-2\omega_3^4)
\right.\no\\&&\left.
-J_0(0) 2 (-8 M^2 t_1+8 M^2 \omega_3^2+3 \omega_3^2 t_1) \right] \Bigg\}\no\\
B_3 & = & -i \frac{g_A^3}{F^5} \Bigg\{
\frac{\omega_1^2-q_1\cdot (q_2+q_3)}{16 \omega_1 (t_3-\omega_1^2)^2} \no\\&&\Bigg[ 
(K_0(t_3,\omega_1)-K_0(t_3,-\omega_1))(t_3^3+2 M^2 t_3^2 -4 t_3^2 \omega_1^2 -
8 t_3 M^4 + 4 M^2 \omega_1^2 t_3 + 8 M^4 \omega_1^2)\no\\&&
-(J_0(\omega_1)-J_0(-\omega_1))(t_3^2-2 M^2 t_3 + 2 \omega_1^2 t_3 -4 M^2 \omega_1^2)
\no\\&&
+I_0(t_3) 4 \omega_1 (t_3^2-2 M^2 t_3 + 2 \omega_1^2 t_3 -4 M^2 \omega_1^2)
+\frac{\omega_1 (t_3-2 M^2)(t_3-\omega_1^2)}{2} \Bigg]\no\\&&
-\frac{1}{\omega_2 (t_2-\omega_2^2)} \left[
(K_0(t_2,\omega_2)-K_0(t_2,-\omega_2))
\frac{t_2(t_2^2-2 \omega_2^2 t_2-2 M^2 t_2 + 4 M^2 \omega_2^2)}{16}\right.\no\\&&\left.
-(J_0(\omega_2)-J_0(-\omega_2))\frac{3 t_2^2-10 M^2 t_2 + 4 \omega_2^2 t_2 -4 \omega_2^4
+4 M^2 \omega_2^2}{48}\right.\no\\&&\left.
+I_0(t_2) \frac{\omega_2^3 (t_2-2 M^2)}{4}
-\Delta_\pi \frac{\omega_2 (t_2-\omega_2^2)}{6}
+\frac{\omega_2}{16 \pi^2}\frac{9 t_2 -30 M^2 + 8 \omega_2^2}{18} \right]\no\\&&
+\frac{M^2(\omega_3^2-q_3\cdot (q_1-q_2))}{16 \omega_3 (t_1-\omega_3^2)^2} \Bigg[
-(J_0(\omega_3)-J_0(-\omega_3)) (t_1+2\omega_3^2)
+I_0(t_1) 4 \omega_3 (t_1+2\omega_3^2)\no\\&&
+(K_0(t_1,\omega_3)-K_0(t_1,-\omega_3))  
(-4\omega_3^2 t_1 + t_1^2 -4 M^2 \omega_3^2+4 M^2 t_1)
+\frac{\omega_3(t_1-\omega_3^2)}{2\pi^2} \Bigg] \no\\
D_1 & = & 0\no\\
D_4 & = & -i \frac{g_A^3}{16 F^5} \Bigg\{
\frac{1}{\omega_1 (t_3-\omega_1^2)} \Bigg[
-(K_0(t_3,\omega_1)-K_0(t_3,-\omega_1))(4 M^2 \omega_1^2+t_3^2-4 M^2 t_3)
\no\\&&
+(J_0(\omega_1)-J_0(-\omega_1))(t_3-2 \omega_1^2)
+I_0(t_3) 4 t_3 \omega_1
+\frac{\omega_1(t_3-\omega_1^2)}{2 \pi^2} \Bigg]\no\\&&
+\frac{1}{\omega_2 (t_2-\omega_2^2)} \Bigg[
-(K_0(t_2,\omega_2)-K_0(t_2,-\omega_2))(4 M^2 \omega_2^2+t_2^2-4 M^2 t_2)
\no\\&&
+(J_0(\omega_2)-J_0(-\omega_2))(t_2-2 \omega_2^2)
+I_0(t_2) 4 t_2 \omega_2
+\frac{\omega_2(t_2-\omega_2^2)}{2 \pi^2} \Bigg]\no\\&&
+\frac{1}{\omega_3 (t_1-\omega_3^2)} \Bigg[
-(K_0(t_1,\omega_3)-K_0(t_1,-\omega_3))(4 M^2 \omega_3^2+t_1^2-4 M^2 t_1)\no\\&&
+(J_0(\omega_3)-J_0(-\omega_3))(t_1-2 \omega_3^2)
+I_0(t_1) 4 t_1 \omega_3
+\frac{\omega_3(t_1-\omega_3^2)}{2 \pi^2} \Bigg]
\Bigg\}
\eea
Diagram s34:\\
\bea
A_1 & = & -i \frac{g_A^3}{F^5} \frac{1}{(q_2+q_3-q_1)^2-M^2}
\left(S^2 +\frac{1}{2}\right) J_2^\prime(0) (t_3-M^2)\no\\
A_2 & = & -i \frac{g_A^3}{F^5} \frac{1}{(q_2+q_3-q_1)^2-M^2}
\left(S^2 +\frac{1}{2}\right) J_2^\prime(0) (t_2-M^2)\no\\
B_3 & = & i \frac{g_A^3}{F^5} \frac{1}{(q_2+q_3-q_1)^2-M^2}
\left(S^2 +\frac{1}{2}\right) J_2^\prime(0) (t_1-M^2)\no\\
A_4 & = & D_1 = D_4 = 0
\eea
Diagram s35: Combine with diagrams R1d and 3r.
\item{\underline{Renormalization diagrams:}}
We now give those contributions
from the first order diagrams, which come in via renormalization
and thus cannot be obtained from the
1/m expansion of the relativistic amplitude. To third order, 
the renormalized pion mass, pion Z-factor, decay constant 
and the axial--vector coupling read:
\bea
Z_{\pi} & = & 1-\frac{2 M^2}{F^2} {\ell}_4 - \frac{\Delta_\pi}{F^2}~, \\
M^2_\pi & = & M^2 \left\{ 1 +\frac{2 M^2}{F^2} {\ell}_3 +\frac{\Delta_\pi}{2 F^2} 
\right\}~,\\
F_\pi & = & F \left\{ 1 + \frac{M^2}{F^2} \ell_4 -\frac{\Delta_\pi}{F^2}\right\}~,  \\
\frac{g_A}{F_\pi} & = &\left( \frac{g_A}{F}\right)_0 \left\{ 1 -\frac{M^2}{F^2} \ell_4 +
\frac{4 M^2}{g_A} d_{16}(\lambda)  + \frac{g_A^2}{4 F^2}
\left( \Delta_\pi -\frac{M^2}{4 \pi^2} \right) \right\}~,
\eea
Here, $M,F$ and $g_{A,0}$ denote the corresponding leading order
values in the chiral expansion. Moreover, one has to take care of the 
renormalization of the nucleon mass when making the heavy baryon transformation:
\be
\frac{i}{ v\cdot p} \rightarrow \frac{i}{v\cdot p} + 
\frac{i}{(v\cdot p)^2} \frac{3 g_A^2 M^3}{32 \pi F^2}~.
\ee
The nucleon Z-factor takes the simple form:
\be
Z_N = 1- \frac{g_A^2}{F^2} \frac{3 M^2}{32 \pi^2}~.
\ee
Note that the familiar ln--terms have been treated as described in~\cite{fmsz}.
With that, we find for the pertinent diagrams:\\
Diagram R1a:\\
\bea
A_1 & = & -i \left( \frac{g_A}{F} \right)^3 (\omega_2\omega_3 - q_2\cdot q_3)
 \frac{3 M^2}{4} 
\left(\frac{1}{\omega_1\omega_3}+\frac{1}{\omega_1\omega_2}-\frac{1}{\omega_2\omega_3}
\right)\no\\&&
\left( \frac{g_A^2}{16\pi^2 F^2} - \frac{8d_{16} (\lambda) }{g_A}
-\frac{\Delta_\pi}{2 F^2}(g_A^2+2) \right)\no\\
A_2 & = & -i \left( \frac{g_A}{F} \right)^3 (\omega_2\omega_3 - q_2\cdot q_3)
 \frac{3 M^2}{4} 
\left(-\frac{1}{\omega_1\omega_3}+\frac{1}{\omega_1\omega_2}+\frac{1}{\omega_2\omega_3}
\right)\no\\&&
\left( \frac{g_A^2}{16\pi^2 F^2} - \frac{8 d_{16}(\lambda) }{g_A}
-\frac{\Delta_\pi}{2 F^2}(g_A^2+2) \right)\no\\
A_4 & = & -i \left( \frac{g_A}{F} \right)^5 (\omega_2\omega_3 - q_2\cdot q_3)
 \frac{3 M^3}{64\pi} \no\\&&
\left[-\frac{1}{\omega_1\omega_3}\left(\frac{1}{\omega_1}+\frac{1}{\omega_3}\right)
+\frac{1}{\omega_1\omega_2}\left(\frac{1}{\omega_1}+\frac{1}{\omega_2}\right)
+\frac{1}{\omega_2\omega_3}\left(-\frac{1}{\omega_2}+\frac{1}{\omega_3}\right)
\right]\no\\
B_3 & = & i \left( \frac{g_A}{F} \right)^3 (\omega_1\omega_3 - q_1\cdot q_3)
 \frac{3 M^2}{4} 
\left(\frac{1}{\omega_1\omega_3}+\frac{1}{\omega_1\omega_2}+\frac{1}{\omega_2\omega_3}
\right)\no\\&&
\left( \frac{g_A^2}{16\pi^2 F^2} - \frac{8 d_{16}(\lambda)}{g_A} d_{16}(\lambda)
-\frac{\Delta_\pi}{2 F^2}(g_A^2+2) \right)\no\\
D_1 & = & -i \left( \frac{g_A}{F} \right)^5 
 \frac{3 M^3}{128\pi} \no\\&&
\left[-\frac{1}{\omega_1\omega_3}\left(\frac{1}{\omega_1}+\frac{1}{\omega_3}\right)
+\frac{1}{\omega_1\omega_2}\left(\frac{1}{\omega_1}+\frac{1}{\omega_2}\right)
+\frac{1}{\omega_2\omega_3}\left(-\frac{1}{\omega_2}+\frac{1}{\omega_3}\right)
\right]\no\\
D_4 & = & -i \left( \frac{g_A}{F} \right)^3 
 \frac{3 M^2}{8} 
\left(\frac{1}{\omega_1\omega_3}+\frac{1}{\omega_1\omega_2}-\frac{1}{\omega_2\omega_3}
\right)\no\\&&
\left( \frac{g_A^2}{16\pi^2 F^2} - \frac{8 d_{16}(\lambda)}{g_A}
-\frac{\Delta_\pi}{2 F^2}(g_A^2+2) \right)
\eea
Diagrams R1b+R1c:\\
\bea
A_2 & = & i \frac{g_A}{2 F^3} \frac{\omega_2-\omega_3}{\omega_1} 
\left( \frac{4 M^2}{g_A} d_{16}(\lambda) +\frac{g_A^2 M^2}{32 \pi^2 F^2} 
+\frac{\Delta_\pi}{F^2}
\left( \frac{7}{2} +\frac{g_A^2}{4} \right) \right)\no\\
A_4 & = & -i \frac{3 g_A^3 M^3}{64 \pi F^5} 
\frac{\omega_2-\omega_3}{\omega_1^2}\no\\
B_3 & = & i \frac{g_A}{2 F^3} \frac{\omega_1+\omega_3}{\omega_2} 
\left( \frac{4 M^2}{g_A} d_{16}(\lambda) +\frac{g_A^2 M^2}{32 \pi^2 F^2} 
+\frac{\Delta_\pi}{F^2}
\left( \frac{7}{2} +\frac{g_A^2}{4} \right) \right)\no\\
A_1 & = & D_1 = D_4 = 0
\eea
Diagram R1d:\\
\bea
A_2 & = & -i \frac{g_A}{2 F^3} \left( \frac{4 M^2}{g_A} d_{16}(\lambda)
+\frac{g_A^2 M^2}{32 \pi^2 F^2} +\frac{\Delta_\pi}{F^2} \left( \frac{7}{2}
+\frac{g_A^2}{4} \right)\right)\no\\
B_3 & = & i \frac{g_A}{2 F^3} \left( \frac{4 M^2}{g_A} d_{16}(\lambda)
+\frac{g_A^2 M^2}{32 \pi^2 F^2} +\frac{\Delta_\pi}{F^2} \left( \frac{7}{2}
+\frac{g_A^2}{4} \right)\right)\no\\
A_1 & = & A_4 = D_1 = D_4 = 0
\eea
Diagrams R1e+3r+s35:\\
\bea
A_1 & = & i \frac{g_A}{F^3} \frac{1}{(q_1-q_2-q_3)^2-M^2}
\left\{ -M^2 \left( \frac{2 M^2}{F^2} \ell_3 +\frac{\Delta_\pi}{2 F^2} \right) 
\right.\no\\&&\left.
+(t_3-M^2) \left( \frac{4 M^2}{g_A} d_{16}(\lambda) +
\frac{g_A^2 M^2}{32 \pi^2 F^2} +\frac{\Delta_\pi}{F^2} \left(\frac{9}{2}
+\frac{g_A^2}{4}\right) +\frac{2 M^2}{F^2} \ell_4 \right)
\right\}\no\\
A_2 & = & i \frac{g_A}{F^3} \frac{1}{(q_1-q_2-q_3)^2-M^2}
\left\{ -M^2 \left( \frac{2 M^2}{F^2} \ell_3 +\frac{\Delta_\pi}{2 F^2} \right) 
\right.\no\\&&\left.
+(t_2-M^2) \left( \frac{4 M^2}{g_A} d_{16}(\lambda) +
\frac{g_A^2 M^2}{32 \pi^2 F^2} +\frac{\Delta_\pi}{F^2} \left(\frac{9}{2}
+\frac{g_A^2}{4}\right) +\frac{2 M^2}{F^2} \ell_4 \right)
\right\}\no\\
B_3 & = & -i \frac{g_A}{F^3} \frac{1}{(q_1-q_2-q_3)^2-M^2}
\left\{ -M^2 \left( \frac{2 M^2}{F^2} \ell_3 +\frac{\Delta_\pi}{2 F^2} \right) 
\right.\no\\&&\left.
+(t_1-M^2) \left( \frac{4 M^2}{g_A} d_{16}(\lambda) +
\frac{g_A^2 M^2}{32 \pi^2 F^2} +\frac{\Delta_\pi}{F^2} \left(\frac{9}{2}
+\frac{g_A^2}{4}\right) +\frac{2 M^2}{F^2} \ell_4 \right)
\right\}\no\\
A_4 & = & D_1 = D_4 = 0
\eea
\item{\underline{Counterterm amplitude at threshold:}}\\
Finally, we spell out the contribution from the various counterterms
to the threshold amplitudes. At threshold, $\sqrt{s} = m + 2 M_\pi$,
$q_2 = q_3 = (M_\pi,0)$ and $p_2 = (0,0)$.
The energy of the incoming pion and the incoming nucleon can be easily
evaluated together with the spinor normalization factor ${\cal N}_1$.
At threshold, the amplitude is fully given by $A_1$ and $A_2$.
The contribution from the different diagrams reads:\\
Diagrams 2a+2b:\\
\bea
A_1 & = & -2 i \frac{g_A}{F^3} \frac{M^2 v\cdot p_1}{\omega_1^2} 
(2 c_1 + c_2 + c_3 ) \no \\
A_2 & = & 0
\eea
Diagrams 3a+3b:\\
\bea
A_1 & = & 0\no \\
A_2 & = & i \frac{2 g_A M}{m F^3} [2 c_1 M - \omega_1 (c_2+c_3) ]
\eea
Diagrams 3c+3d:\\
\bea
A_1 & = & i \frac{g_A}{m F^3} \frac{M^2 (\omega_1^2-M^2)}{\omega_1^2}
(2 c_1 + c_2 + c_3) \no \\
A_2 & = & 0
\eea
Diagrams 3n + 3o:\\
\bea
A_1 & = & i \frac{8 M^2 g_A}{F^3} \left[ \frac{c_2}{2 m} + \tilde{d}_{26}(\lambda)
+ \tilde{d}_{27}(\lambda)+ \tilde{d}_{28}(\lambda) \right] \no \\
A_2 & = & 0
\eea
Diagram 3p:\\
\bea
A_1 & = & i \frac{4 M^2}{F^3} \left[ \frac{g_A c_4}{2 m} -d_{10}(\lambda)
- d_{12}(\lambda) - d_{16}(\lambda)+ \frac{1}{2} d_{18} \right] \no \\
A_2 & = & i \frac{2 M^2}{F^3} \left[ -\frac{g_A c_4}{2 m} -d_{11}(\lambda)
-d_{13}(\lambda) + d_{16}(\lambda)- \frac{1}{2} d_{18} \right]
\eea
Diagram 3q:\\
\bea
A_1 & = & i \frac{3 M^3}{2 F^3 (\omega_1-M)} 
\left[ 2 d_{16}(\lambda) - d_{18} \right] \no \\
A_2 & = & -i \frac{M^2(2 \omega_1-M)}{2 F^3 (\omega_1-M)} 
\left[ 2 d_{16}(\lambda) - d_{18} \right]
\eea
Diagram 3r: Combine with R1e.\\
Diagram 3s:\\
\bea
A_1 & = & i \frac{g_A}{F^5} \frac{M}{\omega_1-M} \left[ 2 \ell_1 M (2 \omega_1-M)
+2 \ell_2 \omega_1 (2 M-\omega_1) +\ell_4 M (2 M+\omega_1) \right] \no
\\
A_2 & = & i \frac{g_A}{F^5} \frac{M}{\omega_1-M} \left[ 2 \ell_1 \omega_1 (2 M-\omega_1)
+ \ell_2 (4 M \omega_1 -\omega_1^2-M^2) -\ell_4 M \omega_1 \right]
\eea
Counterterm contribution at threshold from renormalization diagrams:\\
Diagram R1d:\\
\bea
A_1 & = & 0 \no\\
A_2 & = & -i \frac{2 M^2}{F^3} d_{16}(\lambda)
\eea
Diagrams R1e+3r:\\
\bea
A_1 & = & -i \frac{g_A}{F^3} \frac{M^3}{2(\omega_1-M)} 
\left[ \frac{-\ell_3+3 \ell_4}{F^2} +\frac{6}{g_A} d_{16}(\lambda)
\right] \no\\
A_2 & = & i \frac{g_A}{F^3} \frac{M^2}{2(\omega_1-M)} 
\left[ \frac{M \ell_3 + (2 \omega_1-M) \ell_4}{F^2} 
+\frac{2(2 \omega_1-M)}{g_A}  d_{16}(\lambda) \right]
\eea
Altogether:\\
\bea
A_1 & = & \frac{i}{F^3} \Bigg\{ \frac{g_A M^2}{\omega_1^2} (2 c_1 +
c_2 + c_3) 
\left( -2 v\cdot p_1 + \frac{\omega_1^2-M^2}{m} \right)
+\frac{2 M^2 g_A}{m} (2 c_2 + c_4)\no\\&&
+ 8 M^2 g_A (\tilde{d}_{26}(\lambda)+\tilde{d}_{27}(\lambda)+\tilde{d}_{28}(\lambda))
\no\\&&
-4 M^2 (d_{10}(\lambda)+d_{12}(\lambda)+d_{16}(\lambda)-\frac{1}{2} d_{18})
-\frac{3 M^3}{2(\omega_1-M)} d_{18}\no\\&&
+\frac{g_A}{F^2} \frac{M}{\omega_1-M} \left[2\ell_1 M (2\omega_1-M) 
+ 2\ell_2\omega_1(2M-\omega_1) \right. \no\\&& \left.
+\ell_4M\left(\frac{M}{2}+\omega_1\right)
+\frac{M^2}{2} \ell_3\right] 
\Bigg\}~, \no \\ && \\
A_2 & = & \frac{i}{F^3} \Bigg\{
\frac{g_A M}{m} (4c_1 M - 2\omega_1 (c_2+c_3)-c_4 M)\no\\&&
-2 M^2 (d_{11}(\lambda)+d_{13}(\lambda)+\frac{1}{2} d_{18})
+\frac{(2 \omega_1-M)M^2}{2(\omega_1-M)} d_{18}\no\\&&
+\frac{g_A}{F^2} \frac{M}{\omega_1-M}\left[2\ell_1 \omega_1 (2M-\omega_1)
+\ell_2(4M\omega_1-\omega_1^2-M^2)\right. \no\\&& \left.
+\frac{M^2}{2}(\ell_3-\ell_4) \right]\Bigg\}~.
\eea
\end{enumerate}

\bigskip
%\pagebreak
%%%%%%%%%%%%%%%%%%%%%%%%%%%%%%%%%%%%%%%%%%%%%%%%%%%%%%%%%%%%%%%%%%%%%%%%%%

\newpage
%%%%%%%%%%%%%%%%%%%%%%%%%%%%%%%%%%%%%%%%%%%%%%%%%%%%%%%%%%%%%%%%%%%%%%%%%

$\,$

\section*{Figures}

\vspace{1.5cm}
 
\begin{figure}[h]
   \vspace{0.5cm}
   \epsfysize=8cm
   \centerline{\epsffile{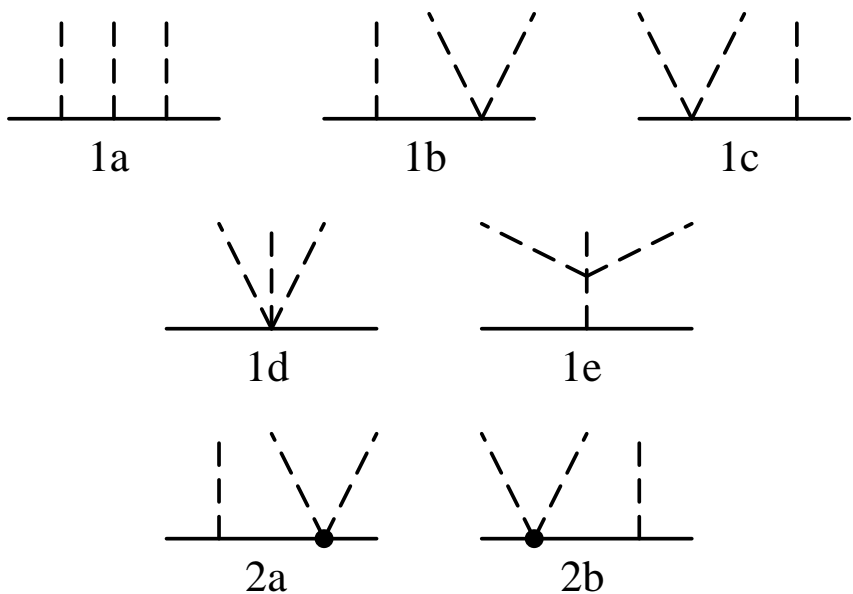}}
   \vspace{1cm}
   \centerline{\parbox{13cm}{\caption{\label{fig1}
Tree graphs with insertions from the dimension one Lagrangian (1a--1e)
and the dimension two Lagrangian as depicted by the dot (2a,b). 
Solid and dashed lines denote nucleons and pions, respectively.
}}}
\end{figure}

\begin{figure}[h]
   \vspace{0.5cm}
   \epsfysize=13cm
   \centerline{\epsffile{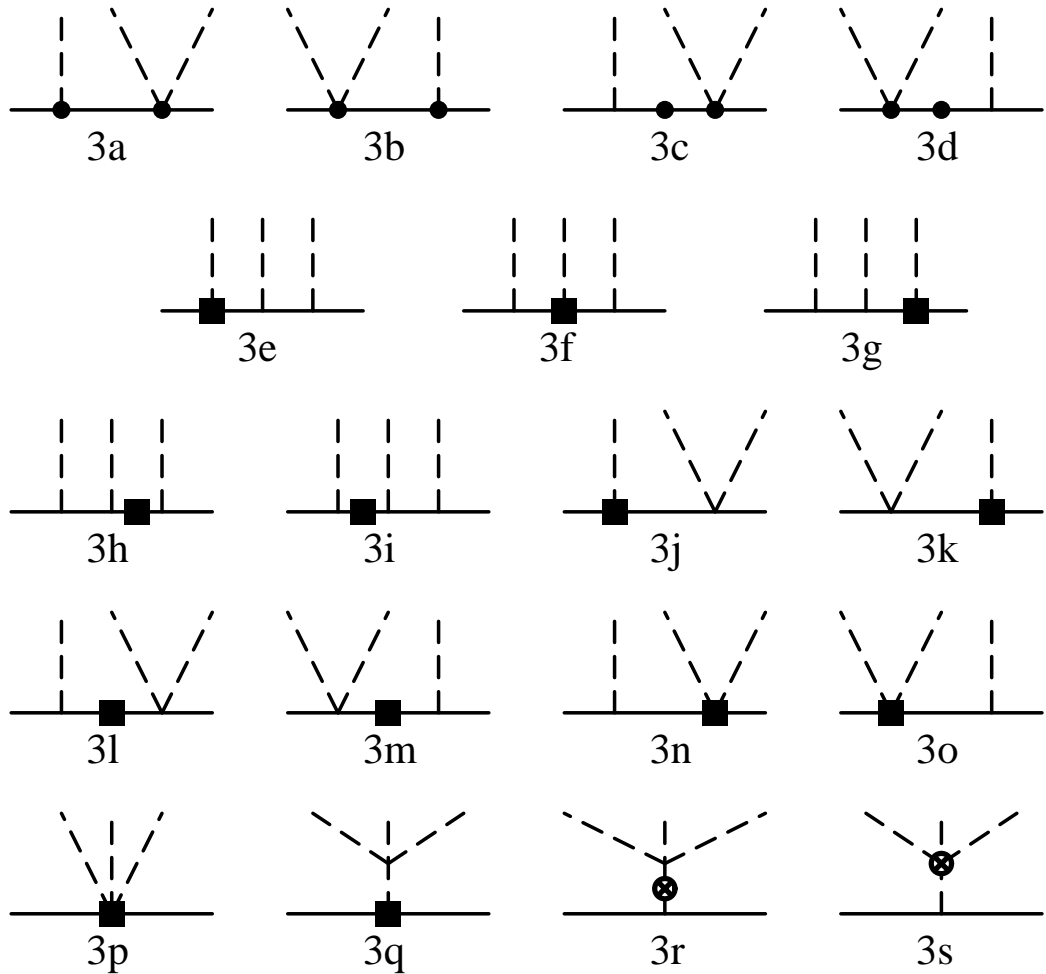}}
   \vspace{1cm}
   \centerline{\parbox{13cm}{\caption{\label{fig2}
Tree graphs of third order in the chiral expansion. The filled box denotes
an insertion from the dimension three pion--nucleon Lagrangian whereas the
circle--cross denotes a fourth order mesonic insertion.
  }}}
\end{figure}

\begin{figure}[h]
   \vspace{0.5cm}
   \epsfysize=13cm
   \centerline{\epsffile{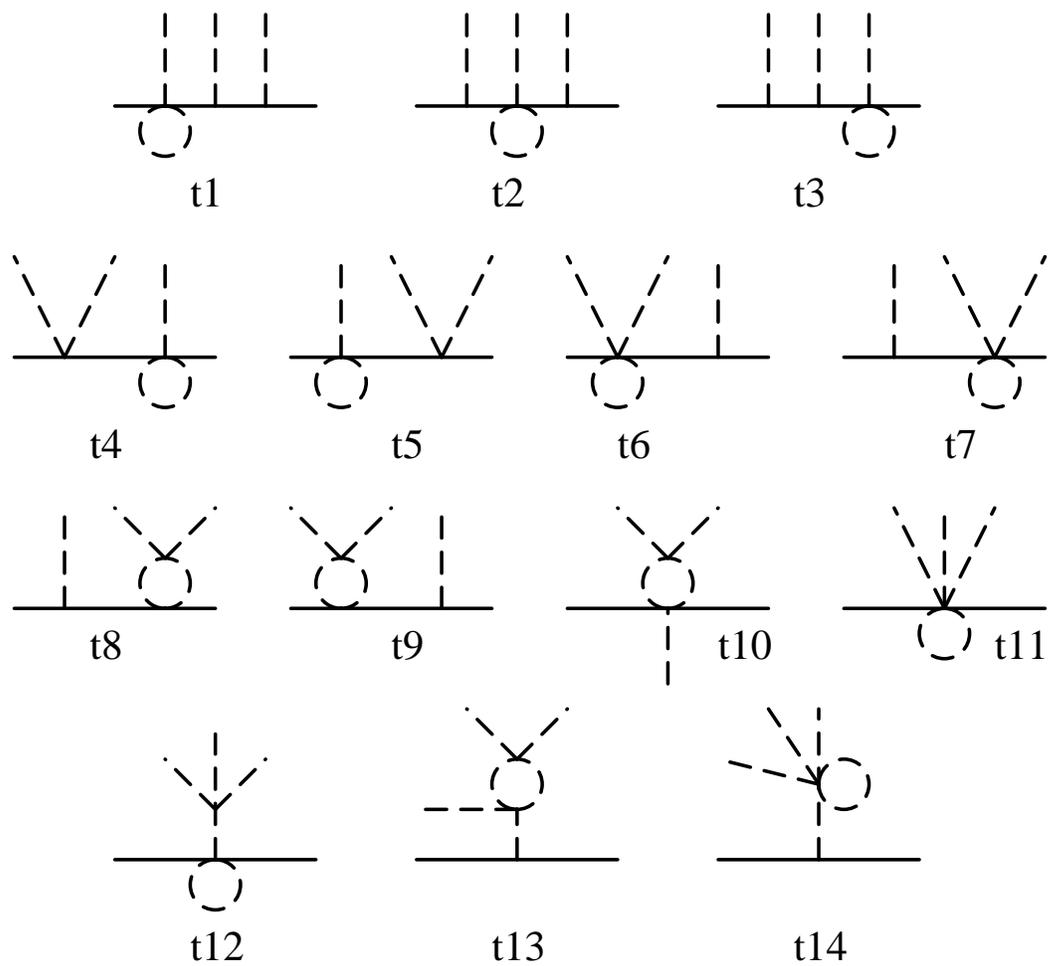}}
   \vspace{1cm}
   \centerline{\parbox{13cm}{\caption{\label{fig3}
One--loop graphs of the tadpole type.
  }}}
\end{figure}

\begin{figure}[h]
   \vspace{0.5cm}
   \epsfysize=13cm
   \centerline{\epsffile{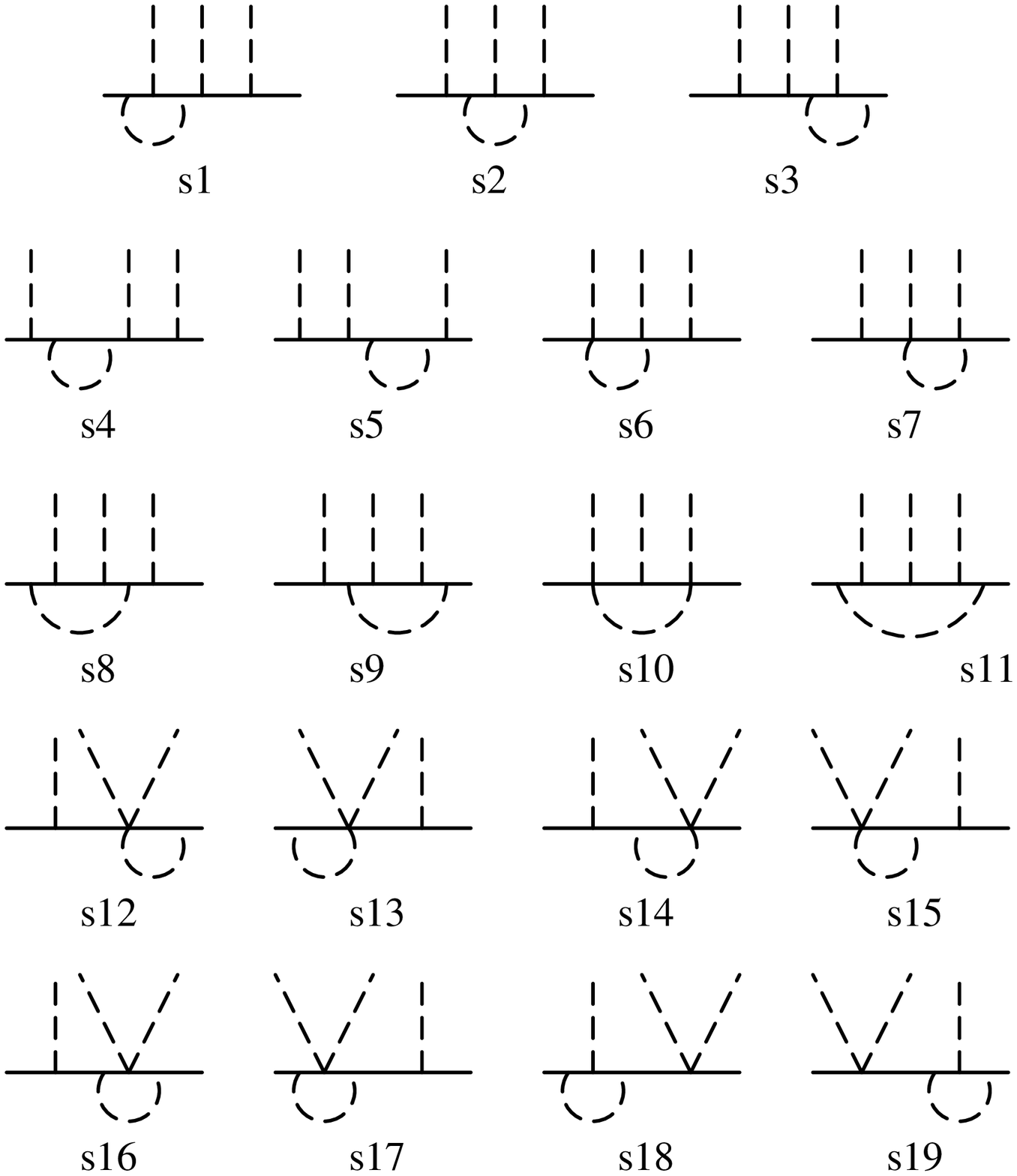}}
   \vspace{1cm}
   \centerline{\parbox{13cm}{\caption{\label{fig4}
One--loop graphs of the self--energy type.
  }}}
\end{figure}

\begin{figure}[h]
   \vspace{0.5cm}
   \epsfysize=13cm
   \centerline{\epsffile{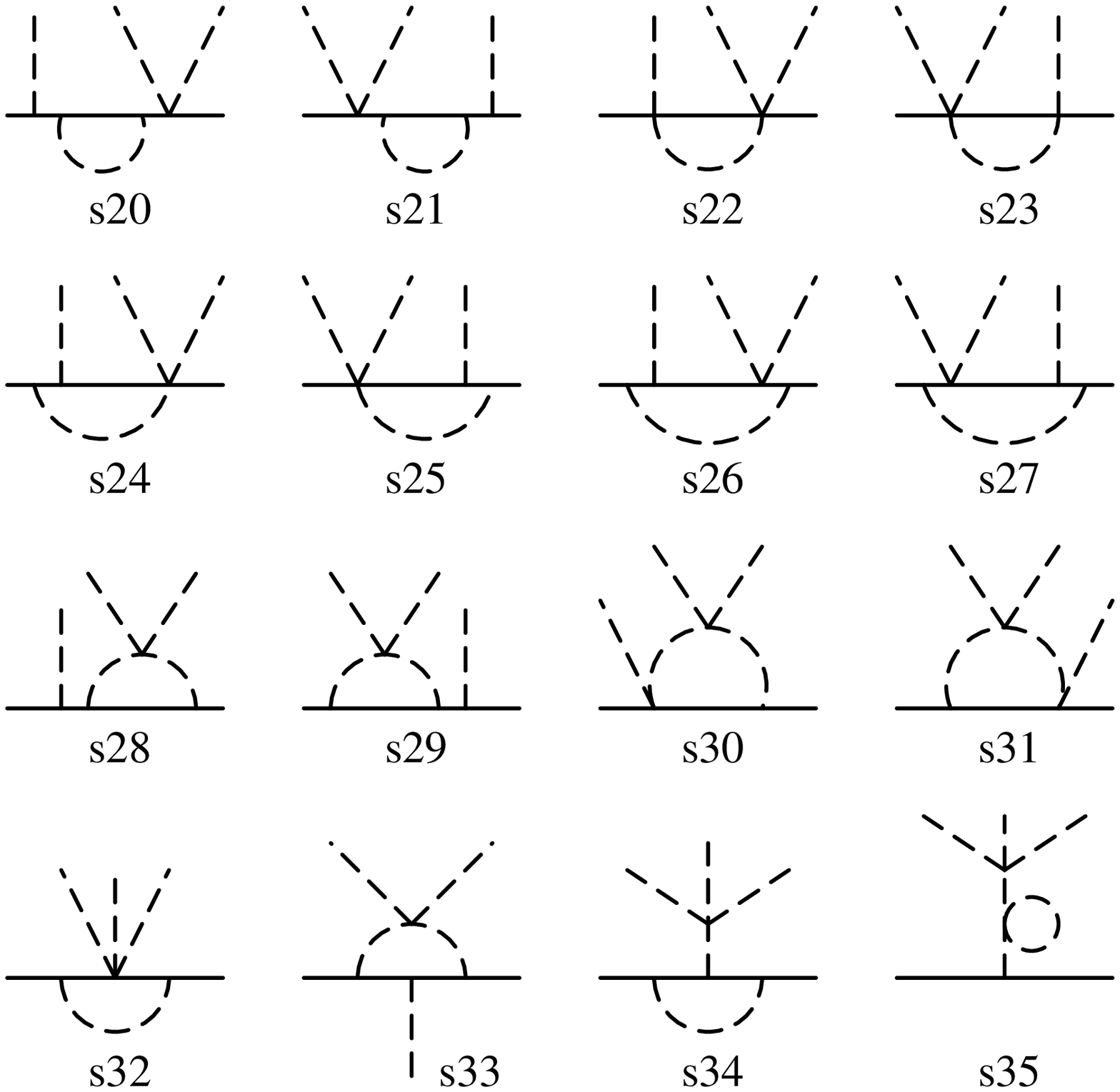}}
   \vspace{1cm}
   \centerline{\parbox{13cm}{\caption{\label{fig5}
Further one--loop graphs of the self--energy type.
  }}}
\end{figure}

\begin{figure}[h]
   \vspace{0.5cm}
   \epsfysize=18cm
   \centerline{\epsffile{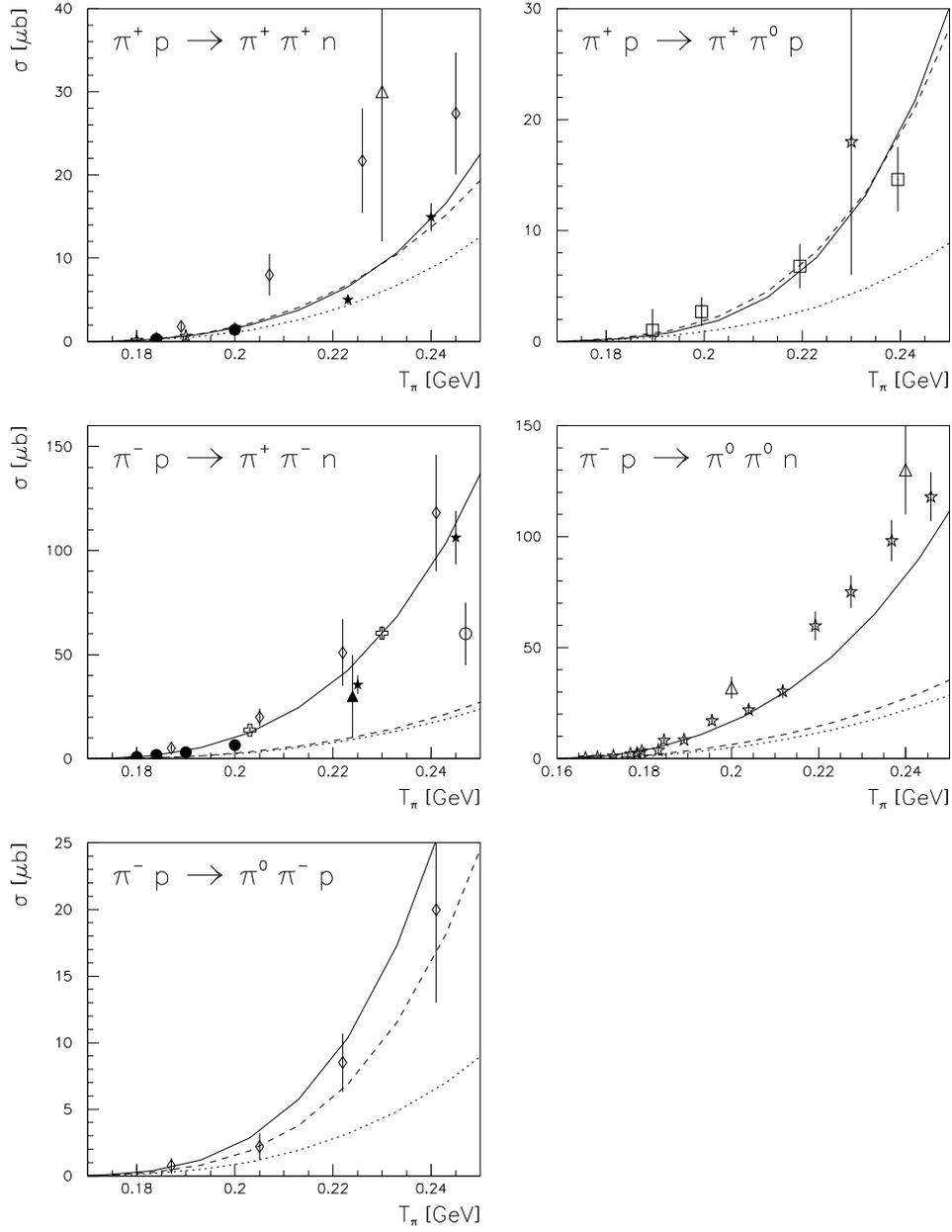}}
   \vspace{1cm}
   \centerline{\parbox{13cm}{\caption{\label{fig6}
Fits to the total cross sections (solid lines). The dashed and dotted
lines refer to the second and first order contributions, respectively.
  }}}
\end{figure}

\begin{figure}[h]
   \vspace{0.5cm}
   \epsfysize=18cm
   \centerline{\epsffile{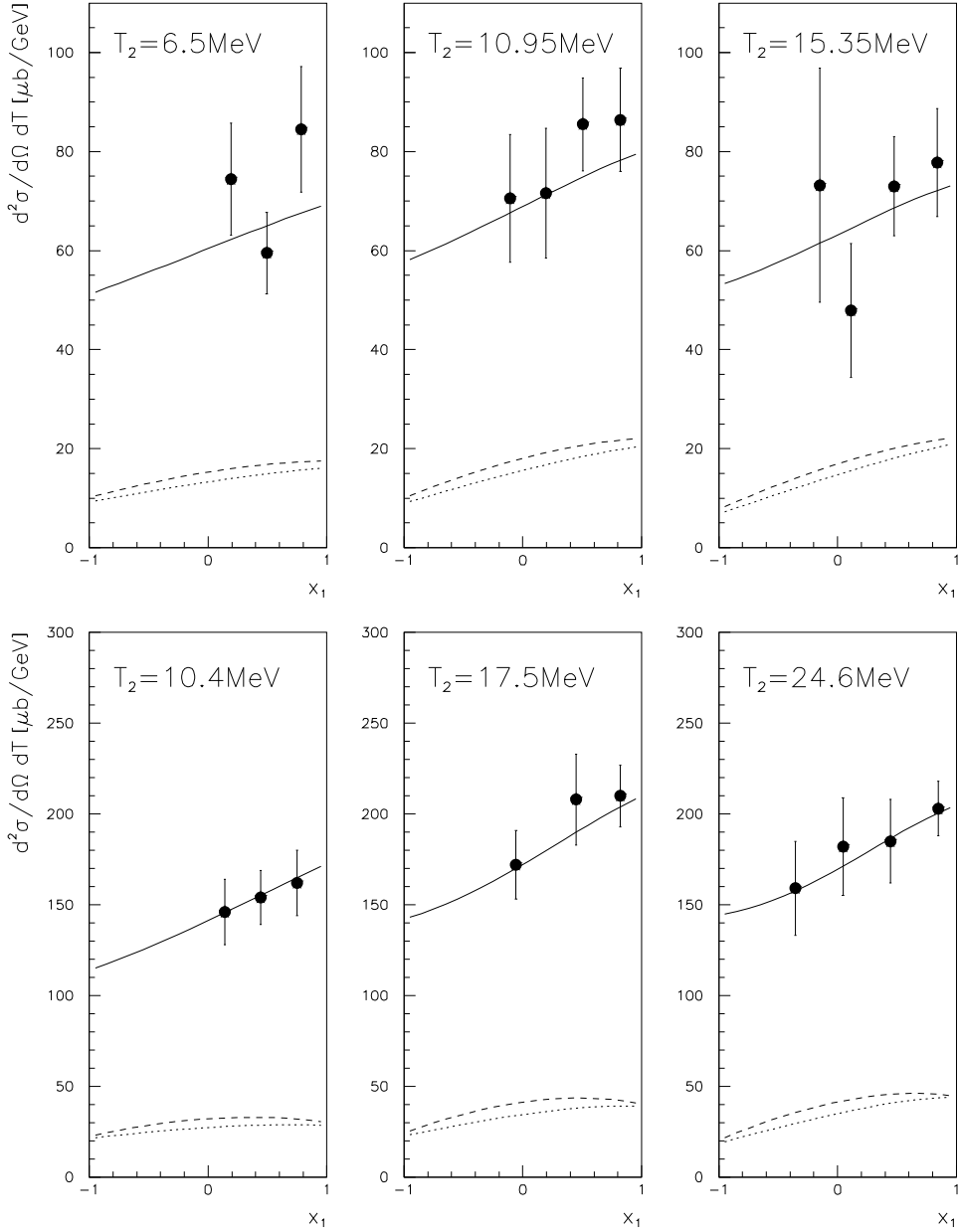}}
   \vspace{1cm}
   \centerline{\parbox{13cm}{\caption{\label{fig7}
Fits to the differential total cross sections for $\pi^- p\to
   \pi^+\pi^- n$ (solid lines)
with respect to the kinetic energy and of solid angle the outgoing $\pi^+$.
$T_2 =\omega_2 - M_\pi$ refers to the outgoing positively charged pion. 
For further notations, see fig.\ref{fig6}.
  }}}
\end{figure}

\begin{figure}[h]
   \vspace{0.5cm}
   \epsfysize=18cm
   \centerline{\epsffile{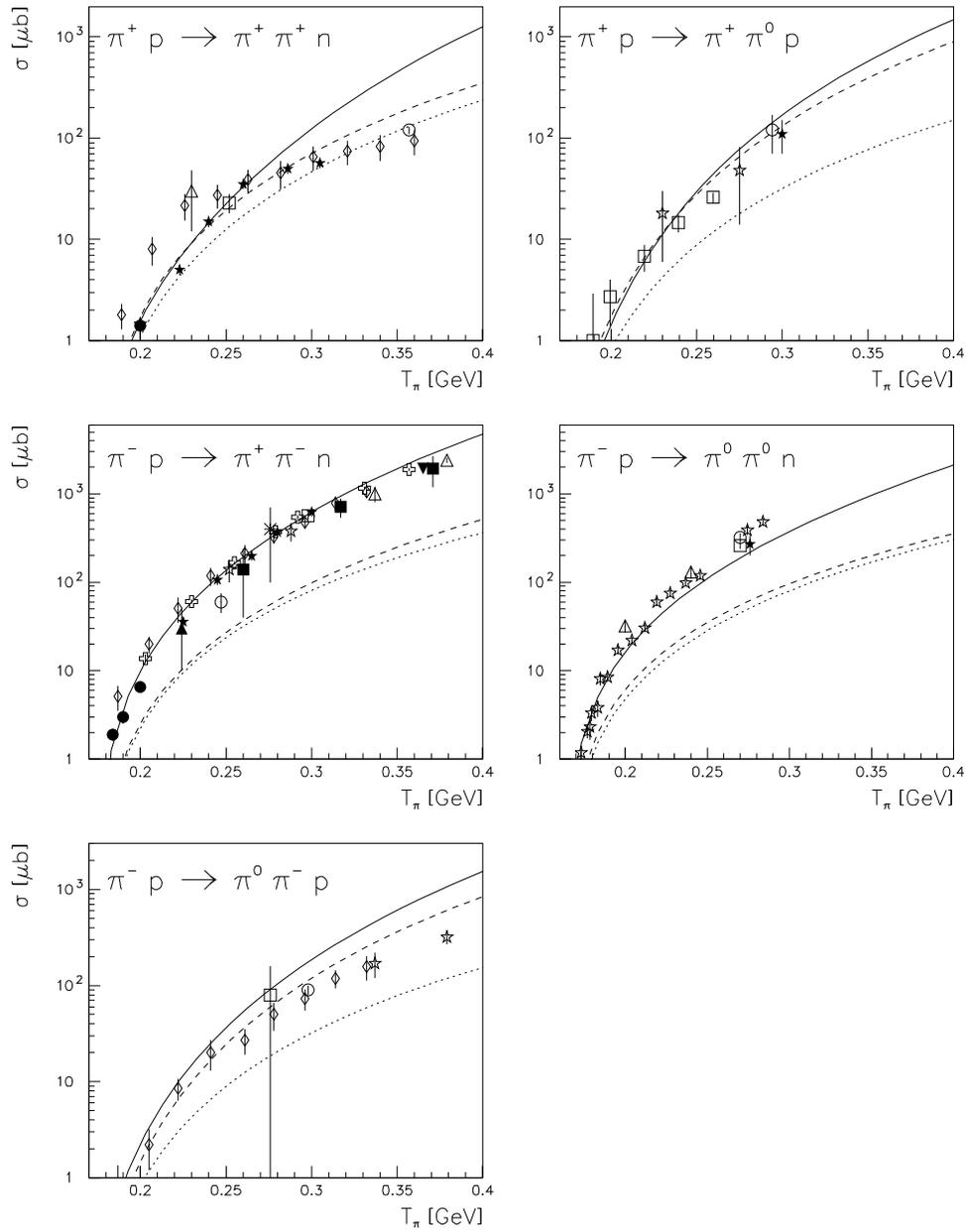}}
   \vspace{1cm}
   \centerline{\parbox{13cm}{\caption{\label{fig8}
Predictions for the total cross sections up to $T_\pi = 400\,$MeV (solid lines). 
For further notations, see fig.\ref{fig6}.
  }}}
\end{figure}

\begin{figure}[h]
   \vspace{0.5cm}
   \epsfysize=16cm
   \centerline{\epsffile{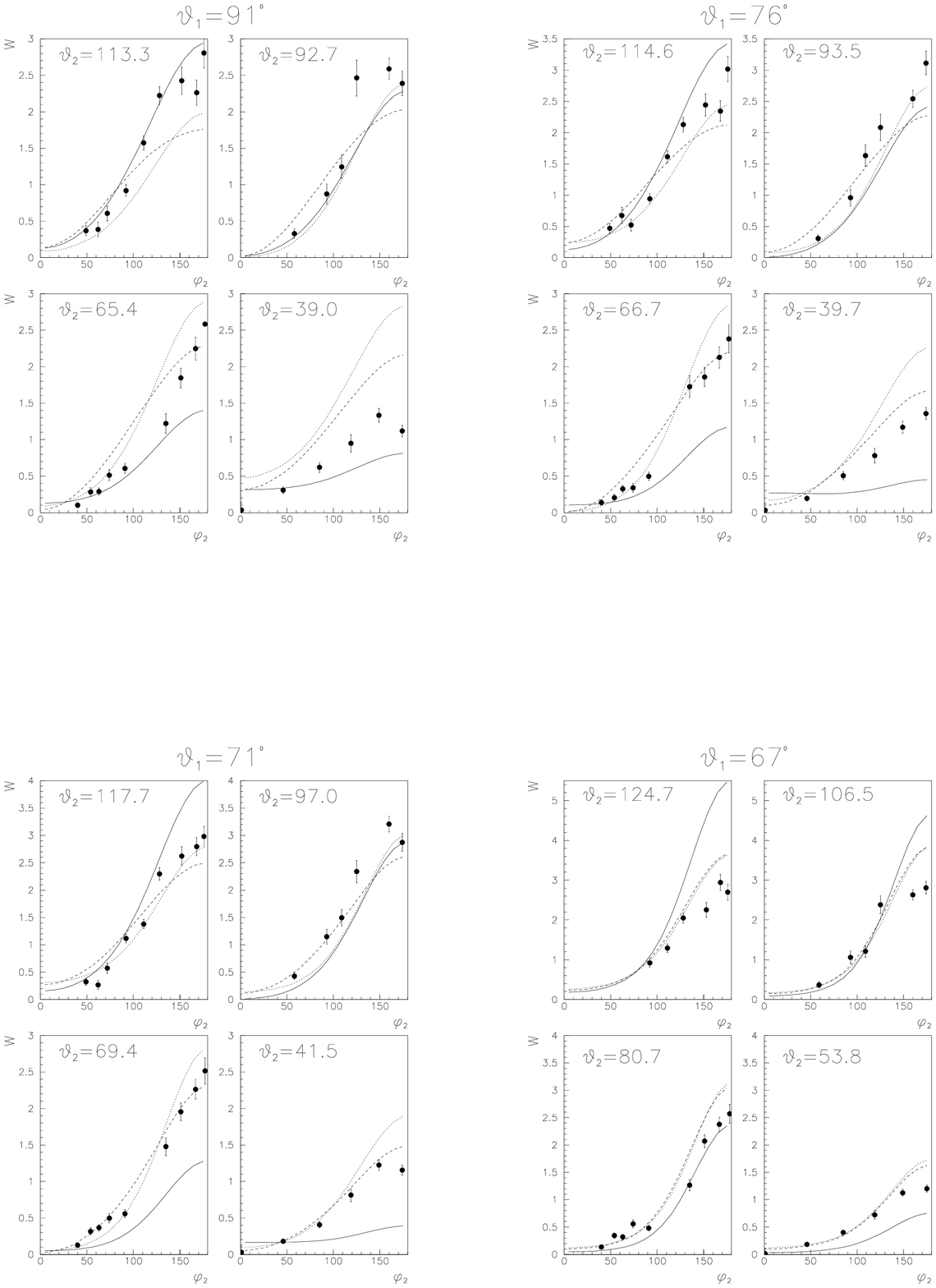}}
   \vspace{1cm}
   \centerline{\parbox{13cm}{\caption{\label{fig9}
Predictions for the angular correlation functions at fixed $\theta_2$.
For further notations, see fig.\ref{fig6}.
  }}}
\end{figure}

\begin{figure}[h]
   \vspace{0.5cm}
   \epsfysize=16cm
   \centerline{\epsffile{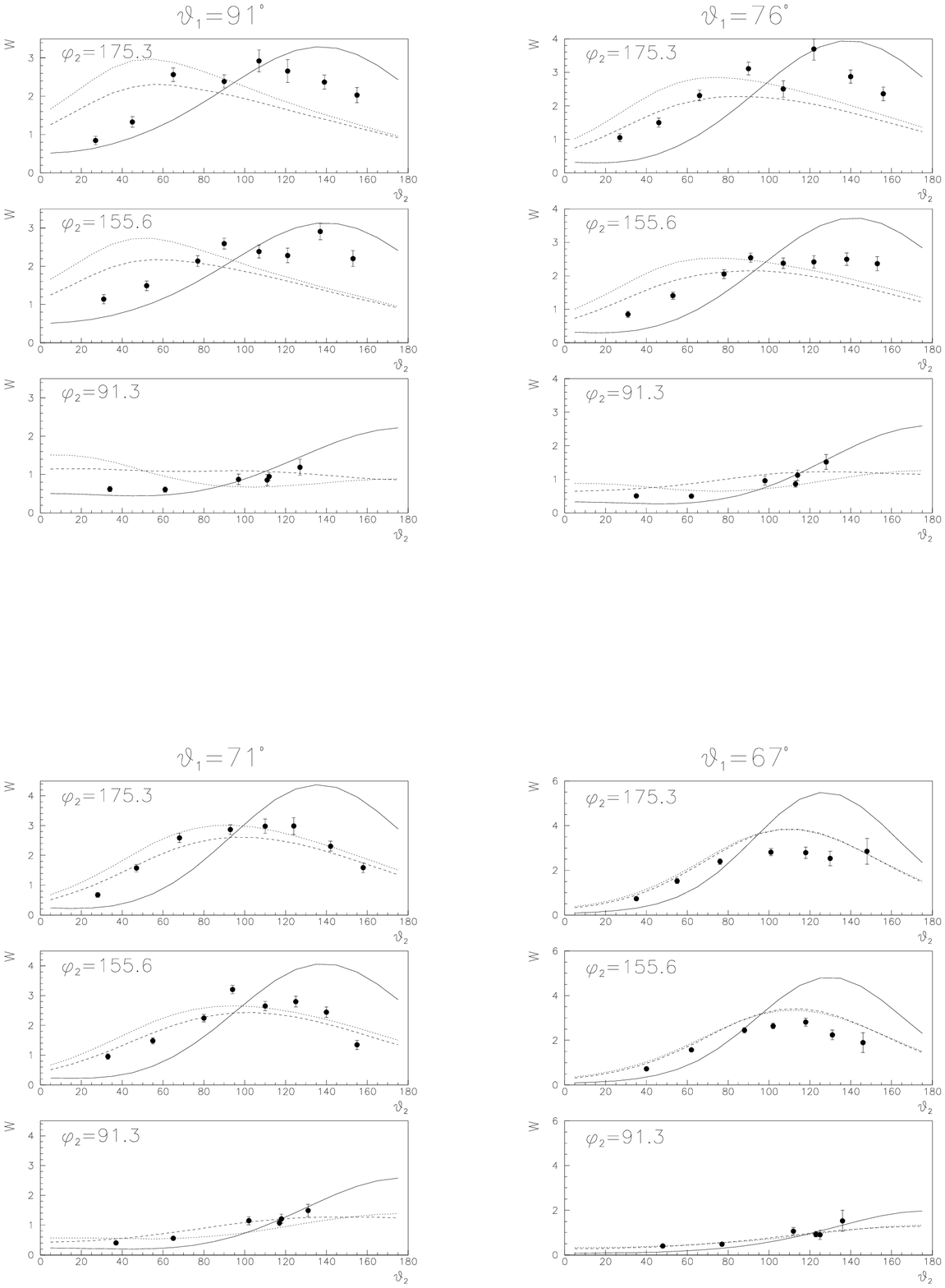}}
   \vspace{1cm}
   \centerline{\parbox{13cm}{\caption{\label{fig10}
Predictions for the angular correlation functions at fixed $\varphi_2$.
For further notations, see fig.\ref{fig6}.
  }}}
\end{figure}

\begin{figure}[h]
   \vspace{0.5cm}
   \epsfysize=16cm
   \centerline{\epsffile{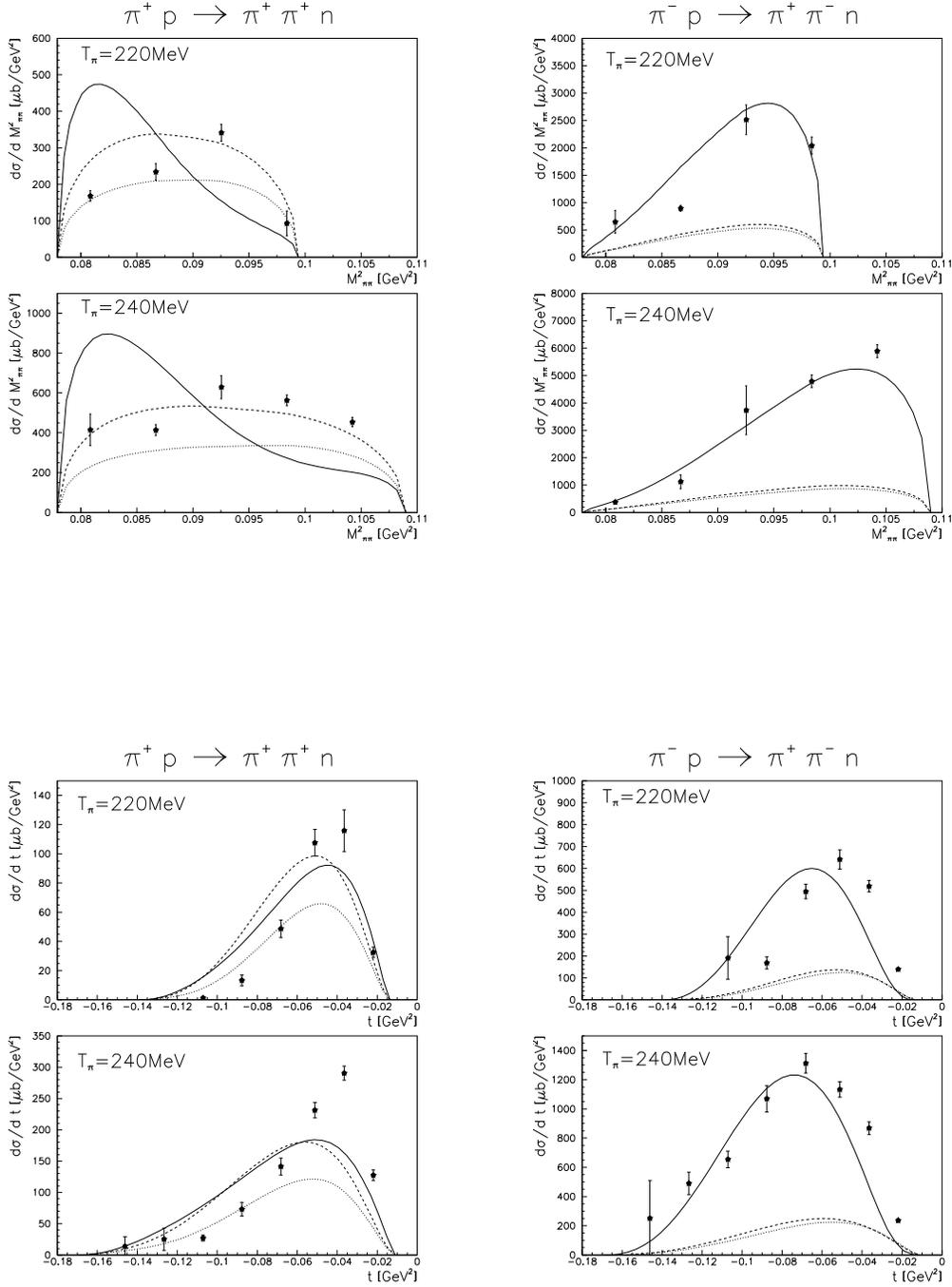}}
   \vspace{1cm}
   \centerline{\parbox{13cm}{\caption{\label{fig11}
Predictions for the differential cross sections $d\sigma / dM_{\pi\pi}^2$ and
$d\sigma / dt$. For further notations, see fig.\ref{fig6}.
  }}}
\end{figure}

%\begin{figure}[h]
%   \vspace{0.5cm}
%   \epsfysize=16cm
%   \centerline{\epsffile{corr2.ps}}
%   \vspace{1cm}
%   \centerline{\parbox{13cm}{\caption{\label{fig12}
%Angular correlation functions.
%  }}}
%\end{figure}

\begin{figure}[h]
   \vspace{0.5cm}
   \epsfysize=16cm
   \centerline{\epsffile{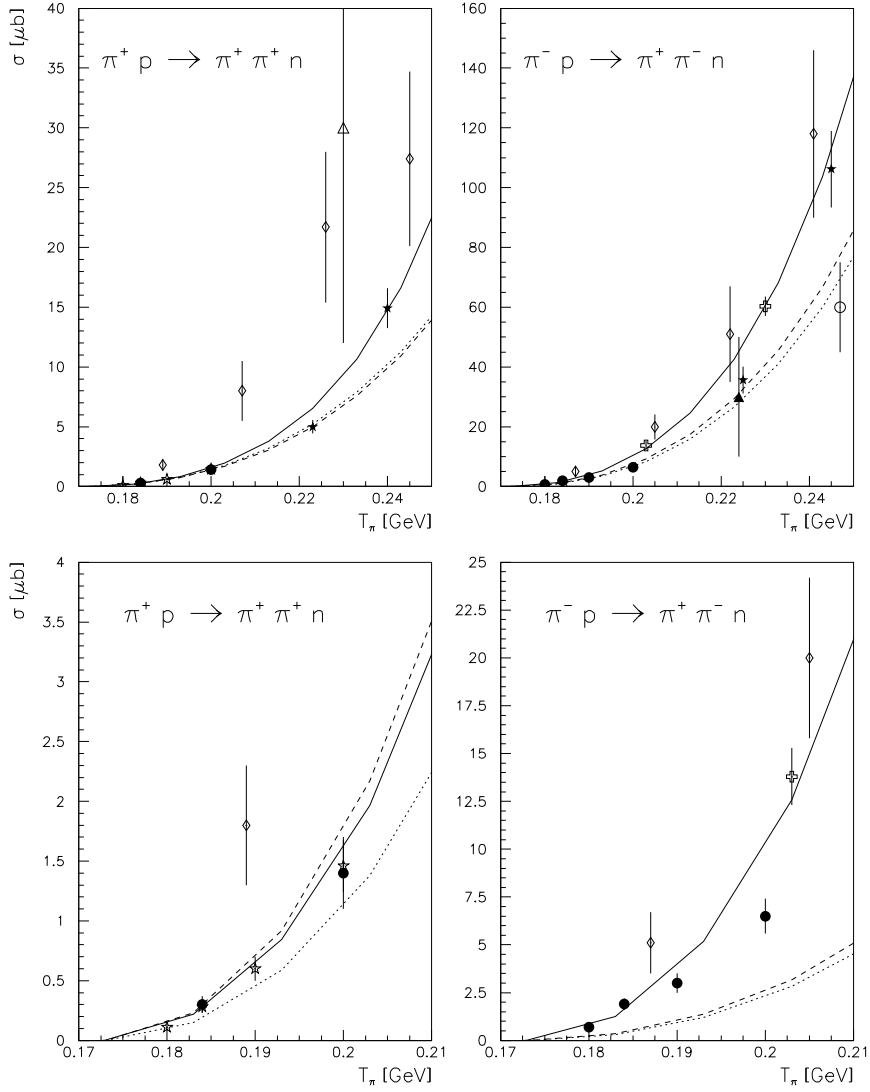}}
   \vspace{1cm}
   \centerline{\parbox{13cm}{\caption{\label{fig13}
Chiral expansion. In the upper two panels, the solid, dashed and dotted lines
refer to the contribution up  to third, second and first order respectively. In the
lower panels, the dashed lines show the third order contribution minus the
ones from the dimension three contact terms. Similarly, for the dotted lines
the loop contributions are in addition subtracted.
  }}}
\end{figure}

\begin{figure}[h]
   \vspace{0.5cm}
   \epsfysize=16cm
   \centerline{\epsffile{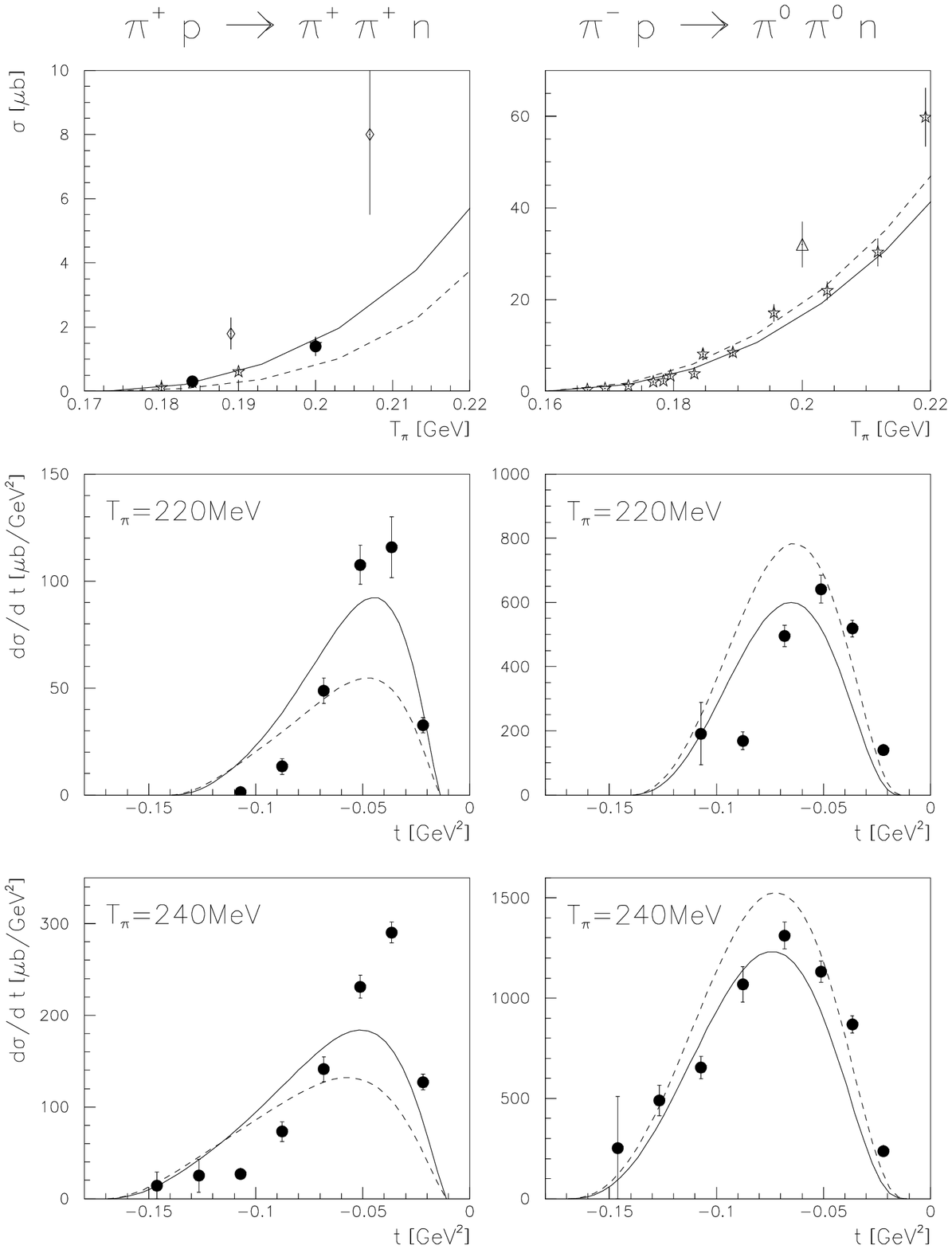}}
   \vspace{1cm}
   \centerline{\parbox{13cm}{\caption{\label{fig14}
Sensitivity to the pion--pion interaction. Considered are the two
channels $\pi^+ p \to  \pi^+\pi^+ n$ (left) and  $\pi^- p \to
\pi^0\pi^0 n$ (right). Upper panels: Total cross sections in the
threshold region. Middle panels: $d\sigma/dt$ at $T_\pi =
220\,$MeV. Lower panels: $d\sigma/dt$ at $T_\pi =  240\,$MeV. The
solid line refers to the standard case, whereas the dashed lines are
obtained by setting $\bar{\ell}_3 = -70$ and keeping all other LECs
fixed.
  }}}
\end{figure}


\begin{thebibliography}{99}
\baselineskip 10pt plus 1pt minus 1pt

\bibitem{orsay} J. Bijnens et al., Nucl. Phys. B508 (1997) 263;
(E) Nucl. Phys. B517 (1998) 639.\vs
\bibitem{bern} M. Knecht et al., Nucl. Phys. B471 (1996) 445.\vs
\bibitem{bkmrev}  V.~Bernard, N.~Kaiser, and Ulf-G.~Mei{\ss}ner,
                  Int. J. Mod. Phys. E4 (1995) 193.\vs
\bibitem{bering} J. Beringer, $\pi N$ Newsletter  7 (1992) 33.\vs
\bibitem{bkmplb} V. Bernard, N. Kaiser and Ulf-G.~Mei\ss ner, Phys. Lett.
B332 (1994) 415; \\ (E)  B338 (1994) 520.\vs
\bibitem{bkmppn} V.~Bernard, N.~Kaiser,  and Ulf-G.~Mei{\ss}ner,
               Nucl. Phys. B457 (1995) 147.\vs
\bibitem{bkmci}  V.~Bernard, N.~Kaiser, and Ulf-G.~Mei{\ss}ner,
                  Nucl. Phys. A619 (1997) 261.\vs
\bibitem{bmz} J. Zhang, N. Mobed and M. Benmerrouche, {\tt nucl-th/9806063}.\vs
\bibitem{em} G. Ecker and M. Moj\v zi\v s, Phys. Lett. B365 (1996) 312.\vs
%(E) (xxx).\vs
\bibitem{fms} N. Fettes, Ulf-G.~Mei\ss ner and S. Steininger,
 Nucl. Phys.  A640 (1998) 199.\vs
\bibitem{oset}E. Oset and M.J. Vicente--Vacas, Nucl. Phys.  A446
      (1985) 584.\vs
\bibitem{jaeck} O. J\"akel, H.W. Ortner, M. Dillig and
        C.A.Z. Vasconcellos, Nucl. Phys. A511 (1990) 733; 
        O. J\"akel,  M. Dillig and C.A.Z. Vasconcellos, Nucl. Phys. A541
        (1992) 675.\vs
\bibitem{jemi} T.S Jensen and A. Miranda, Phys. Rev. C55 (1997) 1039.\vs
\bibitem{kermcl} M. Kermani et al., Phys. Rev. C58 (1998) 3431.\vs
\bibitem{lange} J.B. Lange et al., Phys. Rev. Lett. 80 (1998) 1597.\vs
\bibitem{bkmlec}  V.~Bernard, N.~Kaiser, and Ulf-G.~Mei{\ss}ner,
                  Nucl. Phys. A615 (1997) 483.\vs
\bibitem{kermani} M. Kermani et al., Phys. Rev. C58 (1998) 3419.\vs
\bibitem{gl84}J. Gasser and H.Leutwyler, Ann. Phys. (NY) 158 (1984) 142.\vs
\bibitem{jm} E. Jenkins and A.V. Manohar, Phys. Lett. B255 (1991) 558.\vs
\bibitem{bkkm} V. Bernard, N. Kaiser, J. Kambor and Ulf-G. Mei\ss ner,
Nucl. Phys. B388 (1992) 315.\vs
\bibitem{mm} M. Moj\v zi\v s, Eur. Phys. J. C2 (1998) 181.\vs
\bibitem{fmsz} N. Fettes, Ulf-G.~Mei\ss ner and S. Steininger,
 JHEP 9809 (1998) 008.\vs
\bibitem{ecker} G. Ecker, Phys. Lett.  B336  (1994) 508.\vs
\bibitem{LECval} J. Bijnens, G. Colangelo and J. Gasser,
  Nucl. Phys. B427 (1994) 417.\vs
\bibitem{se91} M.~Sevior et al., Phys. Rev. Lett. 66  (1991) 2569.\vs
\bibitem{ke90} G. Kernel et al., Z.~Phys.  C48 (1990) 201.\vs
\bibitem{kr76} A.V. Kratsov et al., Leningrad Institute of Nuclear Physics
   preprint No. 209, 1976.\vs
\bibitem{ki62} J. Kirz, J. Schwartz and R.D. Tripp, Phys. Rev. 126
  (1962) 763.\vs
\bibitem{poc} D.~Po\v cani\'c et al.,  Phys. Rev. Lett. 72 (1993) 1156.\vs
\bibitem{ba75} Yu.A. Batsuov et al., Sov. J. Nucl. Phys. 21 (1975)
  162.\vs
\bibitem{ar72} M. Arman et al., Phys. Rev. Lett. 29 (1972) 962.\vs
\bibitem{ba63} V. Barnes et al., CERN Report 63-27, 1963.\vs
\bibitem{ke89} G. Kernel et al., Phys. Lett.  B216  (1989) 244; ibid
               B225 (1989) 198.\vs
\bibitem{bj80} C.W. Bjork et al., Phys. Rev. Lett. 44 (1980) 62.\vs
\bibitem{jo74} J.A. Jones, W.W.M. Allison and D.H. Saxon, Nucl. Phys.
               B83 (1974) 93.\vs
\bibitem{bl70} I.M. Blair et al., Phys. Lett.  B32  (1970) 528.\vs
\bibitem{sa70} D.H. Saxon, J.H. Mulvey and W. Chinowsky, Phys. Rev. D2
               (1970) 1790.\vs
\bibitem{ba65} Yu.A. Batsuov et al., Sov. J. Nucl. Phys. 1 (1965) 374.\vs
\bibitem{bl63} T.D. Blokhintseva et al., Sov. Phys. JETP 17 (1963) 340.\vs
\bibitem{de61} J. Deahl et al., Phys. Rev. 124 (1961) 1987.\vs
\bibitem{pe60} W.A. Perkins et al.,  Phys. Rev. 118 (1960) 1364.\vs
\bibitem{lo91a} J. Lowe et al., Phys. Rev. C44 (1991) 956.\vs
\bibitem{be80} A.A. Bel'kov et al., Sov. J. Nucl. Phys. 31 (1980) 96.\vs
\bibitem{be78} A.A. Bel'kov et al., Sov. J. Nucl. Phys. 28 (1978) 657.\vs
\bibitem{bu77} S.A. Bunyatov et al., Sov. J. Nucl. Phys. 25 (1977) 177.\vs
\bibitem{kr75} A.V. Kratsov et al.,Sov. J. Nucl. Phys. 20 (1975) 500.\vs 
\bibitem{manl} M.D. Manley, Phys.Rev. D30 (1984) 536.\vs
\bibitem{man2} M.D. Manley, Phys.Rev. D30 (1984) 904.\vs
\bibitem{mue} R. M\"uller et al., Phys. Rev. C48 (1993) 981.\vs
\bibitem{bohn} U. Bohnert, Thesis, University of Erlangen, 1993.\vs
\bibitem{malz} D. Malz, Thesis, University of Erlangen, 1989.\vs

\end{thebibliography}
\end{document}